%% file: dissertation.tex
\documentclass[phd,lith,english]{liuthesis}

\include{settings}

\department{Institutionen för teknik och naturvetenskap}
\departmentenglish{Department of Science and Technology}
\edition{1:1}

\titleenglish{Data-Driven Smart Maintenance of Historic Buildings}
\division{Division of Physics, Electronics and Mathematics}
\thesisnumber{2444}

\publicationyear{2025}
\isbnprint{8118}{059}{6}
\isbnpdf{8118}{060}{2}
\divanumber{456789}
\author{Zhongjun Ni}

\begin{document}

\chapterstyle{VZ43}


\include{intro}

\printbibliography[title={References}]


\end{document}

%% file: settings.tex
\usepackage[backend=biber,style=ieee,doi=false,hyperref]{biblatex}
\ifxetex 
\fi

\usepackage{amsmath,amssymb,amsfonts}
\usepackage{algorithmic}
\usepackage{graphicx}
\usepackage{booktabs}
\usepackage{multirow}
\usepackage{listings}
\usepackage[font={sf,scriptsize},
            labelfont={bf}, 
            caption=false]
           {subfig}
\usepackage{bm}
\usepackage{arydshln}
\usepackage[ruled, linesnumbered]{algorithm2e}

\usepackage{calc,color}
\newif\ifNoChapNumber

\makeatletter
\makechapterstyle{VZ43}{

}
\makeatother



%% file: intro.tex

\chapter{Introduction}
\label{cha:introduction}

Digital transformation is reshaping how buildings are monitored, operated, and maintained. In this process, digital technologies such as Internet of Things (IoT), cloud computing, edge computing, digital twins, and artificial intelligence (AI) play an essential role. These technologies enable real-time data collection, advanced analytics, and intelligent decision-making. These advances offer new opportunities for proactive and efficient building management by enabling rapid actions and deeper insights. Historic buildings, as unique cultural assets, require continuous preservation to protect their historical, architectural, and social values. Sustainable management of these buildings must balance conservation goals with functional needs, such as providing public access to cultural heritage. This thesis investigates how digital technologies can be integrated to create a data-driven solution to support smart maintenance of historic buildings. This chapter introduces the background, motivation, research scope, and structure of the thesis.

\section{Background}
\label{sec:background}

This section introduces the concept of digital transformation and its relevance to the maintenance of historic buildings. It begins by outlining recent developments in the digitalization of the built environment and the role of key enabling digital technologies. Then, it highlights the importance of historic buildings as cultural heritage assets and the need to preserve their historical, architectural, and social values. The section concludes by reviewing current maintenance practices, identifying common limitations, and emphasizing the growing need for more proactive, data-driven approaches.

\subsection{Digital Transformation in the Built Environment}

Digital transformation involves the integration of digital technologies to fundamentally reshape how organizations operate~\cite{vial_understanding_2019}. As buildings account for 30\% of global final energy use and 26\% of energy-related carbon dioxide (CO\textsubscript{2}) emissions~\cite{iea_2025}, digital transformation plays a critical role in improving operational efficiency and reducing environmental impact. In the built environment, digital transformation enables more efficient and sustainable building operations throughout the life cycle by improving energy performance, occupant comfort, and system responsiveness~\cite{jia_adopting_2019}. For example, heating, ventilation, and air-conditioning (HVAC) systems can be adjusted based on occupancy levels, and lighting systems can respond to natural daylight~\cite{jia_adopting_2019}. Moreover, data from heterogeneous sources, such as occupancy and outdoor weather, can be integrated to support more informed and effective decision-making.

The ongoing digital transformation is driven by technologies such as IoT, cloud computing, edge computing, digital twins, and AI. IoT supports real-time data collection and control by connecting sensors, actuators, and building management systems, allowing intelligent operation of HVAC, lighting, and other subsystems~\cite{gubbi_internet_2013, balaji_brick_2018}. Cloud computing offers scalable resources for processing large volumes of data~\cite{liu_methodology_2021}, while edge computing enhances system responsiveness, resilience to network disruptions, and data privacy by processing information locally near the source~\cite{kong_edge_2023}. Digital twins provide dynamic virtual models of physical buildings, supporting real-time monitoring, predictive maintenance, and simulation-based optimization~\cite{ni_improving_2021, jiang_digital_2021}. AI techniques, including machine learning and deep learning, analyze data streams to support intelligent decision-making, maintain indoor environmental quality, and improve energy efficiency~\cite{zhang_review_2021}. Together, these technologies form the foundation of smart buildings, where digital systems manage building operations across the life cycle to enhance comfort, efficiency, and cost-effectiveness~\cite{jia_adopting_2019}.

\subsubsection{IoT, Cloud Computing, and Edge Computing}

IoT refers to the interconnection of physical objects equipped with sensing, processing, and communication capabilities, enabling them to collect data, exchange information, and respond to their environment~\cite{gubbi_internet_2013, sisinni_industrial_2018}. In the built environment, IoT is a key enabler of digital transformation by linking physical building components, users, and services through intelligent digital interfaces~\cite{atta_digital_2020}. This connection links the physical environment with its digital representation, allowing ongoing monitoring, real-time data processing, and remote system control. A typical IoT system is a comprehensive infrastructure that includes wired or wireless sensors and actuators, communication network connectivity, cloud services, data analytics capabilities, and user interfaces~\cite{jia_adopting_2019}. These components are typically organized into three layers: the perception layer, which collects physical data through sensors and actuators; the network layer, which transmits data using protocols such as Wi-Fi, Bluetooth, Zigbee, or long-term evolution (LTE); and the application layer, which supports analytics, visualization, and control functions~\cite{jia_adopting_2019}. In smart buildings, IoT supports various applications such as HVAC optimization, energy management, occupancy detection, and comfort enhancement~\cite{minoli_iot_2017}.

While IoT facilitates continuous data collection, cloud computing provides scalable resources for storing and processing large volumes of data~\cite{liu_methodology_2021}. Public cloud platforms, such as Amazon Web Services~\cite{amazon_aws_2021}, Microsoft Azure~\cite{microsoft_azure_2021}, and Google Cloud~\cite{google_cloud_2021}, offer flexible, on-demand computing resources, making advanced digital services more accessible across a wide range of applications. These platforms allow integrating sensors, storage systems, analytics engines, and user interfaces into a unified architecture that supports remote access and real-time monitoring. Moreover, cloud services reduce operational costs through pay-as-you-go pricing models, lowering entry barriers for small organizations~\cite{hua_edge_2023}. They also support modular service development, allowing stakeholders to build and deploy specialized tools, such as AI-powered analytics, simulations, and visualization applications, within a shared ecosystem.

Edge computing complements cloud computing by addressing key limitations of cloud-based IoT systems, including high latency, bandwidth constraints, privacy concerns, and transmission costs~\cite{kong_edge_2023, hua_edge_2023}. While cloud platforms offer centralized resources for large-scale computation and storage, they may not meet the low-latency and real-time demands of many smart building applications. Edge computing addresses this by moving data processing closer to the source, typically at or near the sensing device, thus reducing response time, improving data privacy, and decreasing network load~\cite{shi_edge_2016}. Edge computing has been applied in time-sensitive tasks like indoor localization~\cite{liu_edge_2017}, intrusion detection~\cite{dhakal_machine_2017}, and resource management~\cite{singh_gru_2024}. 

\subsubsection{Digital Twins and Ontology-based Data Modeling}

The concept of digital twins refers to virtual representations of physical systems that remain continuously synchronized with their real-world counterparts~\cite{grieves_digital_2014}. By integrating real-time data with computational models, digital twins enable continuous monitoring, diagnostics, simulation, and data-driven decision-making~\cite{wright_how_2020}. They are particularly effective for systems that evolve over time, as they allow for the early detection of deviations, faults, or degradation~\cite{wright_how_2020}. Initially developed for industrial applications such as product design, manufacturing, and predictive maintenance~\cite{tao_digital_industry_2019, delgado_digital_2021}, digital twins have since been adopted across a range of domains, including aerospace, energy, civil infrastructure, and healthcare~\cite{liu_novel_2019, glaessgen_digital_2012, bp_twin_2018, jiang_digital_2021}.

In the built environment, digital twins typically consist of three interconnected components: a virtual representation of the building, real-time data streams from physical systems, and synchronization mechanisms to keep the digital and physical counterparts aligned~\cite{wright_how_2020}. Various modeling approaches have been used, including geometric, informational, and parametric models~\cite{yang_review_2020, luo_overview_2021}. Geometric and information models are mainly static and commonly used for documentation and restoration~\cite{delgado_digital_2021}. In contrast, parametric models incorporate dynamic attributes that reflect functional behaviors and environmental conditions, enabling advanced analytics and supporting preventive maintenance strategies~\cite{luo_overview_2021}. Digital twins have been applied to use cases such as indoor climate monitoring~\cite{khajavi_digital_2019, rosati_air_2020} and anomaly detection in building systems~\cite{lu_digital_2020}. However, the lack of standardized data formats across implementations limits interoperability, scalability, and the transferability of digital twin systems between buildings and platforms~\cite{zhang_automatic_2022, shahzad_digital_2022}.

To address these limitations, ontology-based data modeling has been proposed to support semantically consistent and interoperable digital twins~\cite{boje_towards_2020, balaji_brick_2018}. Ontologies formally define domain-specific entities, their properties, and the relationships among them~\cite{gruber_toward_1995}, providing a structured and machine-readable data schema for digital systems. In the context of buildings, where data originate from multiple stakeholders using heterogeneous tools and formats~\cite{acierno_architectural_2017, noor_modeling_2019}, ontologies enable coherent integration and interpretation of fragmented information. This semantic alignment is essential for scaling digital twin applications and facilitating effective knowledge exchange across disciplines~\cite{noor_modeling_2019}.

\begin{figure}[!tb]
\includegraphics{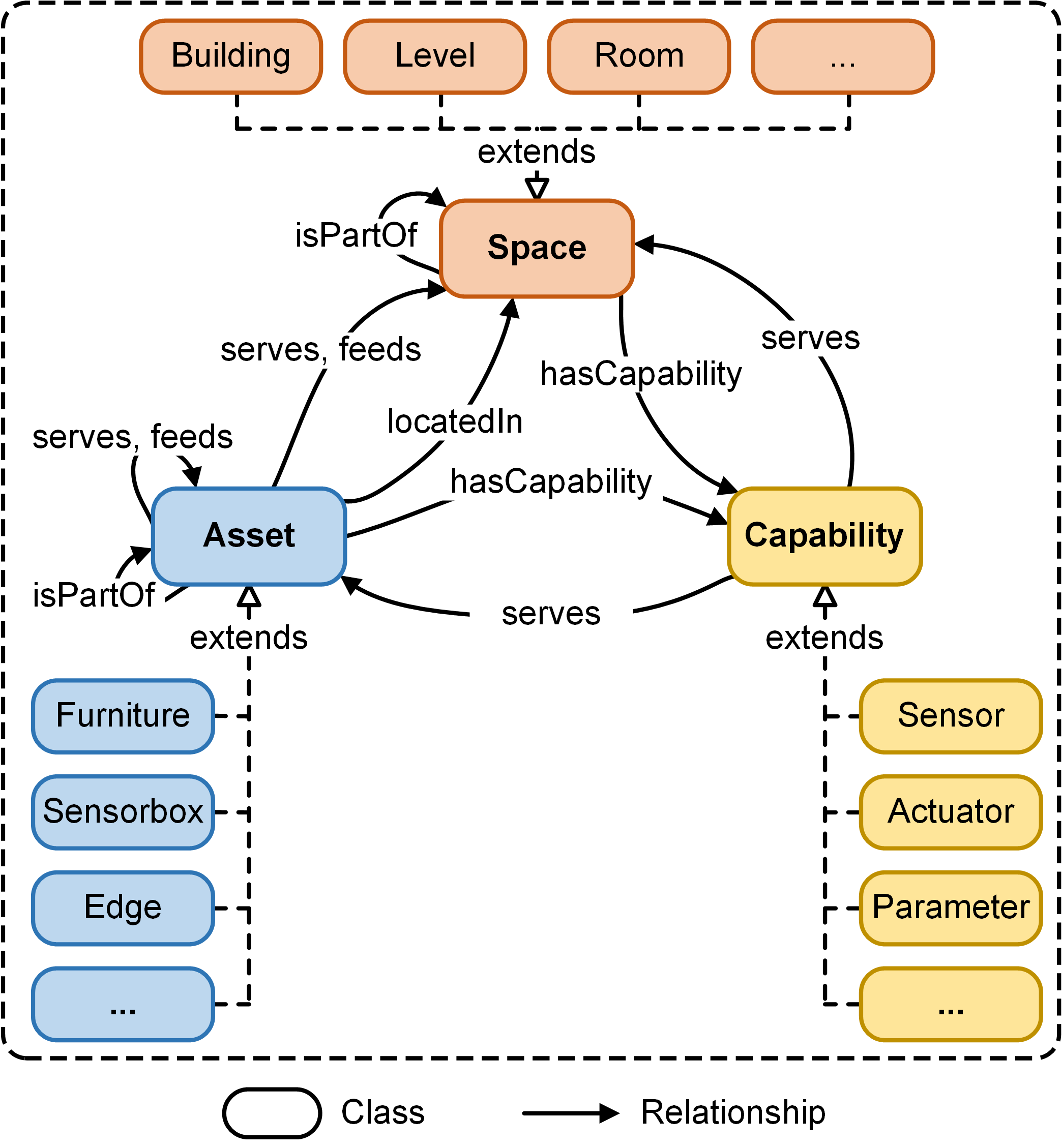}
\centering
\caption{Main classes and relationships in an extended ontology based on RealEstateCore~\cite{ni_enabling_2022}. The ontology organizes entities into three core classes: Space, Asset, and Capability. It also defines semantic relationships between them to support consistent data integration and interpretation.}\label{fig:phd_enabling_fig5}
\end{figure}

Several domain-specific ontologies have been developed to support semantic modeling in the built environment. Brick~\cite{balaji_brick_2018} defines a metadata schema for representing equipment, sensors, and subsystems, enabling consistent tagging and facilitating portable applications across platforms. RealEstateCore~\cite{hammar_realestatecore_2019} focuses on the semantic representation of real estate assets, supporting integration across cloud and edge infrastructures. Fig.~\ref{fig:phd_enabling_fig5} illustrates an extended ontology~\cite{ni_enabling_2022} based on RealEstateCore, where building-specific entities are categorized into three main classes and connected through defined relationships. Both Brick and RealEstateCore are compatible with the Digital Twins Definition Language (DTDL)~\cite{azure_dtdl_2022}, which is used to describe digital models of devices, assets, spaces, and environments. This compatibility enables seamless deployment of digital twin applications in cloud-based systems. Notably, there is an ongoing effort to align Brick and RealEstateCore, promoting interoperability between these leading metadata standards.

The use of ontologies is especially valuable when integrating digital twins with technologies such as IoT, edge computing, and cloud platforms. Ontologies enable semantic interoperability across heterogeneous systems, making it possible to interpret and manage both time-series and contextual data consistently~\cite{gyrard_building_2018, kaed_forte_2017}. Graph-based data models, which are commonly used in ontological structures, are well suited for representing the complex and evolving relationships between sensors, spaces, systems, and control logic~\cite{kleppmann_designing_2017}. This capability is particularly important in smart buildings, where contextual awareness and adaptability are essential for real-time control, system monitoring, and informed long-term decision-making~\cite{noor_modeling_2019}.

\subsubsection{AI, Machine Learning, and Deep Learning}

AI enables buildings to adapt, learn, and improve performance over time. This capability, often called building intelligence, relies on analyzing past behavior, adjusting to changing conditions, and making data-informed decisions that enhance operational efficiency~\cite{alanne_overview_2022}. With ongoing digitalization, buildings generate large volumes of data from sensors, meters, and automation systems~\cite{runge_review_2021}. In 2020, it was estimated that 37 zettabytes of data were generated worldwide~\cite{smart_buildings_data_online_2024}. These continuous data flows form the basis for building data-driven models that enable more efficient building operation and control of systems and equipment~\cite{gunay_data_2019, zhang_review_2021}. AI techniques, particularly machine learning and deep learning, facilitate the transformation of raw data into actionable insights.

Machine learning offers scalable tools for extracting insights from building data. Supervised learning methods are widely used for time-series forecasting, classification, and regression tasks~\cite{alanne_overview_2022}. Techniques such as artificial neural networks (ANNs), support vector regression, decision trees, and random forests have been applied to predict energy consumption, indoor environmental conditions, and equipment performance~\cite{amasyali_review_2018, somu_deep_2021, arulmozhi_machine_2021}. These methods can handle nonlinear relationships and incorporate multiple input features, making them well-suited for the aforementioned complex applications.

Deep learning has emerged as a powerful extension of traditional machine learning, offering advanced capabilities for modeling high-dimensional data and capturing long-term temporal dependencies~\cite{runge_review_2021}. Open-source frameworks such as PyTorch~\cite{paszke_pytorch_2019} have made it easier to develop and deploy deep learning models. Various architectures have been adopted for time series forecasting, including recurrent neural networks (RNNs) such as long short-term memory (LSTM) and gated recurrent units (GRUs)~\cite{lara_temporal_2020}, temporal convolutional networks (TCNs)~\cite{wang_short-term_2021}, and Transformer-based networks~\cite{wang_transformer_2022, dong_short-term_2023}. While point forecasting remains the most common approach, there is increasing attention to probabilistic forecasting to capture predictive uncertainty and support risk-aware decision-making~\cite{oneill_development_2016}.

\subsubsection{Federated Learning for Privacy-Preserving Data Analytics}

Federated learning (FL) is a decentralized machine learning paradigm that allows multiple clients, such as buildings, devices, or organizations, to collaboratively train a shared model without transferring raw data~\cite{mcmahan_communication_2017, kairouz_advances_2021}. In contrast to conventional methods that require centralized data aggregation, FL retains data at the local level and shares only model updates or parameters. Such an approach is especially valuable in privacy-critical environments like smart buildings, where operational data can expose patterns related to occupancy, equipment operation, or user behavior~\cite{guo_towards_2020}. FL has been applied to a range of building-related tasks, including HVAC optimization~\cite{guo_towards_2020}, indoor climate forecasting~\cite{norouzi_applicability_2023}, residential load prediction~\cite{fernandez_privacy_2022}, and building automation~\cite{perifanis_federated_2023}. As sensor deployments continue to grow, FL provides a scalable approach to decentralized analytics that supports data protection regulations and preserves local data ownership.

\begin{figure}[!tb]
\includegraphics{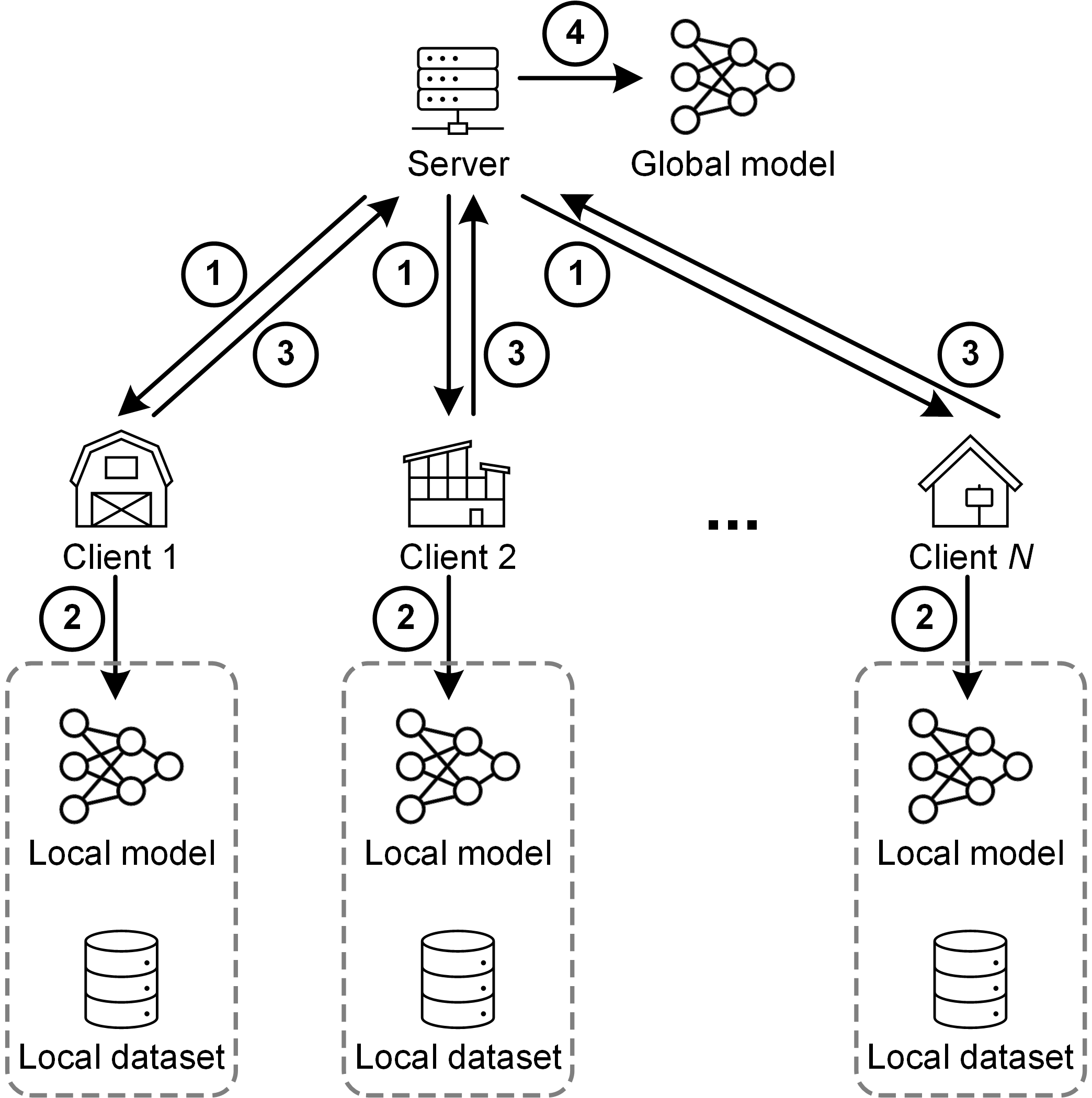}
\centering
\caption{An illustration of federated learning (FL) for training a forecasting model~\cite{ni_federated_2025}. The process typically involves several communication rounds between a central server and multiple clients. Each round consists of four steps. (\textbf{1}) distributes the current global model parameters to participated clients. (\textbf{2}) each client trains the model locally using its private dataset. (\textbf{3}) clients send their locally updated model parameters back to the server. (\textbf{4}) the server aggregates the updates to refine the global model without accessing any local dataset.}\label{fig:phd_federated_fig1}
\end{figure}

FL can be classified into three main types: horizontal, vertical, and transfer learning, based on how data and features are distributed across clients~\cite{yang_federated_2019}. In building analytics, horizontal FL is most commonly used. In this setting, clients share similar feature spaces (e.g., indoor temperature and relative humidity) but collect data from different physical environments. Several frameworks and tools support the practical deployment of FL. For instance, BuildFL~\cite{guo_towards_2020} implements a parameter server architecture in which clients independently train local models and share parameter updates for global aggregation. Flower~\cite{beutel_flower_2020} provides a flexible infrastructure for both research and production environments. A typical FL workflow consists of several communication rounds between a central server and local clients. As illustrated in Fig.~\ref{fig:phd_federated_fig1}, each round begins with the server distributing a global model to clients. Clients then train the model locally on their private datasets and return model updates. The server aggregates these updates to refine the global model without accessing raw data~\cite{mcmahan_communication_2017}. 

However, FL also introduces several technical challenges. Data distributed across clients are often non-independent and identically distributed (non-IID), imbalanced in volume, and subject to bandwidth limitations~\cite{li_federated_learning_2022}. These issues can negatively affect model convergence and overall performance. To mitigate such problems, researchers have proposed various solutions, including personalized local adaptation, client clustering, and differential privacy techniques~\cite{khalil_federated_2021, fernandez_privacy_2022}. These strategies enhance the robustness, scalability, and privacy-preserving capabilities of FL, making it a practical approach for smart building analytics.

\subsection{Historic Buildings and Heritage Conservation}

Historic buildings are important cultural assets that embody the architectural, technological, and social achievements of past societies. Their values often lie in associations with historical events, influential figures, or distinctive design innovations~\cite{historic_2022}. In addition to their symbolic and aesthetic significance, these buildings support community identity, cultural continuity, and social cohesion. Many historic buildings are still in active use as museums, libraries, places of worship, or residences, and often house collections of high cultural or intellectual value~\cite{doerr_ontologies_2009}. As physical records of history, historic buildings are as valuable as the artifacts they house. Preserving them is essential for protecting cultural knowledge and promoting heritage appreciation across generations~\cite{ashrae_handbook_hvac_applications_2011}.

Despite their values, historic buildings face multiple threats. Many suffer from aging infrastructure, delayed maintenance, and inadequate environmental controls~\cite{camuffo_environmental_2001, camuffo_microclimate_2002}. Global climate change has worsened these problems by increasing temperature and humidity fluctuations, which accelerate material degradation~\cite{zarzo_long_2011, camuffo_microclimate_2019}. Meanwhile, an unstable indoor climate can lead to dimensional changes in hygroscopic materials, corrosion in metals, biological growth, and chemical deterioration~\cite{en_15757_conservation_2010, ashrae_handbook_hvac_applications_2011}. In buildings open to the public, human activities, such as visitor movement and ventilation disturbances, introduce further variability, complicating indoor climate control~\cite{camuffo_indoor_1999, yüksel_review_2021}.

Maintaining a stable indoor climate is critical for preserving sensitive materials and ensuring occupant comfort. Key environmental parameters include temperature, relative humidity (RH), and CO\textsubscript{2} concentration~\cite{leijonhufvud_standardizing_2018, en_15757_conservation_2010}. Standards such as EN 15757:2010~\cite{en_15757_conservation_2010} and EN 16242:2012~\cite{en_16242_conservation_2012} define acceptable ranges for these parameters based on material properties and historic climate conditions. Their goal is to minimize mechanical stress, deformation, and microbial growth. In practice, however, maintaining such stability is challenging. Indoor climate of a building is shaped by interactions between external weather, the building envelope, HVAC operation, and occupancy patterns~\cite{luciani_influence_2013, pisello_coupling_2018}. Managing these interdependent factors requires advanced monitoring and control strategies adapted to the specific characteristics of historic buildings.

\subsection{Maintenance Practices and Preventive Conservation}

Preventive conservation and smart maintenance are increasingly important in managing historic buildings. Preventive conservation refers to measures aimed at avoiding or slowing material deterioration through early detection, risk management, and routine maintenance~\cite{balen_preventive_2015, marra_combining_2021}. In contrast, traditional maintenance in historic buildings is often reactive. It typically relies on manual inspections and is only initiated after visible damage appears~\cite{forster_maintenance_2009}. This reactive approach is often the result of limited funding, unclear responsibilities, and a lack of awareness about the long-term value of preventive care~\cite{ashrae_handbook_hvac_applications_2011}. These limitations further highlight the importance of predictive maintenance, which uses real-time data and analytics to forecast failures and schedule interventions based on actual equipment condition~\cite{ferreira_impact_2021}.

Digital transformation has introduced advanced tools that facilitate environmental monitoring, data-driven decision-making, and early intervention in historic buildings. IoT sensors and wireless networks enable real-time tracking of indoor conditions, which is essential for identifying potential risks and responding promptly~\cite{colace_iot_2021}. Simulation-based methods allow for the assessment of moisture behavior, temperature distribution, and HVAC system performance without physically altering the building fabric~\cite{schellen_overview_2007, bay_assessment_2022, schijndel_application_2008}. These technologies also support occupancy-aware climate control, helping to reduce energy use while maintaining conditions that preserve materials and ensure occupant comfort~\cite{hammar_realestatecore_2019, silva_impact_2021}.

Despite the growing availability of digital tools, their adoption in maintenance of historic buildings remains limited. Many current systems depend entirely on centralized cloud infrastructure, which can introduce issues related to latency, platform dependency, and data privacy~\cite{hua_edge_2023}. Edge computing and lightweight machine learning models offer promising alternatives by enabling decentralized, low-latency analysis directly at the building level~\cite{wang_convergence_2020}. Although digital twins have demonstrated potential to improve monitoring accuracy and support preventive conservation~\cite{jouan_digital_2020, angjeliu_development_2020}, broader implementation depends on interdisciplinary collaboration, standardized data schemas, and user-friendly platforms that address the specific needs of heritage professionals.

\section{Motivation}
\label{sec:motivation}

The sustainable preservation of historic buildings presents complex technical and cultural challenges that require innovative, data-driven solutions. In this context, digital transformation opens new opportunities to support preventive conservation, improve energy efficiency, and enable informed decision-making through integrated sensing, analytics, and control. 

The motivation of this thesis lies in four aspects. First, there is a need for adaptable and scalable digital solutions to support long-term heritage preservation. Preventive conservation requires continuous monitoring of indoor environmental parameters. This requirement calls for robust data storage, real-time analytics, and visualization tools. Previous approaches based on offline loggers or private cloud servers often suffer from limited reliability, scalability, and responsiveness~\cite{zarzo_long_2011, guo_ima_2012, corbellini_cloud_2018}. Recent advances in public cloud and edge computing offer more flexible architectures for deploying IoT-based sensing systems tailored to heritage contexts~\cite{microsoft_azure_2021, kong_edge_2023}.

Second, it is essential to represent historic buildings using semantically rich and interoperable data models. Traditional documentation methods focus primarily on geometric or visual representations and do not support dynamic analysis or reasoning~\cite{yang_review_2020, bruno_historic_2018}. At the same time, the digitalization of heritage conservation is producing large volumes of heterogeneous data from diagnostics, monitoring systems, and maintenance records~\cite{noor_modeling_2019}. Without a shared data schema, these resources remain fragmented and difficult to integrate. Ontology-based data modeling provides a structured approach to unify this information, ensure semantic interoperability, and enable collaboration across disciplines and systems~\cite{balaji_brick_2018, hammar_realestatecore_2019, boje_towards_2020}.

Third, more accurate and interpretable energy forecasting techniques are needed, especially in public historic buildings with varying occupancy patterns and operational constraints. Accurate forecasts are critical for optimizing building operations, enabling demand-side management, and reducing environmental impacts~\cite{khalil_machine_2022, chen_probabilistic_2020}. However, most research focuses on residential,  office, or commercial buildings, while historic buildings remain underrepresented with their unique operational profiles and conservation priorities. Additionally, conventional models often rely on point forecasts~\cite{kim_sequence_2021}. Probabilistic forecasting techniques can better account for uncertainty and improve reliability in these complex environments~\cite{wen_multi-horizon_2017}.

Fourth, there is a need to improve indoor climate forecasting using advanced deep learning and privacy-aware analytics. Nevertheless, many existing predictive systems are centralized, raising privacy and latency concerns when sensitive operational data cannot be shared~\cite{guo_towards_2020}. FL presents a compelling solution by allowing model training to be carried out across distributed clients without sharing raw data~\cite{mcmahan_communication_2017}. When combined with edge computing, FL can support localized, real-time decision-making while addressing privacy, bandwidth, and connectivity constraints~\cite{kong_edge_2023, wang_convergence_2020}. However, FL-based data analytics are still largely unexplored in the heritage domain.

In summary, this thesis is driven by the need to develop scalable, interoperable, and intelligent digital solutions for the long-term conservation of historic buildings. By integrating IoT, cloud and edge computing, digital twins, ontology-based data modeling, deep learning, and FL, the research aims to advance smart maintenance and preventive conservation practices that are well-suited to the specific needs of cultural heritage environments.

\section{Aim and Objectives}
\label{sec:aim}

This thesis aims to advance data-driven smart maintenance of historic buildings. The research is grounded in digital transformation of the built environment and focuses on integrating emerging technologies, including IoT, cloud and edge computing, digital twins, ontology-based modeling, machine learning, deep learning, and FL, to create scalable, interoperable, and intelligent digital solutions tailored to the specific needs of heritage conservation. To achieve this goal, the thesis pursues four specific sub-goals:
\begin{itemize}
\item[\textbf{G1:}] Design and implement an IoT system for long-term preservation of historic buildings. This includes addressing challenges related to stable data storage, real-time analytics, and flexible system deployment using both cloud and edge computing infrastructures.
\item[\textbf{G2:}] Develop parametric digital twins using ontology-based data modeling to ensure semantic consistency, enable interoperability, and support discovery of insights from collected data.
\item[\textbf{G3:}] Investigate deep learning techniques for energy forecasting in public historic buildings. This involves developing models for both point and probabilistic forecasts, capable of generating one-step-ahead and multi-horizon predictions based on building-specific conditions and external variables.
\item[\textbf{G4:}] Enhance indoor climate forecasting by applying deep learning in combination with privacy-preserving methods such as FL. This approach supports decentralized model training across multiple buildings while protecting sensitive operational data.
\end{itemize}

\begin{figure}[!tb]
\includegraphics{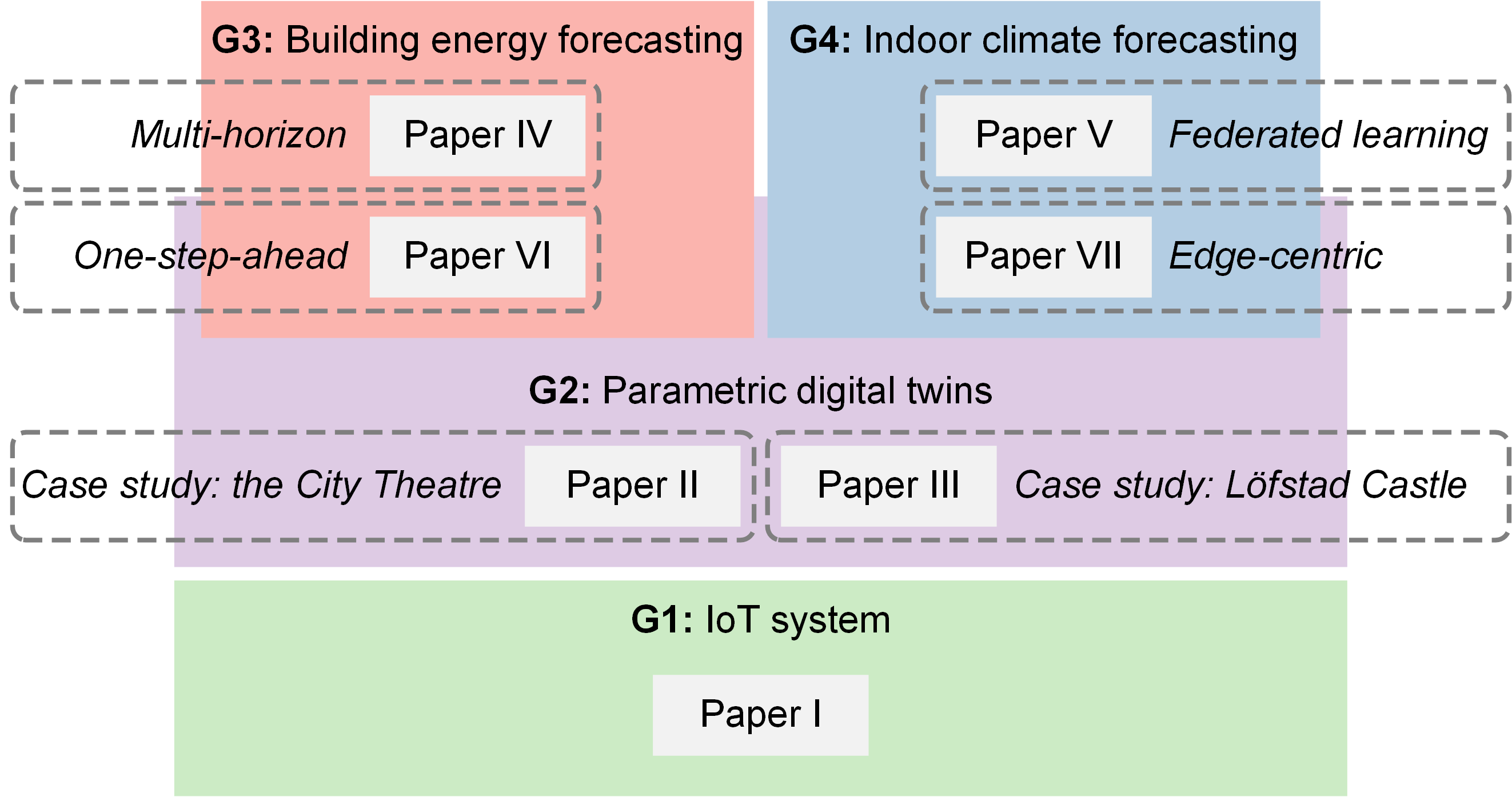}
\centering
\caption{Connections between included papers and their contributions to research objectives.}\label{fig:phd_fig2}
\end{figure}

Each of the seven included papers contributes to the achievement of the defined sub-goals. As illustrated in Fig.~\ref{fig:phd_fig2}, \textbf{Paper I} addresses \textbf{G1} by developing a cloud-based IoT sensing system for historic buildings, offering a flexible solution for stable data storage, real-time analytics, and user-friendly visualization. \textbf{Paper II} supports \textbf{G2} by demonstrating how IoT combined with ontology-based data modeling can be used to construct parametric digital twins. \textbf{Paper III} builds on this work by providing a detailed implementation and evaluation in a different historic building, showing how parametric digital twins can generate insights to support conservation decision-making. \textbf{Paper IV} addresses \textbf{G3} by applying advanced deep learning architectures for multi-horizon energy forecasting in public historic buildings. \textbf{Paper VI} advances this further by integrating deep learning models with parametric digital twins, illustrating how predictive analytics can be embedded within a unified data representation to enable intelligent energy management. \textbf{Paper V} contributes to \textbf{G4} by exploring FL for privacy-preserving, multi-horizon indoor climate forecasting. It evaluates various federated algorithms and deep learning models under real-world data heterogeneity. \textbf{Paper VII} also supports \textbf{G4} by proposing an edge-centric solution that deploys both parametric digital twins and predictive models at the edge, enabling low-latency indoor climate forecasting while reducing reliance on cloud infrastructure.

Together, these contributions establish a coherent research trajectory toward data-driven smart maintenance of historic buildings. By bridging theoretical models, technical implementations, and real-world applications, the thesis offers a practical solution for enabling smart maintenance within the broader context of digital transformation.

\section{Scope and Delimitations}
\label{sec:delimitations}

This thesis focuses on developing and validating data-driven approaches for smart maintenance of historic buildings, with particular emphasis on indoor climate management and energy optimization. The scope covers the integration of digital technologies, including IoT, cloud and edge computing, ontology-based parametric digital twins, deep learning, and federated learning, to support monitoring, predictive analytics, and intelligent decision-making for heritage conservation. The aim is to assist facility managers in maintaining indoor conditions, such as temperature and RH, within recommended ranges in an energy-efficient manner. It also enables experts to achieve conservation targets by aligning environmental parameters with established thresholds. 

The scope of the thesis does not include structural health monitoring or degradation processes unrelated to indoor environmental factors. In addition, non-digital conservation methods and broader economic or life cycle assessments are beyond the focus of this work.

The research is grounded in case studies conducted at selected historic buildings in Sweden. These include the City Museum, the City Theatre, and the Auditorium in the city of Norrköping, as well as Löfstad Castle in Östergötland. Although the methods are designed to be transferable to other heritage sites, technical validation and implementation are limited to these buildings.

\section{Thesis Outline}
\label{sec:outline}

The remainder of this thesis is structured as follows.

Chapter~\ref{cha:methodology} presents the methodological foundations of the research. It introduces the overall system design, IoT-based sensing solution, and ontology-based data modeling approach used to create parametric digital twins. The chapter also formulates the time series forecasting problem and describes the deep learning architectures employed. It explains the integration of deep learning with FL, including federated workflows and optimization strategies.

Chapter~\ref{cha:iot} describes the deployment of an IoT-based sensing solution for long-term indoor environmental monitoring in historic buildings. It evaluates system stability and introduces a suite of cloud-hosted data applications.

Chapter~\ref{cha:parametric} focuses on the creation and application of parametric digital twins. It presents case studies from the City Theatre and Löfstad Castle to demonstrate how semantic modeling and sensor integration support real-time monitoring and preventive conservation.

Chapter~\ref{cha:deep_learning} examines deep learning methods for energy forecasting in historic buildings. It compares model performance for predicting electricity consumption and heating load, evaluates the effect of incorporating future information, and analyzes the computational cost of various architectures.

Chapter~\ref{cha:edge_federated} investigates the deployment of predictive models on edge devices and evaluates FL approaches for indoor climate forecasting. It assesses inference costs on resource-constrained hardware and compares model performance across local, centralized, and federated learning cases.

Chapter~\ref{cha:summary} summarizes the author's key contributions in the seven included research papers.

The last chapter concludes the thesis by synthesizing its contributions, reflecting on how the research objectives were achieved, and discussing broader implications for the field of heritage conservation.

\chapter{Methodology}
\label{cha:methodology}

This chapter outlines the methodology used to investigate, develop, and validate data-driven approaches for smart maintenance of historic buildings. It begins with an overview of the integrated solution and its overall design. The next section describes the development of the IoT-based sensing system, including its architecture and deployment on both cloud and edge platforms to enable scalable data collection and real-time analytics. The chapter then explains the creation of parametric digital twins using ontology-based data modeling to organize and integrate heterogeneous building data. It then explores the use of advanced deep learning techniques for multi-horizon forecasting of energy consumption and indoor climate conditions. The combination of deep learning and FL is further introduced to enable privacy-aware forecasting across different buildings. The chapter concludes with the design and implementation of case studies in four historic buildings in Sweden, covering the deployment of sensing systems, the collection of real-world data, and the validation of the proposed forecasting and analysis methods.

\section{Overall System Design}
\label{sec:system_design}

\begin{figure}[!tb]
\includegraphics{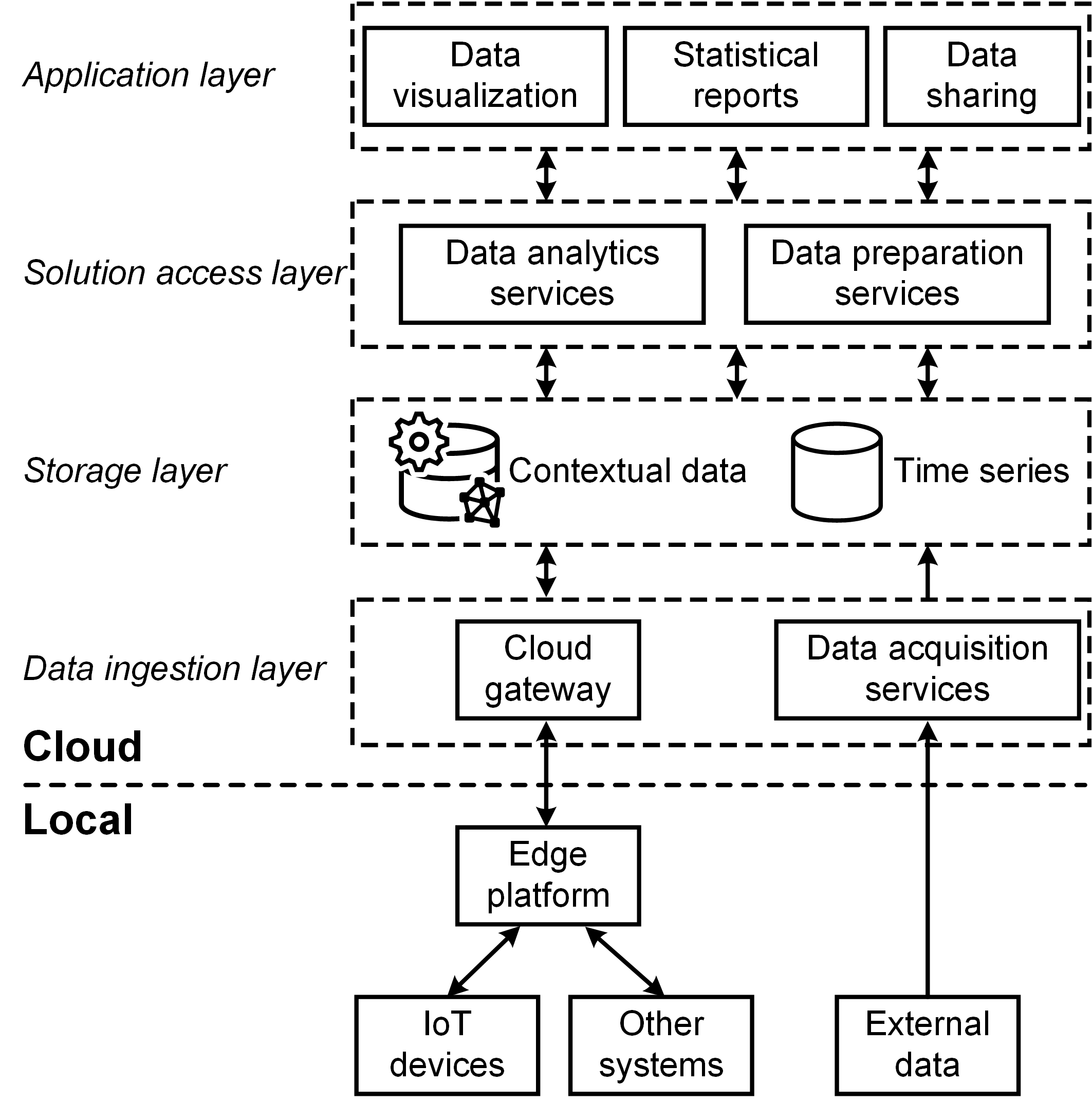}
\centering
\caption{System architecture of the integrated solution supporting data-driven smart maintenance of historic buildings~\cite{ni_parametric_2025}.}\label{fig:phd_parametric_fig1}
\end{figure}

The system is designed to support smart maintenance of historic buildings by creating an integrated digital environment for data-driven monitoring, analysis, and decision-making, making it easier to manage settings and respond to changing conditions. It incorporates key functions such as data collection, modeling, storage, querying, analytics, and visualization. As illustrated in Fig.~\ref{fig:phd_parametric_fig1}, the system architecture consists of two main parts: a local part and a cloud part.

\subsection{The Local Part}

The local part serves as the primary data source, collecting both internal and external information related to building operations and environmental conditions. Typical internal data include indoor environmental parameters, energy consumption, and equipment status. These data are acquired using IoT-based sensing devices. The edge platform manages these data streams, performs initial preprocessing, and uploads selected data to the cloud. With built-in storage, computing, and networking capabilities, the edge platform supports real-time processing and privacy-sensitive analytics, making it suitable for latency-critical or bandwidth-constrained scenarios. It can also integrate with existing building systems, such as building management systems, to access additional data from legacy infrastructures. External data, such as weather information and energy prices, are gathered through open application programming interfaces (APIs).

\subsection{The Cloud Part}

The cloud part provides scalable remote storage, analytics, and intelligent services, while also serving as a secure and reliable backup. It is structured into four layers: data ingestion, storage, solution access, and application. The data ingestion layer manages communication between the edge and cloud platforms. It includes cloud gateways that translate protocols and relay data to downstream services, ensuring bi-directional communication and facilitating the synchronization of digital twin models. 

The storage layer handles two data types: contextual and time series. Contextual data, which describe spatial layouts and relationships among entities such as rooms and sensors, are stored using ontology-based graph models to ensure semantic consistency. Time series data, including dynamic measurements such as temperature and energy consumption, are stored in relational or time series databases. These two types of data are linked through unique identifiers, allowing integrated spatial and temporal analysis.

The solution access layer offers interfaces for querying and analyzing both contextual and time series data. It supports exploratory data analysis (EDA) to identify patterns and guide the selection of machine learning algorithms for forecasting, anomaly detection, and decision support. Validated predictive models can be deployed for real-time inference on streaming data. This enables proactive maintenance by detecting environmental deviations or system failures.

At the top of the stack, the application layer provides visualization and user interaction. It delivers real-time dashboards and historical data views to assist facility managers, researchers, and conservation professionals in monitoring and managing building performance. 

\section{IoT-based Sensing Solution}
\label{sec:iot}

The implemented IoT-based sensing solution provides a systematic approach for collecting data and enabling the creation of parametric digital twins of historic buildings. It is designed for long-term deployment, scalable integration with a cloud-edge architecture, and real-time data processing to support smart maintenance. The solution integrates multiple environmental sensors for collecting data, uses an edge platform for communication, and leverages Microsoft Azure services for data storage and application deployment. The solution is scalable, allowing future expansion and integration of additional features as needed.

\subsection{The Local Part}

\subsubsection{First Version: Arduino + Raspberry Pi CM3+}

\begin{figure}[!tb]
\includegraphics{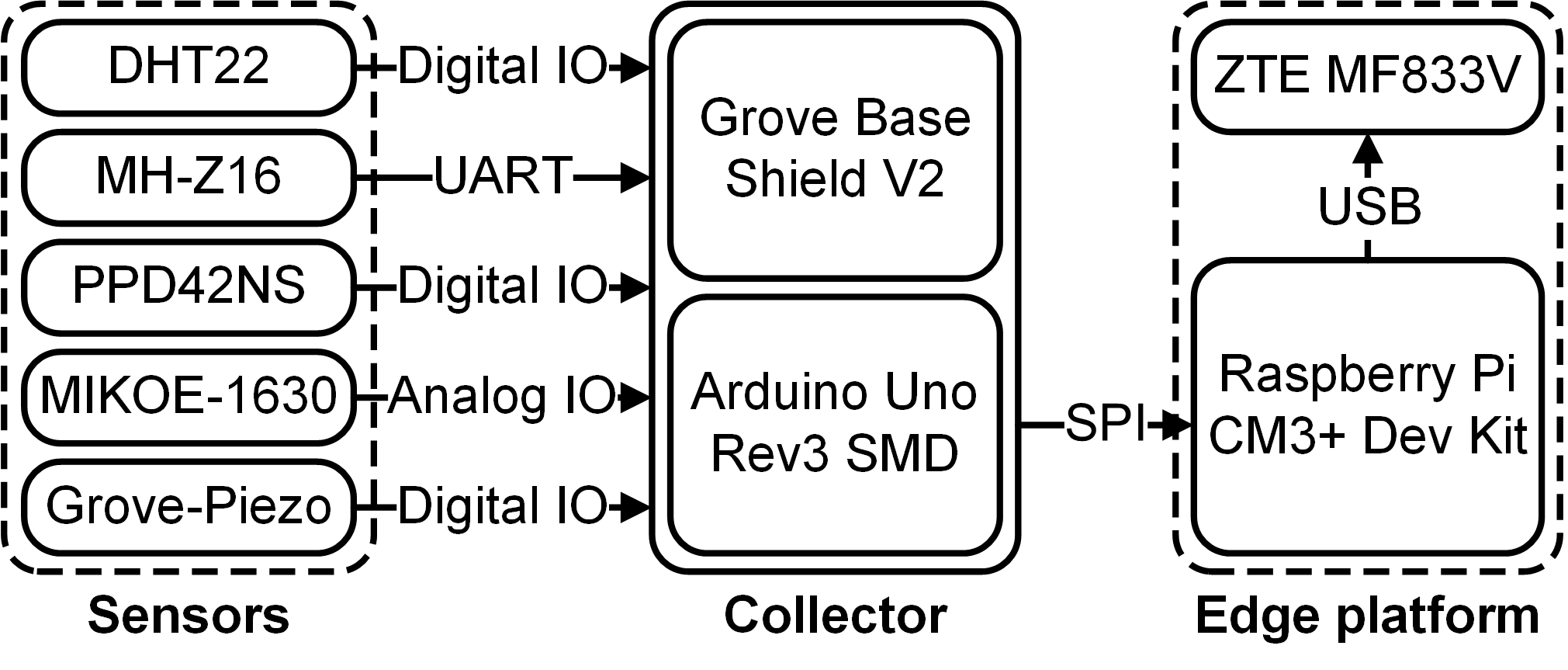}
\centering
\caption{Overview of used hardware components and their connections in the first version~\cite{ni_enabling_2022}. Arrows mark the direction of data flow.}\label{fig:phd_enabling_fig3}
\end{figure}

The local sensing system was initially implemented and deployed in three historic buildings in Norrköping, Sweden: the City Museum, the City Theatre, and the Auditorium. It features a modular design (see Fig.~\ref{fig:phd_enabling_fig3}), with separate hardware components for data acquisition and processing. This modularity enables easy integration, replacement, or expansion of sensors, supporting flexible deployment across varied historic buildings. The implementation provides early validation of a cloud-connected IoT sensing system tailored to the requirements of historic building conservation, as presented in Papers I and II.

An Arduino Uno microcontroller (Arduino, Somerville, United States), connected via a Grove Base Shield (Seeed Technology, Shenzhen, China), is used to collect environmental data from various sensors. The collected data are then transmitted to a Raspberry Pi Compute Module (CM) 3+ Development Kit (Raspberry Pi Foundation, Cambridge, United Kingdom), which serves as the core of the edge platform. A ZTE MF833V 4G USB modem (ZTE, Shenzhen, China) provides mobile broadband connectivity. The edge platform performs local preprocessing and facilitates communication with cloud services. 

To monitor key environmental conditions relevant to preservation of historic buildings, five sensing devices are employed to capture six critical parameters. These parameters are essential for assessing risks of material degradation and ensuring occupant comfort in heritage environments.
\begin{itemize}
\item Temperature and RH are measured using DHT22 (Seeed Technology, Shenzhen, China).
\item CO\textsubscript{2} concentration is monitored using MH-Z16 (Winsen Electronics Technology, Zhengzhou, China). 
\item Particulate matter (PM) with a diameter greater than 1~\textmu{m} is detected by PPD42NS (Shinyei Corporation, New York, United States). 
\item Several gases that affect air quality, such as ammonia, nitrogen oxides, benzene, and CO\textsubscript{2}, are measured by MIKOE-1630 (MikroElektronika, Beograd, Serbia).
\item Vibration, flexibility, impact, and touch are captured by the Grove-Piezo sensor (Seeed Technology, Shenzhen, China).
\end{itemize}

\subsubsection{Second Version: Grove Base Hat + Raspberry Pi 3 Module B+}

\begin{figure}[!tb]
\includegraphics{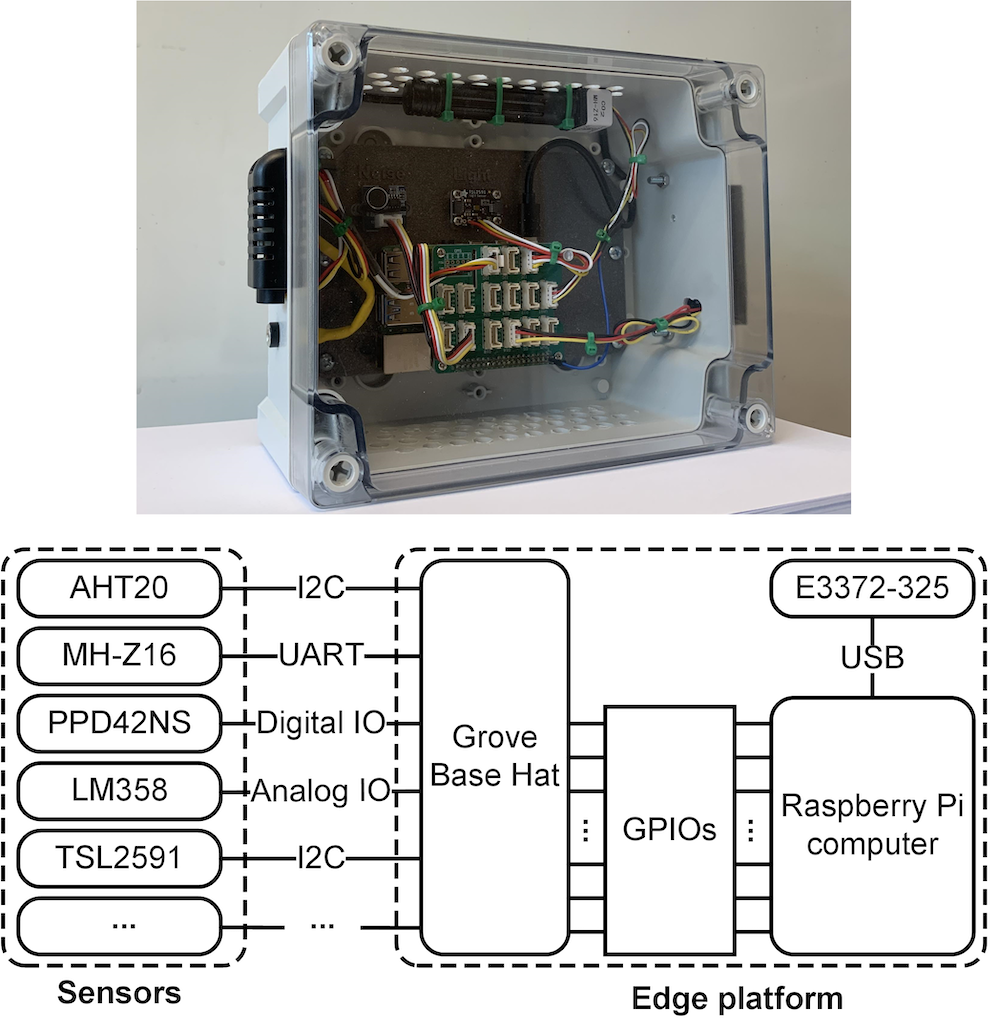}
\centering
\caption{The upper image shows a sensor box containing an edge platform and six environmental sensors. The lower diagram presents used hardware components and their connections in the second version~\cite{ni_parametric_2025}.}\label{fig:phd_parametric_fig2}
\end{figure}

In later deployments, beginning with the case study at Löfstad Castle (Paper III) and continuing in subsequent work, the sensing system was redesigned to simplify the hardware architecture and enhance operational reliability. A Raspberry Pi 3 Model B+ (Raspberry Pi Foundation, Cambridge, United Kingdom) equipped with a Grove Base Hat (Seeed Technology, Shenzhen, China) is used to interface directly with environmental sensors, removing the need for a separate microcontroller (see Fig.~\ref{fig:phd_parametric_fig2}). A 4G USB modem (E3372-325, Huawei, Shenzhen, China) provides cellular connectivity for sites without stable Wi-Fi access. This integrated and compact setup reduces system complexity and improves data transmission reliability.

Sensor selection is adapted to the specific conservation needs of each site. Sensors are chosen for their reliability, simplicity, and compatibility with the edge platform. A typical configuration includes five sensing devices: AHT20 (Adafruit, New York, United States) for temperature and RH, MH-Z16 for CO\textsubscript{2} concentration, PPD42NS for particulate matter, LM358 (Seeed Technology, Shenzhen, China) for noise levels, and TSL2591 (Adafruit, New York, United States) for light intensity. These sensors, along with the Raspberry Pi and Grove Base Hat, are packaged in a plastic enclosure (201~$\times$~163~$\times$~98~mm) for on-site deployment, as shown in upper part of Fig.~\ref{fig:phd_parametric_fig2}. The system supports modular expansion. For example, in the basement of the main building at Löfstad Castle, additional ultrasonic sensors (UR18.DA0.2-UAMJ.9SF, Baumer, Frauenfeld, Switzerland) were installed to monitor groundwater level (GWL) variations, addressing site-specific moisture risks in the conservation strategy.

\subsection{The Cloud Part}

The cloud infrastructure was implemented using several public services provided by Microsoft Azure~\cite{microsoft_azure_2021}. Although Azure serves as the primary cloud provider in this research, the system is designed to be platform-agnostic, ensuring compatibility with alternative services such as Amazon Web Services~\cite{amazon_aws_2021} or with on-premise solutions when required by data privacy regulations, institutional policies, or infrastructure constraints. The architecture is organized into two main components: the back end and the front end. The back end manages core domain logic, including data ingestion, storage, and processing. The front end offers a user-friendly interface for accessing, visualizing, and interacting with the collected data.

Core services used in the cloud infrastructure include Azure IoT Hub, Azure Functions, Azure SQL Database, and Azure App Service. Azure IoT Hub acts as the cloud gateway, enabling secure and scalable communication between edge platforms and the cloud. Azure Functions provides serverless, event-driven computing, allowing lightweight tasks, such as storing incoming data or retrieving external datasets, to be executed efficiently. Time series data collected from deployed sensors are stored in Azure SQL Database, a cloud-based relational storage service offering built-in scalability and high availability. Azure App Service hosts web applications, providing an integrated platform for deploying and managing cloud-based user interfaces. Service tier selection was based on experimental requirements but can be scaled to support more complex applications or larger data volumes as needed.

To support visualization and real-time data interaction, a web-based application was developed using Python (v3.8.18) along with supporting libraries, including Dash (v2.3.1), Dash Bootstrap Components (v0.13.1), and pandas (v1.4.1). The application enables facility managers and researchers to explore both real-time and historical data through interactive charts, facilitating insight generation and data-driven decision-making.

\section{Parametric Digital Twins and Ontology-based Data Modeling}
\label{sec:parametric_dt}

\begin{figure}[!tb]
\includegraphics{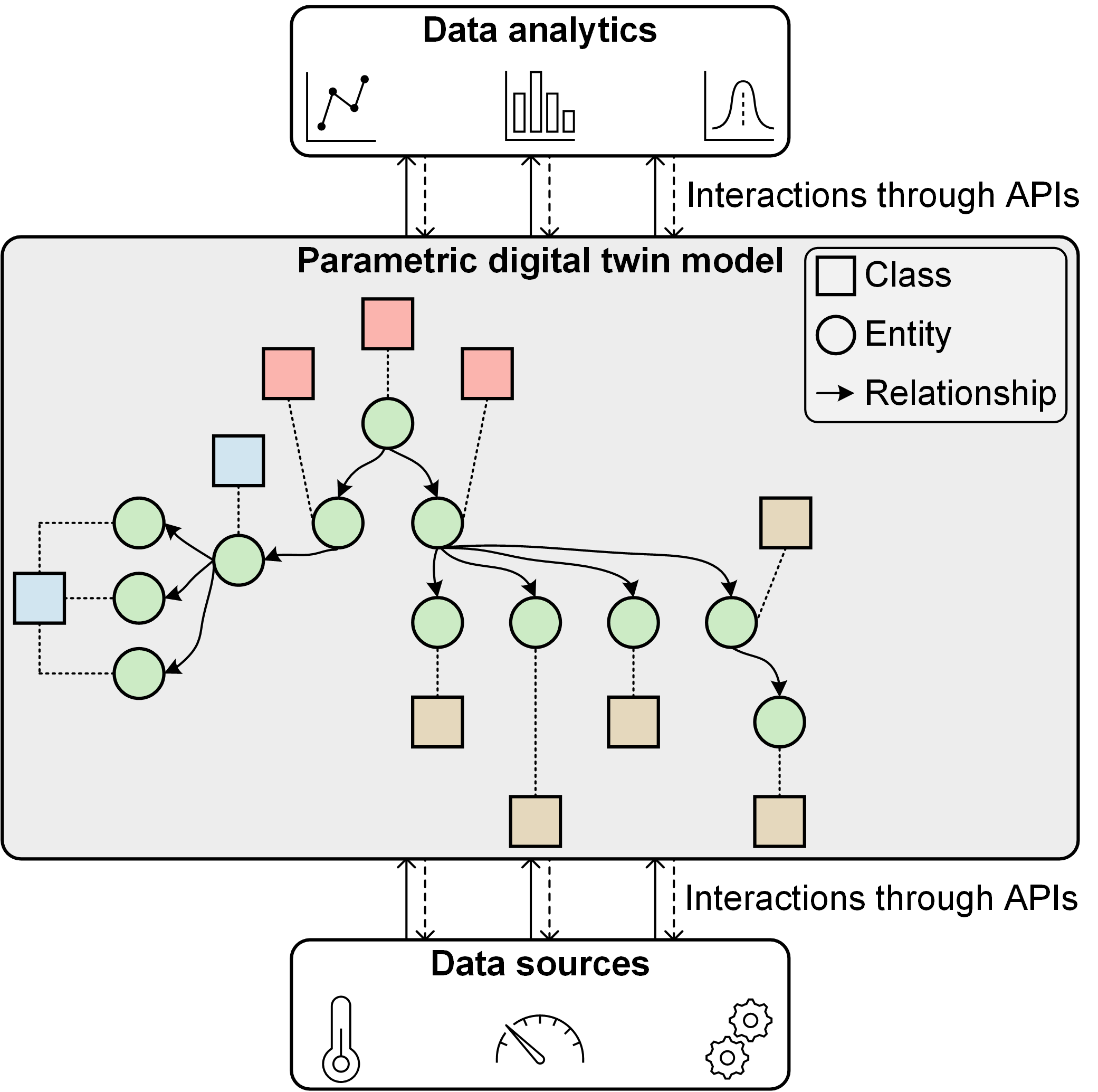}
\centering
\caption{Illustration of the parametric digital twin model~\cite{ni_leveraging_2023}. The model encapsulates essential information about physical entities and their relationships, and exposes read/write application programming interfaces (APIs) to support integration with other modules, such as data updates from sensors and data access for analytics.}\label{fig:phd_leveraging_fig1}
\end{figure}

A parametric digital twin of a physical building encapsulates both the static structure and dynamic behavior of the building. It functions as a semantic data model representing key physical entities and their relationships while integrating real-time and historical data streams to support monitoring, analytics, and control. As shown in Fig.~\ref{fig:phd_leveraging_fig1}, the design of a parametric digital twin includes two main components: modeling of contextual data to describe the physical and logical structure of a building and continuous updating of dynamic status information, primarily time series data, through standardized interfaces such as APIs.

To model contextual data, the RealEstateCore ontology~\cite{hammar_realestatecore_2019} was initially adopted and extended in combination with the Azure DTDL~\cite{azure_dtdl_2022}, as presented in Paper II. Given the growing alignment between the RealEstateCore and Brick ontologies~\cite{balaji_brick_2018}, later works, i.e., Papers III, VI, and VII, shifted to using the Brick ontology. Brick was extended to include additional entity types relevant to heritage conservation, such as a wider range of sensors and weather-related components. As illustrated in Fig.~\ref{fig:phd_leveraging_fig1}, the ontological graph uses square nodes to represent classes such as \textit{Location}, \textit{Equipment}, \textit{Point}, and \textit{Resource}. Circular nodes represent entities, which are instances of these classes. Examples of entities under each class include locations (e.g., rooms and floors), equipment (e.g., fans and HVAC units), and resources (e.g., electricity and air). Time-varying data, such as sensor readings or control signals, are modeled as points. These points may be physical (e.g., a CO\textsubscript{2} sensor) or virtual (e.g., a derived average temperature), and they serve as interfaces between the contextual model and real-time data streams. Relationships among entities define hierarchical structures (e.g., floors and rooms), functional dependencies (e.g., sensors linked to equipment), and spatial associations. The resulting model can be queried using the SPARQL protocol and RDF query language (SPARQL), enabling structured access to building metadata.

Time series data from IoT devices, such as environmental sensors, are stored in a relational database optimized for temporal queries. Each time series is uniquely identified by a universally unique identifier (UUID), which links the data to its corresponding entity in the ontology. The data table follows a simple schema with three columns: TIME, UUID, and VALUE, and uses a composite primary key on TIME and UUID. This structure ensures efficient storage, retrieval, and analysis of large-scale sensor data while preserving traceability to the semantic model.

By integrating ontology-based contextual modeling with structured time series storage, a parametric digital twin provides a consistent, extensible, and data-rich representation of a historic building. This approach facilitates seamless integration with analytical processes, allows for real-time monitoring, and improves the flexibility of the solution for deployment in other buildings.

\section{Deep Learning for Time Series Forecasting}
\label{sec:dl_ts_forecasting}

This section presents the methodology for applying deep learning methods to time series forecasting tasks relevant to smart maintenance of historic buildings. Both building energy use and indoor environmental conditions exhibit strong temporal dependencies and are naturally modeled as time series. Deep learning has demonstrated strong capabilities in capturing complex, nonlinear patterns in time series data. This section begins by formulating the forecasting problem, followed by an overview of recent deep learning architectures designed for time series forecasting. Finally, it introduces the loss functions and evaluation metrics employed to train and assess forecasting models.

\subsection{Problem Formulation}

\begin{figure}[!tb]
\includegraphics{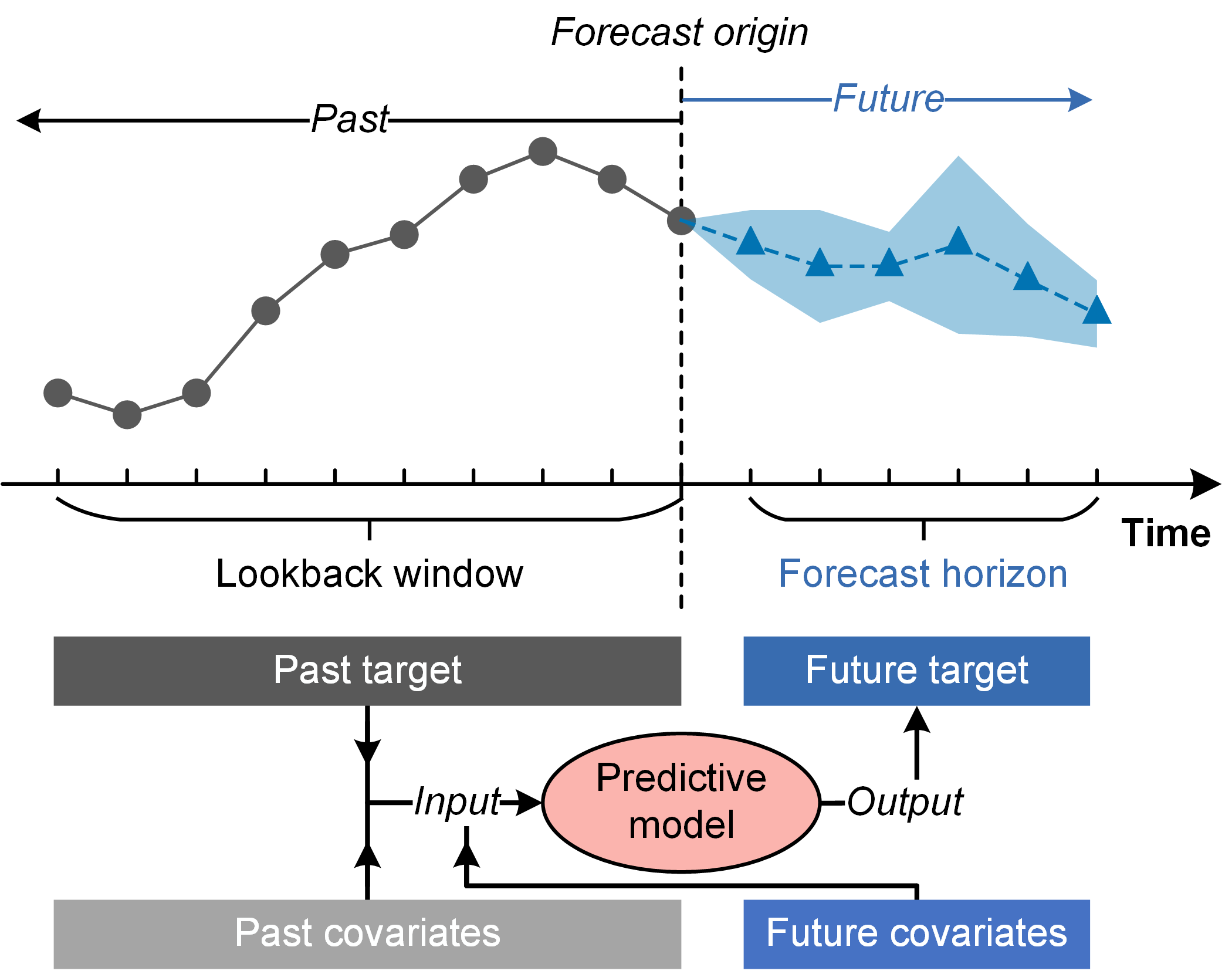}
\centering
\caption{Illustration of multi-horizon time series forecasting, adapted from~\cite{ni_edge_2024}. The black dots indicate observed values of a target variable over a historical lookback window, and the blue triangles represent forecasts of the target variable over a future forecast horizon. The shaded region denotes a predicted uncertainty range. The model uses past values of the target and associated covariates, along with future covariates, to generate forecasts across the prediction horizon.}\label{fig:phd_fig1}
\end{figure}

The forecasting tasks addressed in this thesis are formulated as supervised learning problems using time series data. While the formulation is presented for a univariate target variable, it can be readily extended to multivariate cases. Let the target variable be a real-valued time series $y\in \mathbb{R}$, representing quantities such as energy consumption or indoor environmental conditions. Predictor variables that may influence the target are divided into two categories: those observed in the past (including the forecast origin) and those known in the future (after the forecast origin), as illustrated in Fig.~\ref{fig:phd_fig1}. The past predictor variables are denoted by a real row vector $\mathbf{x_{b}} \in \mathbb{R}^{k}$, and the future predictor variables by a real row vector $\mathbf{x_{f}} \in \mathbb{R}^{m}$.

At time step $t$, the observed target value is ${y}_{t}$, and the corresponding predictor variables are represented as $\mathbf{x_{b}}_{t} =[{x_{b}}_{1,t} ,{x_{b}}_{2,t} ,...,{x_{b}}_{k,t}]$ and $\mathbf{x_{f}}_{t} =[{x_{f}}_{1,t} ,{x_{f}}_{2,t} ,...,{x_{f}}_{m,t}]$. All variables are sampled at fixed intervals and organized chronologically.

A point forecasting model predicts the future values of the target variable over a forecast horizon $h$ using historical data from a loopback window of length $w$. The model takes the following general form:
\begin{equation}
\hat{y}_{t+1:t+h} =f_{\mathbf{\theta }}(y_{t-w+1:t} ,\mathbf{x_{b}{}}_{t-w+1:t} ,\mathbf{x_{f}{}}_{t+1:t+h}), \label{eqn:phd_point_forecast}
\end{equation}
where $\hat{y}_{t+1:t+h} = \left[\hat{y}_{t+1},\hat{y}_{t+2},...,\hat{y}_{t+h}\right] \in \mathbb{R}^{h}$ are the predicted values of the target variable over the forecast horizon $h$, $y_{t-w+1:t} = [y_{t-w+1} ,y_{t-w+2} ,...,y_{t}] \in \mathbb{R}^{w}$ as well as $\mathbf{x_{b}}_{t-w+1:t} = \{\mathbf{x_{b}}_{t-w+1} ,\mathbf{x_{b}}_{t-w+2} ,...,\mathbf{x_{b}}_{t}\}$ are observations of the target and predictor variables over the loopback window $w$, $\mathbf{x_{f}}_{t+1:t+h} =\{\mathbf{x_{f}}_{t+1} ,\mathbf{x_{f}}_{t+2} ,...,\mathbf{x_{f}}_{t+h}\}$ are observations of predictor variables over the forecast horizon, and $f_{\mathbf{\theta }}{(.)}$ is the forecasting model with parameters $\mathbf{\theta }$. When $h=1$, the task is a one-step-ahead forecasting problem; when $h>1$, it becomes a multi-horizon forecasting problem.

Probabilistic forecasting aims to estimate a full predictive distribution rather than a single expected value. Instead of assuming a predefined distribution, quantile regression~\cite{koenker_regression_1978} is used in this research to directly predict quantiles of the target variable. The $p$th quantile represents the value below which a proportion $p \in (0, 1)$ of the distribution lies~\cite{hao_quantile_2007}. Given a set of quantile levels $\mathcal{Q} \subset (0,1)$, the model learns to predict each quantile independently:
\begin{equation}
\hat{y}_{t+1:t+h}^{(p)} =g_{\mathbf{\theta }}(y_{t-w+1:t} ,\mathbf{x_{b}{}}_{t-w+1:t} ,\mathbf{x_{f}{}}_{t+1:t+h}), \label{eqn:phd_quantile_forecast}
\end{equation}
where $p$ is an element of the set $\mathcal{Q}$, $\hat{y}_{t+1:t+h}^{(p)} = \left[\hat{y}_{t+1}^{(p)} ,\hat{y}_{t+2}^{(p)},...,\hat{y}_{t+h}^{(p)}\right] \in \mathbb{R}^{h}$ denotes the predicted $p$th quantile across the forecast horizon $h$, $y_{t-w+1:t}$, $\mathbf{x_{b}}_{t-w+1:t}$ and $\mathbf{x_{f}}_{t+1:t+h}$ have the same definition as in the point forecasting model, and $g_{\mathbf{\theta }}{(.)}$ is the prediction function learned by the model.

This formulation provides a unified framework for both point and probabilistic deep learning-based time series forecasting models, enabling flexible applications to predict building energy use and indoor climate.

\subsection{Deep Learning Architectures}

\begin{figure}[!tb]
\includegraphics{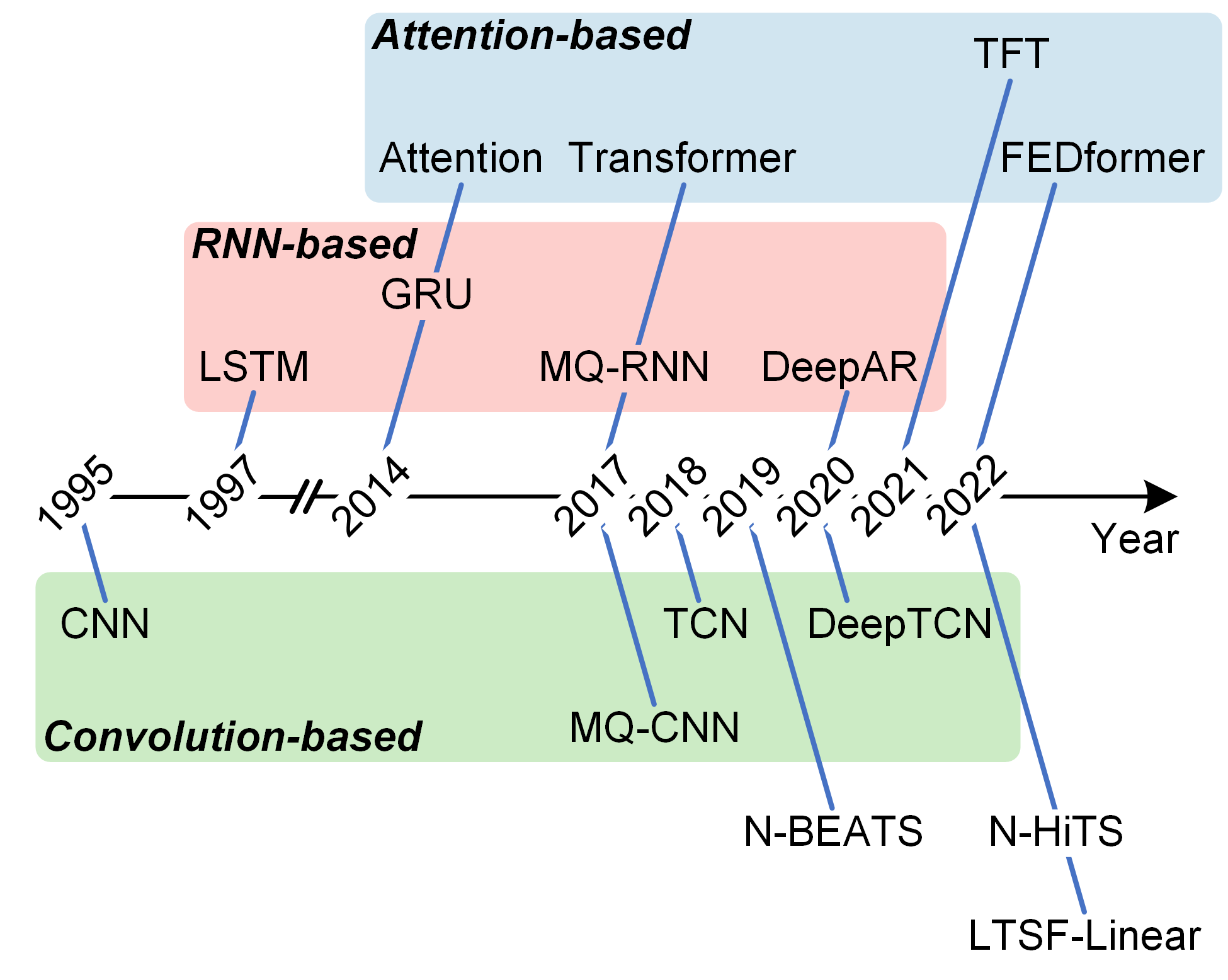}
\centering
\caption{Key developments in the evolution of deep learning architectures for time series forecasting~\cite{ni_deep_2024}.}\label{fig:phd_deep_fig1}
\end{figure}

Deep learning has significantly advanced time series forecasting by enabling flexible, nonlinear modeling of temporal patterns. Traditional fully connected networks, such as ANNs, are limited in capturing temporal dependencies. To overcome this limitation, specialized architectures (see Fig.~\ref{fig:phd_deep_fig1}) have been developed, including RNN-based, convolution-based, attention-based networks, and more recent multilayer perceptron (MLP)-based dense networks. These architectures are designed to handle sequential data effectively while balancing forecasting accuracy, computational efficiency, and interpretability.

Recurrent architectures such as vanilla RNNs use hidden states to summarize historical inputs, updating them recursively at each time step. However, RNNs suffer from vanishing or exploding gradients, which restrict their ability to learn long-term dependencies. Two widely used variants, LSTM~\cite{hochreiter_long_1997} and GRU~\cite{cho_properties_2014}, address these issues by introducing gating mechanisms. LSTMs incorporate input, forget, and output gates to control information flow, while GRUs simplify the architecture with two gates, improving training speed. Recurrent-based forecasting models such as MQ-RNN~\cite{wen_multi-horizon_2017} and DeepAR~\cite{salinas_deepar_2020} have been proposed to scale RNNs to large datasets by learning global models across many time series. These architectures are particularly effective for short- to medium-range dependencies but are limited by sequential computation, which limits parallelization.

Convolutional neural networks (CNNs) offer an alternative approach by capturing local temporal patterns through convolutional filters. TCNs~\cite{bai_empirical_2018} combine causal and dilated convolutions to extend the receptive field without recursion. Other convolution-based networks include MQ-CNN~\cite{wen_multi-horizon_2017} and DeepTCN~\cite{chen_probabilistic_2020}, which follow an encoder-decoder design for sequence modeling. These architectures benefit from efficient parallelization, making them well-suited for large-scale forecasting tasks. However, their performance can degrade when modeling long-range dependencies, as doing so typically requires increased depth and architectural complexity.

Attention-based architectures have become prominent in time series forecasting due to their ability to capture global dependencies within sequences. The original transformer (TF) architecture~\cite{vaswani_attention_2017} introduced a self-attention mechanism that eliminates recurrence, enabling parallel computation and dynamic weighting of input features across time. The temporal fusion transformer (TFT)~\cite{lim_temporal_2021} extends this approach by combining LSTM encoders with attention layers and gating mechanisms, allowing it to process both static covariates and dynamic temporal inputs within a unified framework. The frequency enhanced decomposed transformer (FEDformer)~\cite{zhou_fedformer_2022} enhances long-range forecasting by incorporating Fourier-based frequency decomposition. Despite their effectiveness, attention-based models are computationally intensive and may struggle to preserve temporal order in long sequences~\cite{zeng_transformers_2023}.

MLP-based dense networks have recently emerged as simpler alternatives that retain competitive performance. The neural basis expansion analysis time series (N-BEATS)~\cite{oreshkin_n-beats_2019} and its successor neural hierarchical interpolation for time series (N-HiTS)~\cite{challu_nhits_2023} use stacks of fully connected layers to decompose time series into trend and seasonality components. These architectures benefit from high interpretability and efficient training. Time-series dense encoder (TiDE)~\cite{das_long_2023} uses dense MLP layers combined with temporal decomposition modules, achieving accurate forecasts without relying on recurrence, convolution, or attention mechanisms. Time-series mixer (TSMixer)~\cite{chen_tsmixer_2023}, inspired by the MLP-Mixer architecture in computer vision, mixes temporal and feature dimensions separately through stacked MLPs. These architectures emphasize scalability, low computational cost, and ease of deployment, making them attractive for forecasting tasks in resource-constrained environments.

The design of deep learning architectures for time series forecasting continues to evolve. Some studies have questioned the necessity of complex attention mechanisms in time series tasks, arguing that simpler, well-designed networks can outperform transformers in certain settings~\cite{zeng_transformers_2023}. Ultimately, the choice of architecture should consider specific characteristics of the data, such as sequence length, noise level, seasonality, and the requirements for interpretability or scalability.

\subsection{Loss Functions and Evaluation Metrics}

To train point forecasting models, the squared loss function was used to minimize the total error between predicted and actual values. This approach estimates the conditional mean of the target variable~\cite{hyndman_forecasting_2018}. Given a training set $\mathcal{S} =\{(y_{t-w+1:t} ,\mathbf{x_{b}{}}_{t-w+1:t} ,\mathbf{x_{f}{}}_{t+1:t+h} ,y_{t+1:t+h})\}_{t=w}^{n+w-1}$, the training loss is defined as:
\begin{equation}
L_{s}(\mathbf{\theta }) = \sum _{t=w}^{n+w-1}\sum _{i=1}^{h} \left(\hat{y}_{t+i} -y_{t+i}\right)^{2}, \label{eqn:phd_square_loss}
\end{equation}
where $n$ is the number of training samples, $h$ is the forecast horizon, and $\hat{y}_{t+i}$ and $y_{t+i}$ are the predicted and actual values of the target variable at time step $t+i$, respectively.

For probabilistic forecasting, quantile regression~\cite{koenker_regression_1978} was adopted to directly estimate conditional quantiles. The quantile loss function~\cite{wen_multi-horizon_2017} for a single forecast at quantile level $p$ is defined as:
\begin{equation}
\ell(\hat{y} ,y,p)=(1-p)(\hat{y} -y)_{+} + p(y-\hat{y})_{+}, \label{eqn:phd_deep_quantile_loss}
\end{equation}
where $(.)_{+}=max(0,.)$. The full quantile loss over the training set $\mathcal{S} =\{(y_{t-w+1:t} ,\mathbf{x_{b}{}}_{t-w+1:t} ,\mathbf{x_{f}{}}_{t+1:t+h} ,y_{t+1:t+h})\}_{t=w}^{n+w-1}$ is denoted by $L_{q}(\mathbf{\theta})$, and
\begin{equation}
L_{q}(\mathbf{\theta }) =\sum _{t=w}^{n+w-1}\sum _{j=1}^{| \mathcal{Q}| }\sum _{i=1}^{h} \ell \left(\hat{y}_{t+i}^{( p_{j})} ,y_{t+i} ,p_{j}\right),
\end{equation}
where $\mathcal{Q}$ is the set of quantile levels and $p_{j}$ is an element of $\mathcal{Q}$.

Model performance was evaluated using two main criteria: computational cost and prediction accuracy. Computational cost was measured by the total model training time. Prediction accuracy for point forecasting was assessed using coefficient of variation of the root mean square error (CV-RMSE) and normalized mean bias error (NMBE), as recommended by ASHRAE Guideline 14-2014~\cite{ashare_measurement_2014}. These two scale-independent metrics are defined as:
\begin{equation}
RMSE=\sqrt{\frac{1}{n}\sum\limits _{t=1}^{n}(\hat{y}_{t} -y_{t})^{2}}, \label{eqn:phd_deep_rmse}
\end{equation}
\begin{equation}
CV\textrm{-}RMSE=\frac{RMSE}{\overline{y}} \times 100, \label{eqn:phd_deep_cv-rmse}
\end{equation}
\begin{equation}
MBE=\frac{1}{n}\sum\limits _{t=1}^{n}(\hat{y}_{t} -y_{t}), \label{eqn:phd_deep_mbe}
\end{equation}
\begin{equation}
NMBE=\frac{MBE}{\overline{y}} \times 100,  \label{eqn:phd_deep_nmbe}
\end{equation}
where $n$ denotes the size of a forecast horizon, $y_{t}$ is the actual value of a target variable at time step $t$, $\hat{y}_{t}$ is the predicted value of the target variable at time step $t$, and $\overline{y}$ is the mean actual value of the target variable over the forecast horizon.

CV-RMSE quantifies the relative dispersion between predicted and observed values~\cite{ashare_measurement_2014}, while NMBE measures systematic bias~\cite{ramos_ruiz_validation_2017}. A positive NMBE indicates overestimation, whereas a negative value reflects underestimation. In model evaluation, primary emphasis was placed on CV-RMSE, provided that NMBE remained within acceptable bounds. Following ASHRAE guidelines, point forecasting models for whole-building energy use are considered acceptable if they achieve a CV-RMSE $\leq$ 30\% and an NMBE within $\pm$10\% when evaluated using hourly data~\cite{ashare_measurement_2014}. For indoor temperature and RH forecasting, commonly accepted thresholds are a CV-RMSE $\leq$ 20\% and an NMBE within $\pm$5\%~\cite{odonovan_predicting_2019}. Although no formal standard exists for CO\textsubscript{2} forecasting, this study adopts the same ASHRAE thresholds (CV-RMSE $\leq$ 30\%, NMBE within $\pm$10\%) as a reasonable benchmark, given the occupancy-driven nature of CO\textsubscript{2} dynamics, which are similar to those of energy use.

For probabilistic forecasting, model performance was evaluated using the $\rho$-risk metric~\cite{salinas_deepar_2020, lim_temporal_2021}, which normalizes the total quantile loss by the cumulative actual values:
\begin{equation}
\rho\textrm{-}risk(p) =\frac{2\times \sum\limits _{t=1}^{n} \ell \left(\hat{y}_{t}^{(p)}, y_{t}, p\right)}{\sum\limits _{t=1}^{n} y_{t}}, \label{eqn:phd_deep_rho_risk}
\end{equation}
where $n$ denotes the size of a forecast horizon, $y_{t}$ is the actual value of a target variable at time step $t$, $\hat{y}_{t}^{(p)}$ denotes the predicted $p$th quantile value at time step $t$, and $\ell \left(\hat{y}_{t}^{(p)} ,y_{t} ,p\right)$ is the $p$th quantile loss calculated by Eq.~\ref{eqn:phd_deep_quantile_loss}. This metric allows performance comparison across different quantile levels while accounting for the scale of the data.

\section{Integration of Deep Learning and Federated Learning}

Integrating deep learning with FL enables privacy-preserving, decentralized training of forecasting models across multiple buildings or devices without requiring central access to raw data. This is particularly important in the context of historic buildings for scenarios where data related to occupancy, environmental conditions, and operations may be sensitive. As the problem formulation for time series forecasting remains consistent with Section~\ref{sec:dl_ts_forecasting}, this section focuses on presenting how various FL algorithms aggregate model parameters to support robust forecasting in heterogeneous, non-IID data environments.

FL aims to collaboratively train a shared global model $f_\theta$ across a set of distributed clients $\mathcal{C}=\{1, 2, ..., N\}$, where each client corresponds to a measurement point or edge node collecting local time series data. In contrast to traditional centralized learning, which requires transferring all data to a central repository, FL preserves privacy by keeping data local. The FL process involves iterative communication rounds between the central server and participating clients. As illustrated in Fig.~\ref{fig:phd_federated_fig1}, each round begins with the server distributing the current global model parameters to selected clients. Each client performs local training using its private dataset and then transmits model updates, often in the form of gradients or parameter weights, back to the server\cite{mcmahan_communication_2017, beutel_flower_2020}. The server aggregates these updates to refine an updated global model, which is redistributed in the next round. This decentralized workflow reduces communication costs, improves system scalability, and enhances data privacy.

However, FL introduces several optimization challenges due to communication constraints, partial client participation, and the non-IID nature of data across clients~\cite{mcmahan_communication_2017}. To address these challenges, various federated optimization algorithms have been proposed to improve convergence, robustness, and training efficiency. One of the earliest and most widely used methods is federated averaging (FedAvg)~\cite{mcmahan_communication_2017}, which performs local stochastic gradient descent updates on each client and aggregates the resulting model parameters through weighted averaging. While FedAvg demonstrates strong performance in IID settings, it may suffer from degraded convergence and instability in heterogeneous environments.

To address limitations of FedAvg, several variants and alternative algorithms have been proposed. Federated median (FedMedian)~\cite{yin_byzantine_2018} replaces mean aggregation with a coordinate-wise median, improving robustness against outliers and adversarial updates. Federated averaging with server momentum (FedAvgM)~\cite{hsu_measuring_2019} enhances FedAvg by introducing momentum on the server side to stabilize learning in highly non-IID settings. SCAFFOLD~\cite{karimireddy_scaffold_2020} employs control variates to correct client drift, aligning local updates with global optimization directions. Similarly, FedProx~\cite{li_federated_optimization_2020} modifies the local objective function by adding a proximal term to penalize divergence from the global model. FedNova~\cite{wang_tackling_2020} normalizes client updates based on local step sizes, promoting fairer aggregation and improved convergence.

In parallel, adaptive optimization algorithms have been adapted for federated settings. Federated adaptive optimization using Adagrad (FedAdagrad), Adam (FedAdam), and Yogi (FedYogi)~\cite{reddi_adaptive_2020} incorporate adaptive learning rates on the server using first- and second-moment estimates of gradients. FedAdagrad and FedAdam accelerate convergence in scenarios with high gradient variance or sparse updates, while FedYogi provides greater stability by updating second-moment estimates more conservatively. These adaptive algorithms are particularly effective in handling the heterogeneous and dynamic conditions common in FL environment.

This study conducts a comparative evaluation of six representative FL algorithms: FedAvg~\cite{mcmahan_communication_2017}, FedMedian~\cite{yin_byzantine_2018}, FedAvgM~\cite{hsu_measuring_2019}, FedAdam, FedAdagrad, and FedYogi~\cite{reddi_adaptive_2020}. These algorithms were selected for their widespread adoption, solid theoretical foundations, and diverse optimization strategies. Their computational procedures are outlined in Algorithm~\ref{algo:phd_federated_fl}. To ensure consistency and fairness in evaluation, algorithms that require client-side modifications, such as SCAFFOLD and FedProx, were excluded. This constraint allows the comparison to focus exclusively on server-side aggregation and optimization methods.

\begin{algorithm}
\begin{footnotesize}
\SetKwInOut{Input}{Input}\SetKwInOut{Output}{Output}
\caption{Summary of FL algorithms~\cite{ni_federated_2025}: FedAvg, FedMedian, FedAvgM, FedAdam, FedAdagrad, and FedYogi.}\label{algo:phd_federated_fl}
\Input{Local datasets $\mathcal{D}^{i}$, number of clients $N$, number of communication rounds $T$, number of local epochs $E$, client-side learning rate $\eta_c$, server-side learning rate $\eta_s$, additional hyperparameters (specific to each algorithm)}
\Output{The final global model parameters $\theta^{T}$}
\textbf{Server executes:}\\
initialize $\theta^{0}$\;
\For{$t=0, 1,..., T-1$}{
Sample a set of clients $S_t$\;
$n \gets \sum _{i \in S_{t}} |\mathcal{D}^{i} | $\;
\For{$i \in S_{t}$ \textbf{in parallel}}{
Send global model parameters $\theta^t$ to client $i$\;
Collect model updates $\Delta \theta _{i}^{t}$ from the client\;
}
Aggregate model updates to compute new global model parameters $\theta^{t+1}$:\\
$\Delta \theta^t \gets \sum _{i\in S_{t}}\frac{|\mathcal{D}^{i} |}{n} \Delta \theta _{i}^{t}$\;
For FedAvg:\\
$\theta^{t+1} \gets \theta^{t} - \Delta \theta^t$\;
For FedMedian:\\
$\theta^{t+1} \gets \theta^{t} - $median$\left(\left\{\Delta \theta _{i}^{t}\right\}_{i\in S_{t}}\right)$\;
For FedAvgM:\\
$m^{t+1} \gets \beta m^t + \Delta \theta _{i}^{t} $\;
$\theta^{t+1} \gets \theta^{t} - \eta_s m^{t+1}$\;
For FedAdam:\\
$m^{t+1} \gets \beta_1 m^t + (1-\beta_1)\Delta \theta _{i}^{t} $\;
$v^{t+1} \gets \beta_2 v^t + (1-\beta_2)(\Delta \theta _{i}^{t})^2 $\;
$\theta^{t+1} \gets \theta^{t} - \frac{\eta_s }{\sqrt{v^{t+1}} +\tau } m^{t+1}$\;
For FedAdagrad:\\
$m^{t+1} \gets \beta_1 m^t + (1-\beta_1)\Delta \theta _{i}^{t} $\;
$v^{t+1} \gets v^t + (\Delta \theta _{i}^{t})^2$\;
$\theta^{t+1} \gets \theta^{t} - \frac{\eta_s }{\sqrt{v^{t+1}} +\tau } m^{t+1}$\;
For FedYogi:\\
$m^{t+1} \gets \beta_1 m^t + (1-\beta_1)\Delta \theta _{i}^{t} $\;
$v^{t+1} \gets v^t + (1-\beta_2)(\Delta \theta _{i}^{t})^2$sign$( v^t - (\Delta \theta _{i}^{t})^2)$\;
$\theta^{t+1} \gets \theta^{t} - \frac{\eta_s }{\sqrt{v^{t+1}} +\tau } m^{t+1}$\;
}
return $\theta^{T}$\;
\textbf{Client $i$ executes:}\\
Receive $\theta^t$ from the server, $\theta _{i}^{t} \gets \theta^t$\;
\For{$k=1, 2,..., E$}{
\For{each batch $\mathcal{B} \subset \mathcal{D}^i$}{
$\theta _{i}^{t} \gets \theta _{i}^{t} - \eta_c \nabla L(\theta _{i}^{t};\mathcal{B})$\;
}
}
$\Delta \theta _{i}^{t} \gets \theta _{i}^{t} - \theta^t$\;
return $\Delta \theta _{i}^{t}$ to the server\;
\end{footnotesize}
\end{algorithm}

\section{Case Studies and Validation}

To evaluate the effectiveness and applicability of the proposed approaches, a series of case studies were conducted in four historic buildings in Sweden. These studies aimed to validate the integrated system for data-driven conservation by assessing its performance under real-world conditions and across varied building types. The case studies focus on demonstrating the feasibility of deploying IoT-based sensing systems, creating parametric digital twins, and developing deep learning models within operational heritage environments. Multiple types of data were collected, including environmental parameters, energy consumption, and weather conditions. The collected data support several investigations: evaluating the stability and reliability of the sensing infrastructure, analyzing indoor climate according to conservation guidelines, and predicting energy use and environmental conditions. Collectively, these case studies provide practical evidence for robustness, flexibility, and practical value of the proposed data-driven approaches in supporting smart maintenance strategies for historic buildings.

\subsection{Description of Case Study Buildings}

Four public historic buildings in Sweden were selected as case studies. Three are located in the city of Norrköping---the City Museum, the City Theatre, and the Auditorium---and one, Löfstad Castle, is situated in the Östergötland region. These buildings represent a range of architectural characteristics, functions, and conservation challenges. They are protected under national heritage legislation and remain in active use, making them well-suited for evaluating the applicability and robustness of the proposed data-driven approaches.

\begin{figure}[!tb]
\includegraphics{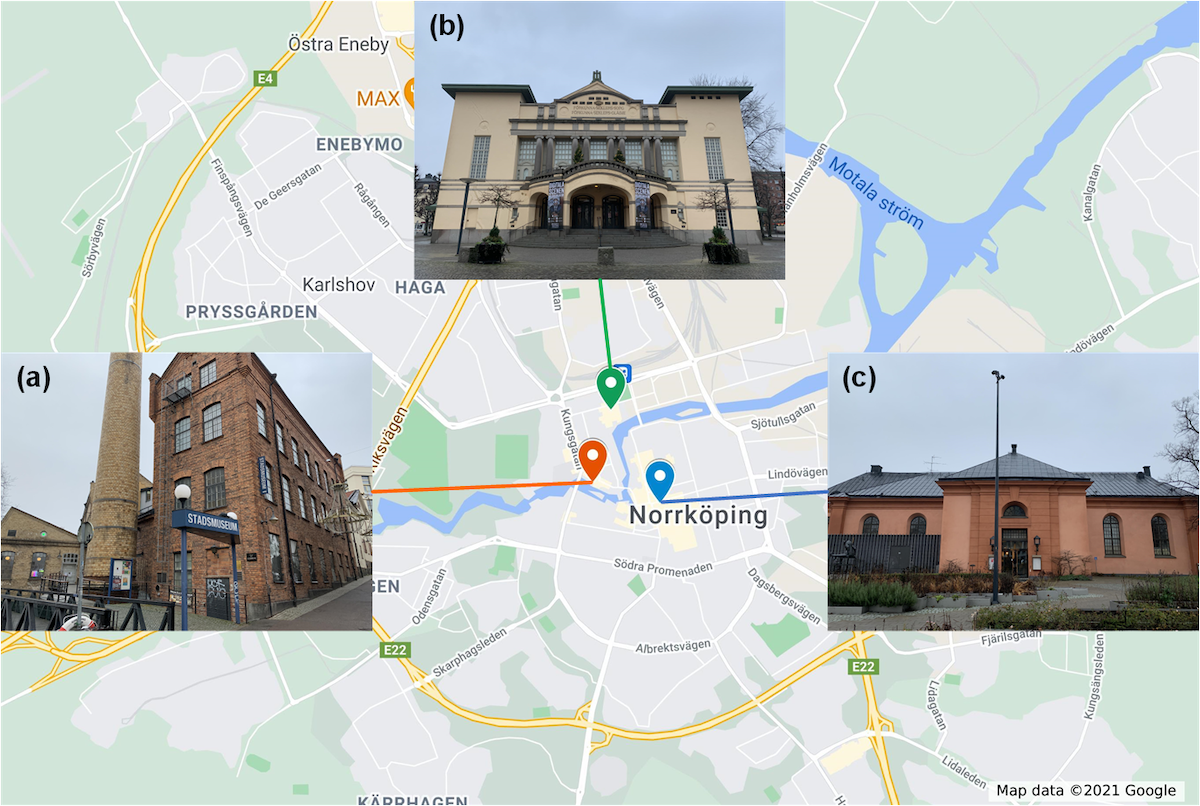}
\centering
\caption{Three historic buildings in Norrköping, Sweden, selected as case studies: (\textbf{a}) the City Museum, (\textbf{b}) the City Theatre, and (\textbf{c}) the Auditorium~\cite{ni_sensing_2021}.}\label{fig:phd_sensing_fig6}
\end{figure}

The City Museum (Fig.~\ref{fig:phd_sensing_fig6}{a}) is located in Norrköping's former industrial district, housed in 19th- and 20th-century factory buildings along the Motala River~\cite{nkp_city_musuem_buildings_2023}. The museum buildings were officially designated as protected in 1990. Its collections include nearly 40,000 artifacts reflecting Norrköping's industrial and cultural heritage, such as weaving and spinning machines, tools, printed fabrics, and advertising signs. Conservation efforts here require balancing public access with the need to preserve historically sensitive materials.

The City Theatre (Fig.~\ref{fig:phd_sensing_fig6}{b}), completed in 1908, is a prominent example of Art Nouveau architecture in central Norrköping~\cite{nkp_city_theatre_2023}. Protected since 1990, the theatre remains in active use for contemporary performing arts. The large open spaces, fluctuating occupancy, and complex lighting, heating, and ventilation systems make the building ideal for evaluating environmental monitoring and forecasting solutions in heritage buildings that host regular public events.

The Auditorium (Fig.~\ref{fig:phd_sensing_fig6}{c}) was originally built as a church in 1827 and served as the city's concert hall from 1913 to 1994~\cite{swedish_national_heritage_board_horsalen_2023}. It now functions as a venue for concerts and lectures. The building was designated as protected in 1978. Like the City Theatre, the Auditorium presents challenges for indoor climate control due to its continued public use and variable occupancy.

\begin{figure}[!tb]
\includegraphics{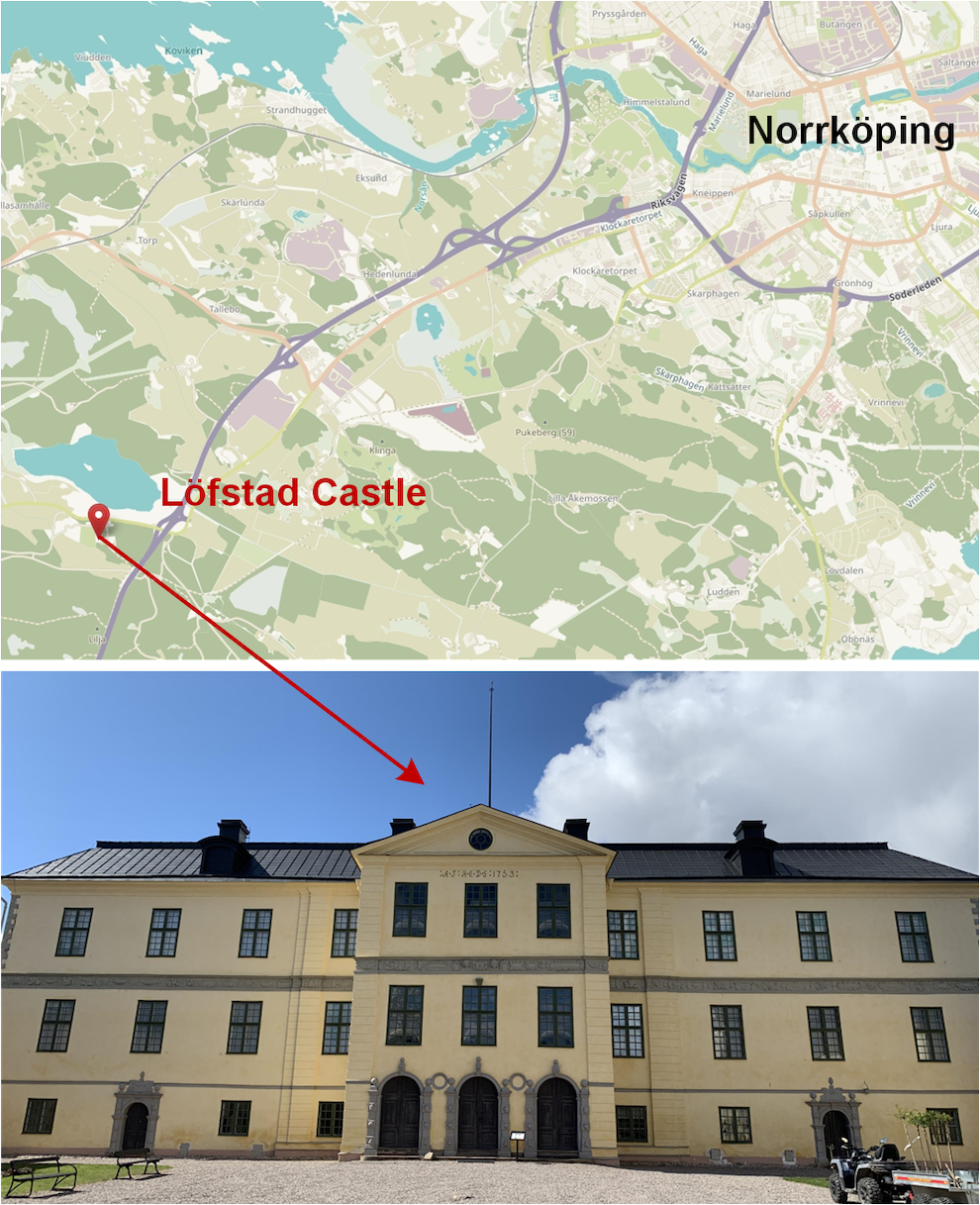}
\centering
\caption{The upper map shows the geographical location of Löfstad Castle, and the lower image depicts the main building of the castle~\cite{ni_parametric_2025}.}\label{fig:phd_parametric_fig3}
\end{figure}

Löfstad Castle (Fig.~\ref{fig:phd_parametric_fig3}), located approximately 9~km southwest of the city of Norrköping, is one of Sweden’s most well-preserved examples of a 17th--18th century manor house. The main building of the castle comprises a basement, three upper floors, and an attic, flanked by two wings and an enclosed courtyard. The main building features thick masonry walls and is naturally ventilated, with only limited electric heating on the first floor. It has been classified as a historic monument since 1983~\cite{lofstad_castle_swedish_national_heritage_board_online}. Its architecture blends Baroque and Rococo styles~\cite{hedlund_2013}. The final private owner, Miss Emilie Piper (1857--1926), donated the majority of the collections in the castle to Östergötland's Museum, and the estate property to Riddarhuset in her will. The property has since been preserved with minimal interior changes~\cite{en_bok_om_lofstad_slott_2022}.

\subsection{Deployment of Local Devices}

\begin{figure}[!tb]
\includegraphics{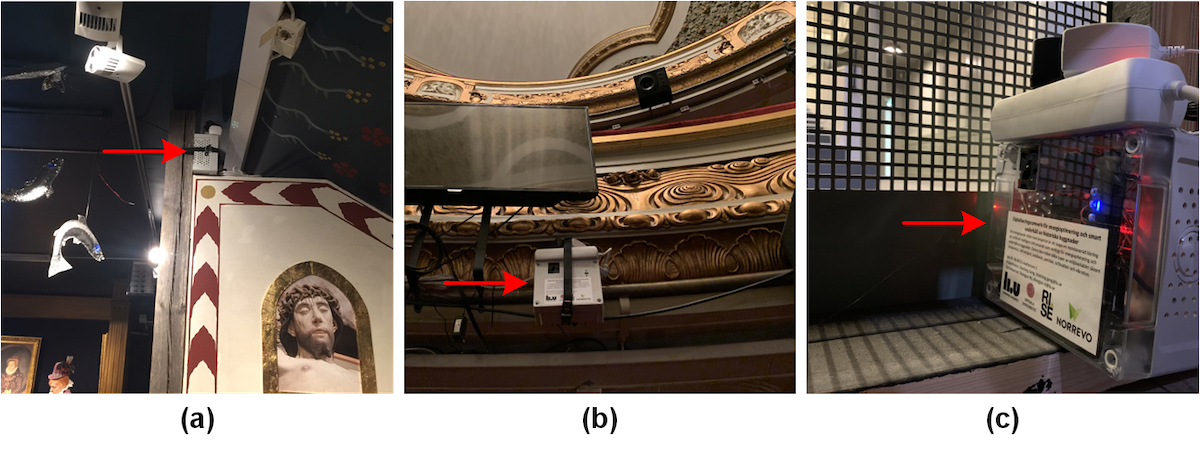}
\centering
\caption{The deployments of sensor boxes in: (\textbf{a}) the City Museum, (\textbf{b}) the City Theatre, and (\textbf{c}) the Auditorium~\cite{ni_sensing_2021}.}\label{fig:phd_sensing_fig7}
\end{figure}

The first phase of deployment began on 16 March 2021, with one sensor box installed in each of the three buildings: the City Museum, the City Theatre, and the Auditorium (see Fig.~\ref{fig:phd_sensing_fig7}). This phase aims to evaluate the robustness of the sensing system under real-world conditions and to gather preliminary insights into the indoor environmental dynamics of the selected historic buildings.

In the City Museum, the sensor box (Fig.~\ref{fig:phd_sensing_fig7}{a}) was installed in an exhibition room on the third floor, where delicate historical objects are displayed. This placement enables continuous monitoring of preservation-critical environmental conditions. In the City Theatre, the sensor box (Fig.~\ref{fig:phd_sensing_fig7}{b}) was positioned beneath the upper grandstand fence, near the spatial center of the main hall, to capture representative ambient conditions during shows. In the Auditorium, the sensor box (Fig.~\ref{fig:phd_sensing_fig7}{c}) was installed beneath the stage, allowing for the measurement of environmental parameters as well as the detection of occupancy-related activities. In all three buildings, environmental data were sampled every 15 seconds, corresponding to four samples per minute.

\begin{figure*}[!tb]
\includegraphics[width=\textwidth]{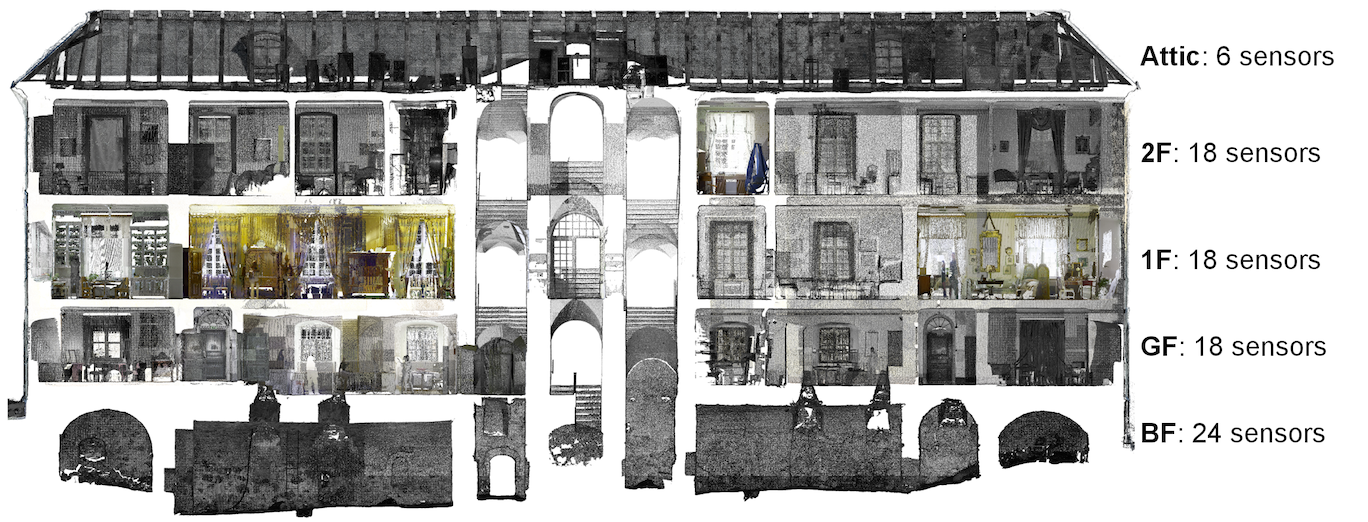}
\centering
\caption{Section view of the 3D model of the main building, showing the distribution of sensor deployments across floors~\cite{ni_parametric_2025}. Floor levels are labeled as BF (basement), GF (ground floor), 1F (first floor), and 2F (second floor).}\label{fig:phd_parametric_fig5}
\end{figure*}

\begin{figure*}[!tb]
\includegraphics[width=\textwidth]{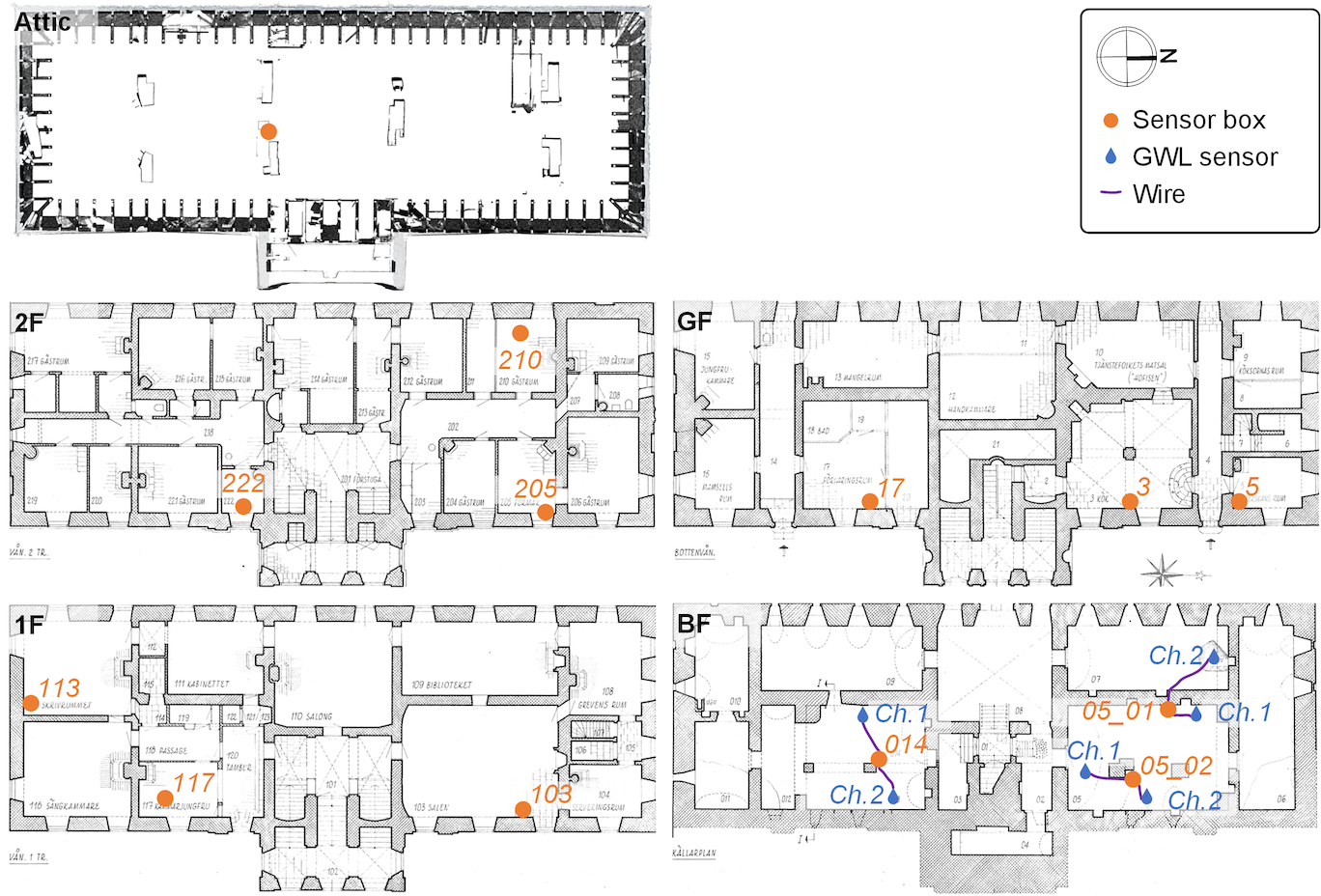}
\centering
\caption{Sensor box and groundwater level (GWL) sensor placement~\cite{ni_parametric_2025}. Orange dots indicate sensor box locations, annotated with room numbers, while blue droplets mark the positions of GWL sensors. In the basement, each sensor box is connected to two additional GWL sensors, labeled as Channel 1 (Ch. 1) and Channel 2 (Ch. 2).}\label{fig:phd_parametric_fig6}
\end{figure*}

A more extensive deployment was carried out at Löfstad Castle, starting in January 2023, with 13 sensor boxes and a total of 84 sensors distributed across all floors of the main building (see Fig.~\ref{fig:phd_parametric_fig5} and Fig.~\ref{fig:phd_parametric_fig6}). Three sensor boxes were installed in the basement (BF), ground floor (GF), first floor (1F), and second floor (2F), respectively, with one placed in the attic. This configuration enables continuous environmental monitoring from the basement to the attic. In addition, six groundwater level (GWL) sensors were installed in the basement to monitor fluctuations in groundwater and identify potential moisture sources that may influence indoor humidity. At Löfstad Castle, environmental data were sampled every 30 seconds to capture short-term fluctuations. This high-frequency data collection supports fine-grained analysis and allows flexible downsampling for visualization and modeling. Further details on the deployment are presented in Paper III.

\begin{figure}[!tb]
\includegraphics{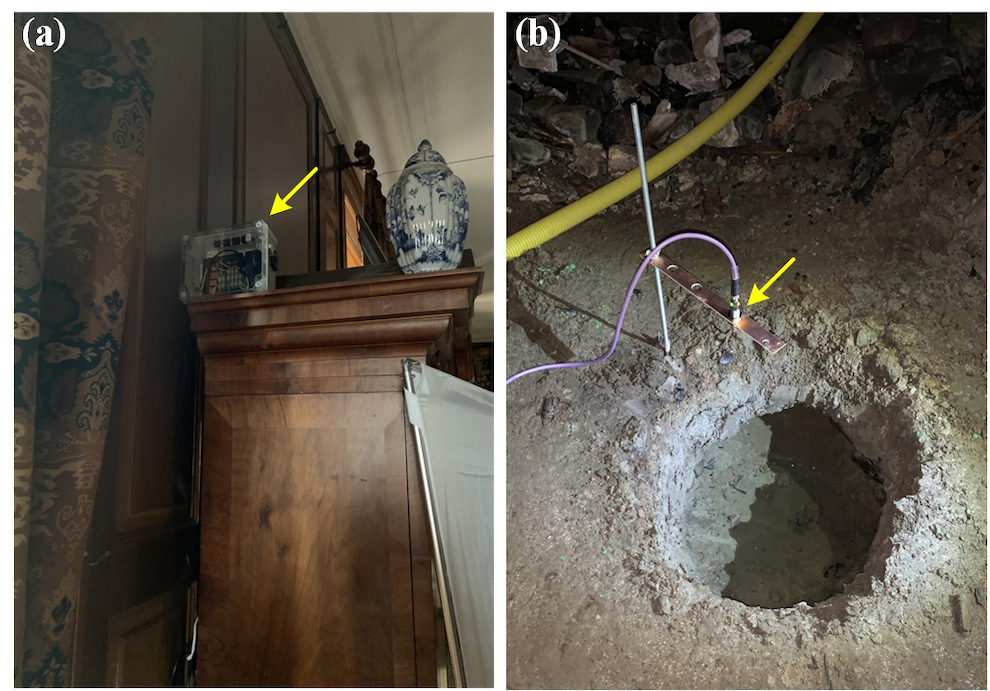}
\centering
\caption{Deployment examples~\cite{ni_parametric_2025}: (\textbf{a}) sensor box installed in Room 103 on the first floor, and (\textbf{b}) GWL sensor (Ch. 1) connected to sensor box 05\_01 in the basement.}\label{fig:phd_parametric_fig7}
\end{figure}

At Löfstad Castle, examples of deployed local devices are shown in Fig.~\ref{fig:phd_parametric_fig7}. In Room 103, the sensor box is positioned on top of a cabinet (see Fig.~\ref{fig:phd_parametric_fig7}{a}). This placement enables monitoring of both indoor climate conditions and occupancy patterns. Fig.~\ref{fig:phd_parametric_fig7}{b} shows a GWL sensor that uses ultrasound to measure the distance from the probe to the water surface. GWL measurements are integrated into the parametric digital twin and analyzed with outdoor precipitation data to identify correlations between rainfall and GWLs. Understanding this relationship supports the assessment of moisture-related risks, such as rising damp, and informs proactive conservation strategies within the digital twin solution.

\subsection{Collection of Other Data}

In addition to local sensor measurements, two supplementary datasets, i.e., energy consumption and outdoor weather conditions, were collected to support analysis and modeling tasks in this study. These datasets are essential for understanding the influence of external factors on indoor environmental dynamics and energy usage in historic buildings.

\begin{figure*}[!tb]
\includegraphics[width=\textwidth]{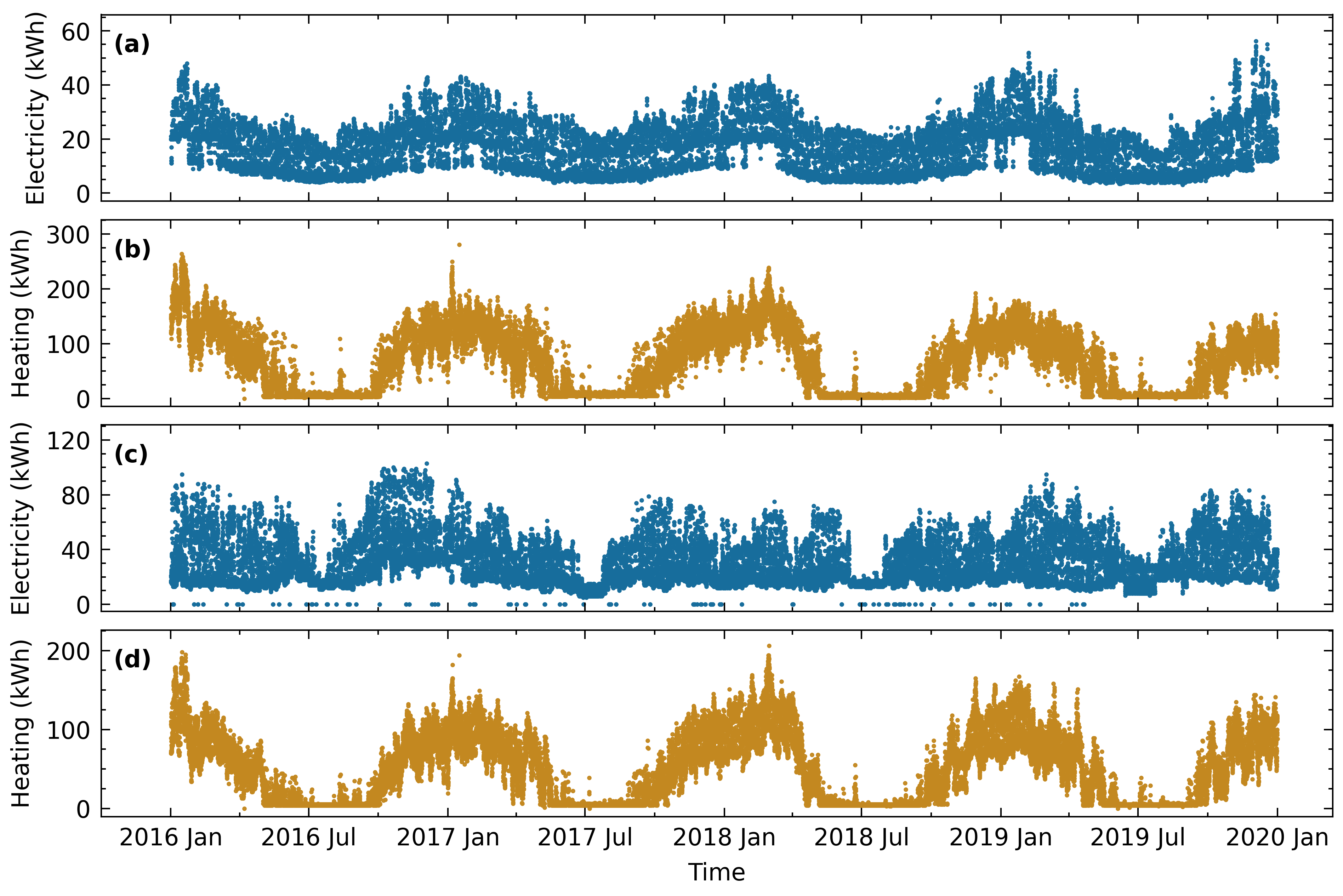}
\centering
\caption{Historical hourly energy use data for the City Museum and City Theatre in Norrköping, Sweden, from 01:00 on January 1, 2016 to 00:00 on January 1, 2020~\cite{ni_deep_2024}: (\textbf{a}) electricity consumption and (\textbf{b}) heating load for the City Museum; (\textbf{c}) electricity consumption and (\textbf{d}) heating load for the City Theatre. All timestamps are in 24-hour format and follow Central European Time (CET).}\label{fig:phd_deep_fig5}
\end{figure*}

Historical hourly energy consumption data were provided by the facility manager responsible for the City Museum, the City Theatre, and the Auditorium in Norrköping. These records span periods before the COVID-19 pandemic, ensuring that variations in public activity caused by pandemic-related restrictions did not distort typical consumption patterns. An example of the energy data from the City Museum and the City Theatre is shown in Fig.~\ref{fig:phd_deep_fig5}. These two buildings represent different operational modes. The City Museum maintains fixed opening hours and aims to provide a stable indoor environment for both heritage conservation and occupant comfort. In contrast, the City Theatre is active primarily during scheduled shows and rehearsals. Its energy consumption fluctuates with show schedules, which vary seasonally and often involve concentrated periods of high activity. These differences offer valuable conditions for evaluating the adaptability and generalization capabilities of predictive models.

Outdoor weather data were obtained through open APIs provided by the Swedish Meteorological and Hydrological Institute (SMHI)~\cite{smhi_open_data_api_2023}. Hourly measurements of dry-bulb temperature, RH, dew point temperature, precipitation, and air pressure were collected from a nearby weather station, which is located approximately 2~km from the three buildings in Norrköping and 7~km from Löfstad Castle. Hourly global irradiance data were retrieved using the geographic coordinates of each building. These measurements are considered sufficiently representative of local ambient conditions, with minimal deviation expected due to the short distances involved.

\subsection{System Stability Assessment}

To evaluate the stability of the IoT sensing system, the end-to-end data transmission loss rate was used as the primary performance metric. This metric reflects the combined reliability of all components involved in the data transmission pipeline, including sensors, edge platforms, communication protocols, and cloud services. Thus, the loss rate provides a comprehensive measure of the system’s overall operational reliability.

The loss rate was evaluated over a 56-day period for three sensor boxes deployed in the City Theatre, the City Museum, and the Auditorium. The period is from 5 April 2021 at 00:00 to 31 May 2021 at 00:00, using Central European Time (CET) as the reference timezone. This time window was chosen to ensure a sufficiently long and continuous period of operation, allowing for a robust evaluation of system performance under real-world conditions.

\subsection{Analysis of Indoor Climate}

The analysis of indoor climate provides critical insights into the environmental dynamics of historic buildings and supports evidence-based strategies for preventive conservation. This study examines indoor climate across four key aspects: temporal patterns of temperature and RH at each monitoring point; correlations between environmental variables; the influence of occupancy, particularly during public events; and seasonal and short-term RH fluctuations that may pose risks to material preservation.

To explore temporal patterns, indoor and outdoor temperature and RH were jointly analyzed. At Löfstad Castle, additional comparisons were conducted across rooms on the same floor and between different floors to identify spatial variations. Correlations between environmental variables were evaluated using the Pearson correlation coefficient, while the Mann–Whitney U test~\cite{mann_test_1947} was applied to assess differences in distributions or medians between groups. The influence of occupants, such as audience presence during shows at the City Theatre, was examined using a combination of EDA and classical statistical methods. EDA captured changes in indoor environmental conditions before and during events, while statistical tests quantified the impact of occupancy levels. Seasonal and short-term RH fluctuations were assessed following the method defined in EN 15757:2010~\cite{en_15757_conservation_2010}. This involved calculating three key statistics: the annual average RH, seasonal cycles using a 30-day centered moving average (CMA), and short-term deviations from the seasonal baseline. Further details of the indoor climate analysis are presented in Papers I and II.

At Löfstad Castle, further analysis was conducted to investigate potential moisture sources and their impact since persistently high RH was observed in the basement. A detailed discussion of this work is presented in Paper III. Two additional specific research questions were explored: how GWLs vary across locations in the basement, and to what extent high RH in the basement influences conditions on the upper floors.

To address the first question, variations in GWL were examined across multiple monitoring points in the basement. The influence of precipitation on GWL dynamics was also analyzed to identify potential correlations between rainfall and moisture accumulation in the basement.

To investigate the second question, indoor and outdoor humidity mixing ratios (MRs)~\cite{brostrom_climate_2015} were compared over a full calendar year. MR, expressed in grams of water vapor per kilogram of dry air (g/kg), quantifies the moisture content in air~\cite{en_16242_conservation_2012}. A consistently higher indoor MR relative to the outdoor MR may indicate ongoing evaporation from moisture sources within the building, such as damp walls or floors. MR was calculated using the following equation:
\begin{equation}
\text{MR}\ =\ 38.015\times \frac{10^{\frac{7.65t}{243.12+t}} \times \text{RH}}{p-\left( 0.06112\times 10^{\frac{7.65t}{243.12+t}} \times \text{RH}\right)},\label{eq:phd_mr}
\end{equation}
where $t$ is temperature ($^{\circ}$C), RH is relative humidity (\%), $p$ is atmospheric pressure (hPa). As indoor pressure was not measured, a standard atmospheric pressure of 1013 hPa was assumed, following EN 16242:2012~\cite{en_16242_conservation_2012}. This approximation introduces an error of less than 1\%, which does not affect the validity of the MR comparisons.

To evaluate risks associated with high RH, mold growth potential was estimated using the isopleth system for substrate category I (LIM\textsubscript{I})~\cite{sedlbauer_prediction_2001}, implemented in WUFI-Bio~\cite{lim_expression_online}. LIM\textsubscript{I} was calculated as follows:
\begin{equation}
    \text{LIM}_{\text{I}} =\ \text{cosh}( 0.128324\times ( 30-t)) +75,\label{eq:phd_rh_lim}
\end{equation}
where $t$ is temperature ($^{\circ}$C). The LIM\textsubscript{I} value, expressed as a percentage, represents the RH threshold above which the risk of biological degradation increases. RH values exceeding this threshold indicate an elevated risk of mold growth and other conservation concerns.

\chapter{IoT-based Sensing System for Historic Buildings}
\label{cha:iot}

This chapter presents selected highlights of the IoT-based sensing system developed for long-term indoor environmental monitoring in historic buildings. It demonstrates practical usability by summarizing its operational stability and associated data applications. These components provide the foundation for continuous data collection, visualization, and analysis, which are critical elements for supporting data-driven maintenance practices. While this chapter outlines key aspects of the system, further details are available in Papers I, II, and III.

\section{System Stability}

The stability of the system was evaluated using data collected between 5 April 2021 at 00:00 and 31 May 2021 at 00:00 (CET). During this 56-day period, three sensor boxes deployed in the City Museum, the City Theatre, and the Auditorium were expected to transmit a total of 967,680 data samples, based on a sampling interval of 15 seconds per sensor box. As shown in Table~\ref{tab:phd_sensing_tab1}, the average data loss rate across all sensor boxes was approximately 2\%, with no apparent variation among individual devices. A total of 19,330 samples were lost. The losses were evenly distributed rather than occurring in large blocks, meaning that there were no extended gaps in the dataset.


\begin{table}[!tb]
\centering
\caption{The loss rate of data samples per sensor box and the average loss rate of data samples for three sensor boxes.}
\label{tab:phd_sensing_tab1}
\end{table}
\begin{figure}[!h]
\includegraphics[width=\textwidth]{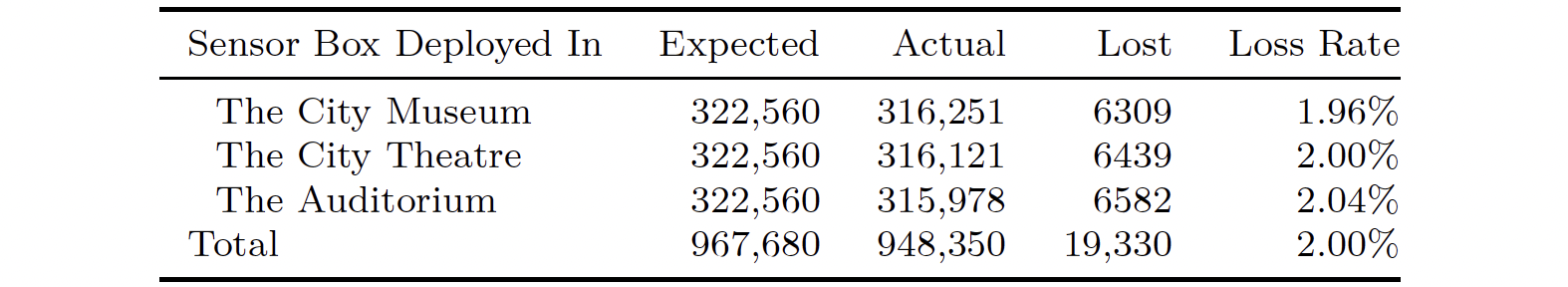}
\centering
\end{figure}

\begin{figure}[!tb]
\includegraphics{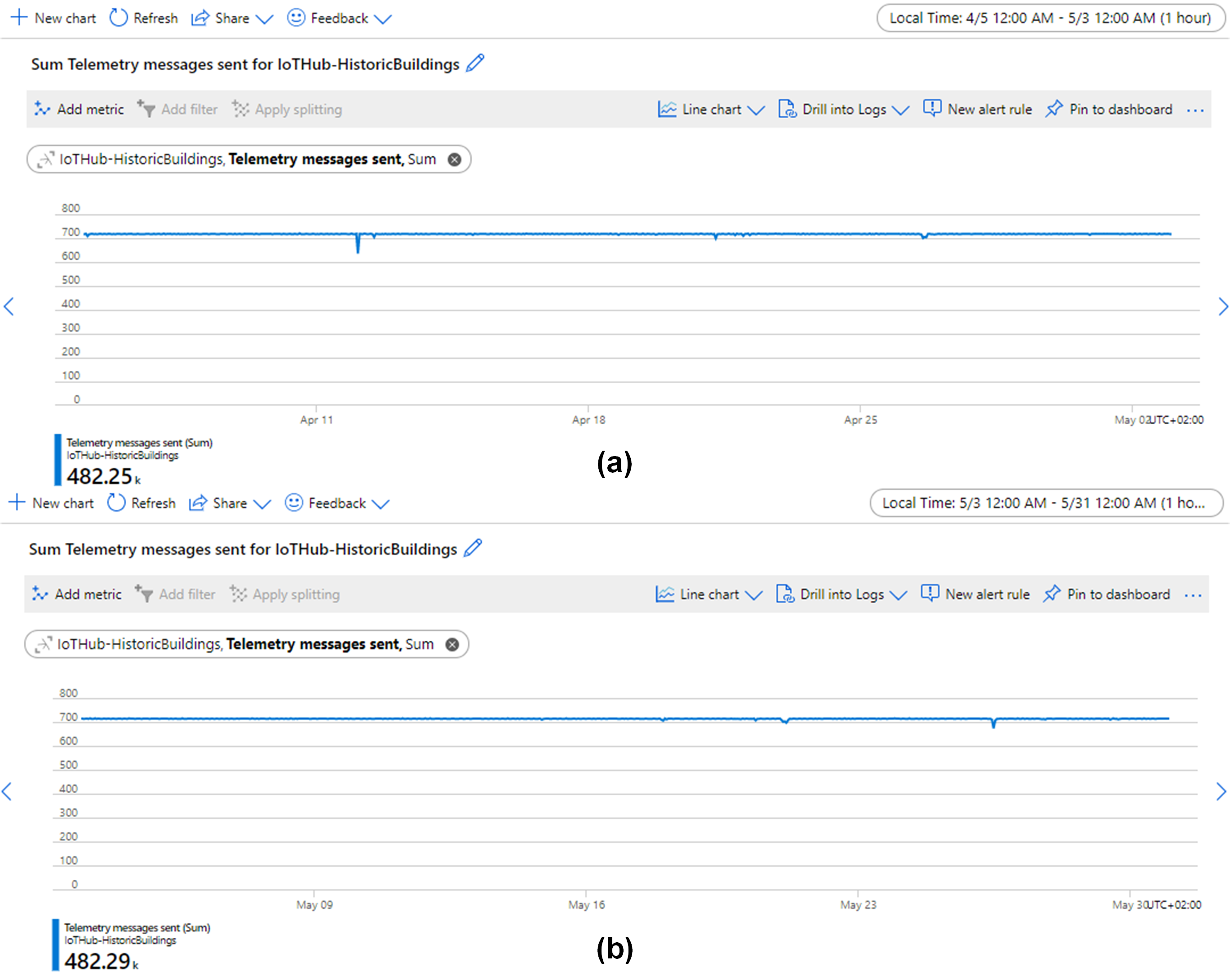}
\centering
\caption{Hourly aggregated message counts transmitted from the edge platform to the IoT Hub between 5 April and 31 May~\cite{ni_sensing_2021}. A total of approximately 964,540 messages were sent: (\textbf{a}) around 482,250 messages from 5 April to 3 May, and (\textbf{b}) about 482,290 messages from 3 May to 31 May.}\label{fig:phd_sensing_fig12}
\end{figure}

To identify the causes of these losses, system-level diagnostics from Azure IoT Hub and Azure Functions were analyzed. As depicted in Fig.~\ref{fig:phd_sensing_fig12}, the IoT Hub received approximately 964,540 messages, suggesting that around 3,140 data samples were lost during transmission from the sensor boxes and the cloud gateway. This loss is likely due to intermittent Internet connectivity, which is expected in real-world historic buildings where network condition may be inconsistent.

\begin{figure}[!tb]
\includegraphics{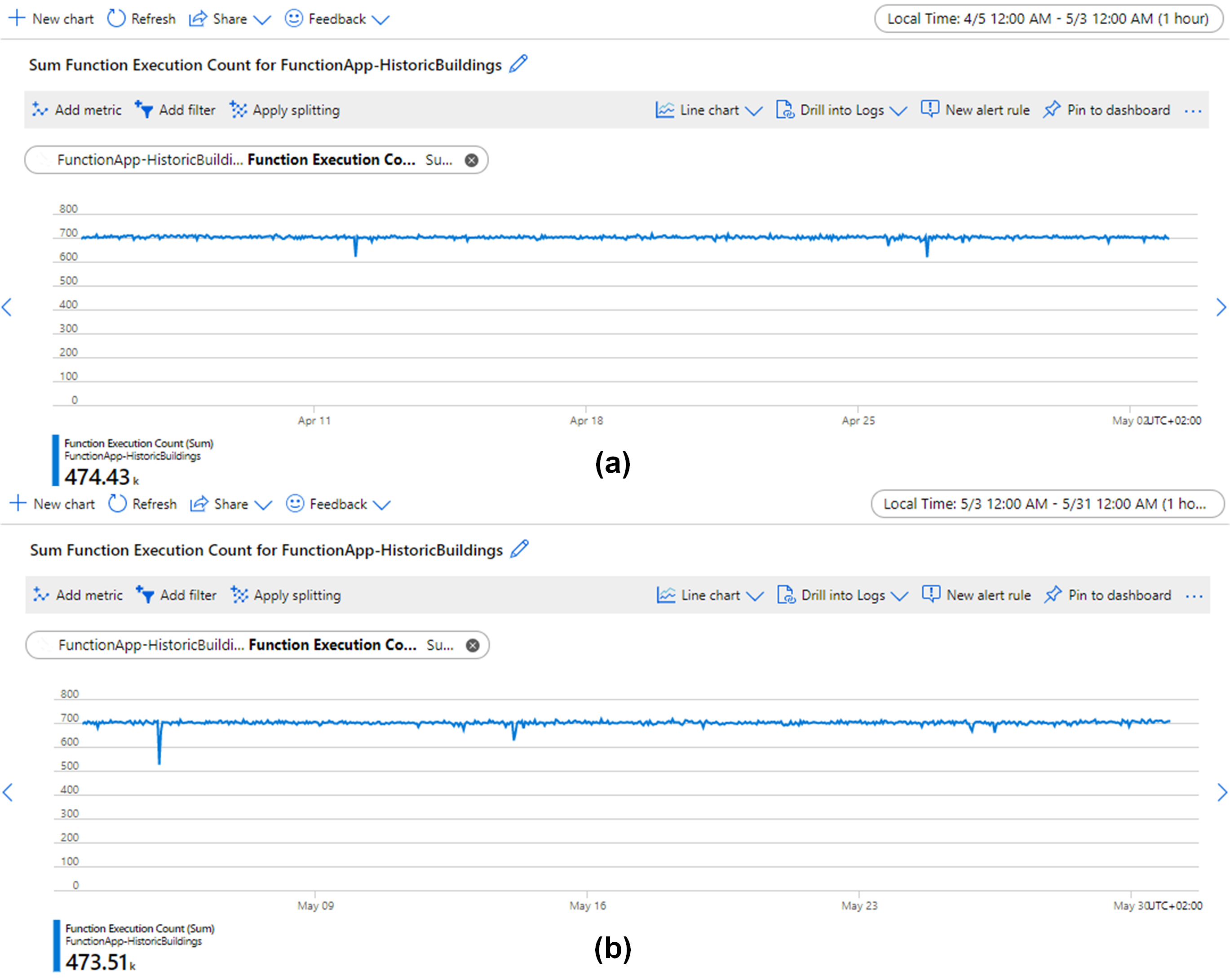}
\centering
\caption{Hourly aggregated counts of executed functions between 5 April and 31 May~\cite{ni_sensing_2021}. A total of approximately 947,940 executions were recorded: (\textbf{a}) about 474,430 between 5 April and 3 May, and (\textbf{b}) around 473,510 between 3 May and 31 May.}\label{fig:phd_sensing_fig13}
\end{figure}

Fig.~\ref{fig:phd_sensing_fig13} shows that Azure Functions were executed approximately 947,940 times. This count is slightly lower than the actual number of received data samples (948,350), indicating minor inconsistencies in metric reporting. After correcting for this, it was estimated that approximately 16,190 samples were lost between the IoT Hub and Azure Functions. As summarized in Table~\ref{tab:phd_sensing_table2}, this segment of the cloud pipeline accounted for roughly 84\% of the total data loss. The primary cause was the use of a shared service plan for hosting the Function App, which lacks a guaranteed service level agreement (SLA)~\cite{azure_subscription_2021}.

\begin{table}[!tb]
\centering
\caption{Summary of data loss by location, including the number and percentage of missing samples~\cite{ni_sensing_2021}.}
\label{tab:phd_sensing_table2}
\end{table}
\begin{figure}[!tb]
\includegraphics[width=\textwidth]{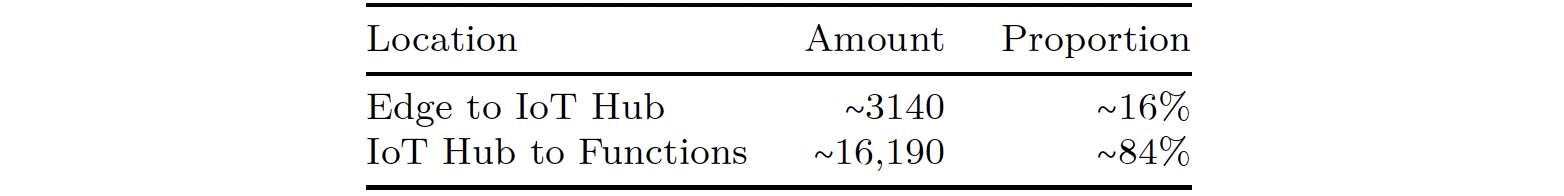}
\centering
\end{figure}

Despite these losses, the system demonstrated robust performance and is well-suited for long-term environmental monitoring in historic buildings. A 2\% loss rate is acceptable for this application and does not compromise the analytical integrity of the dataset. For example, standards such as EN 15757:2010~\cite{en_15757_conservation_2010} typically require hourly data, which can still be reliably derived. For scenarios requiring higher reliability, the cloud service plan can be upgraded to one with a 99.95\% SLA, reducing the expected loss rate to approximately 0.3\% and further improving system suitability for critical heritage monitoring tasks.

\section{Developed Data Applications}

To support data-driven decision-making in heritage conservation, four data applications were developed, focusing on the visualization, analysis, and sharing of collected environmental data. These tools are designed to enhance the accessibility, interpretability, and practical utility of collected data for researchers, facility managers, and other stakeholders involved in the preservation of historic buildings. Real-time visualization facilitates the rapid detection of changes in environmental conditions and their potential causes. Importantly, the applications feature user-friendly interfaces that enable non-experts to access and interpret key insights without requiring advanced technical skills. This promotes timely, informed decision-making for preventive conservation and maintenance, and encourages broader adoption among users with diverse backgrounds.

\begin{figure*}[!tb]
\centering
\subfloat[]{\includegraphics[width=0.8\textwidth]{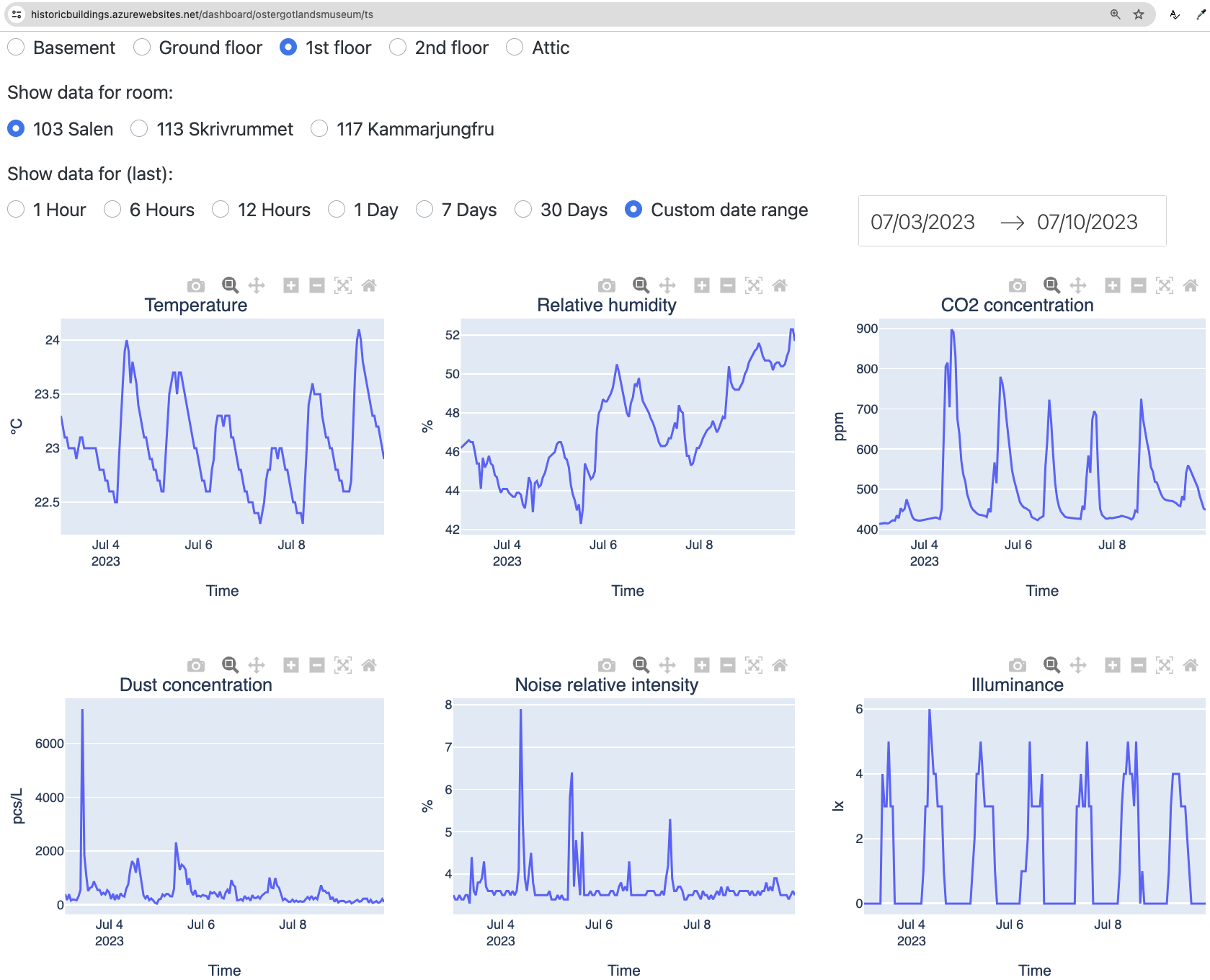}}
\hfill
\subfloat[]{\includegraphics[width=0.8\textwidth]{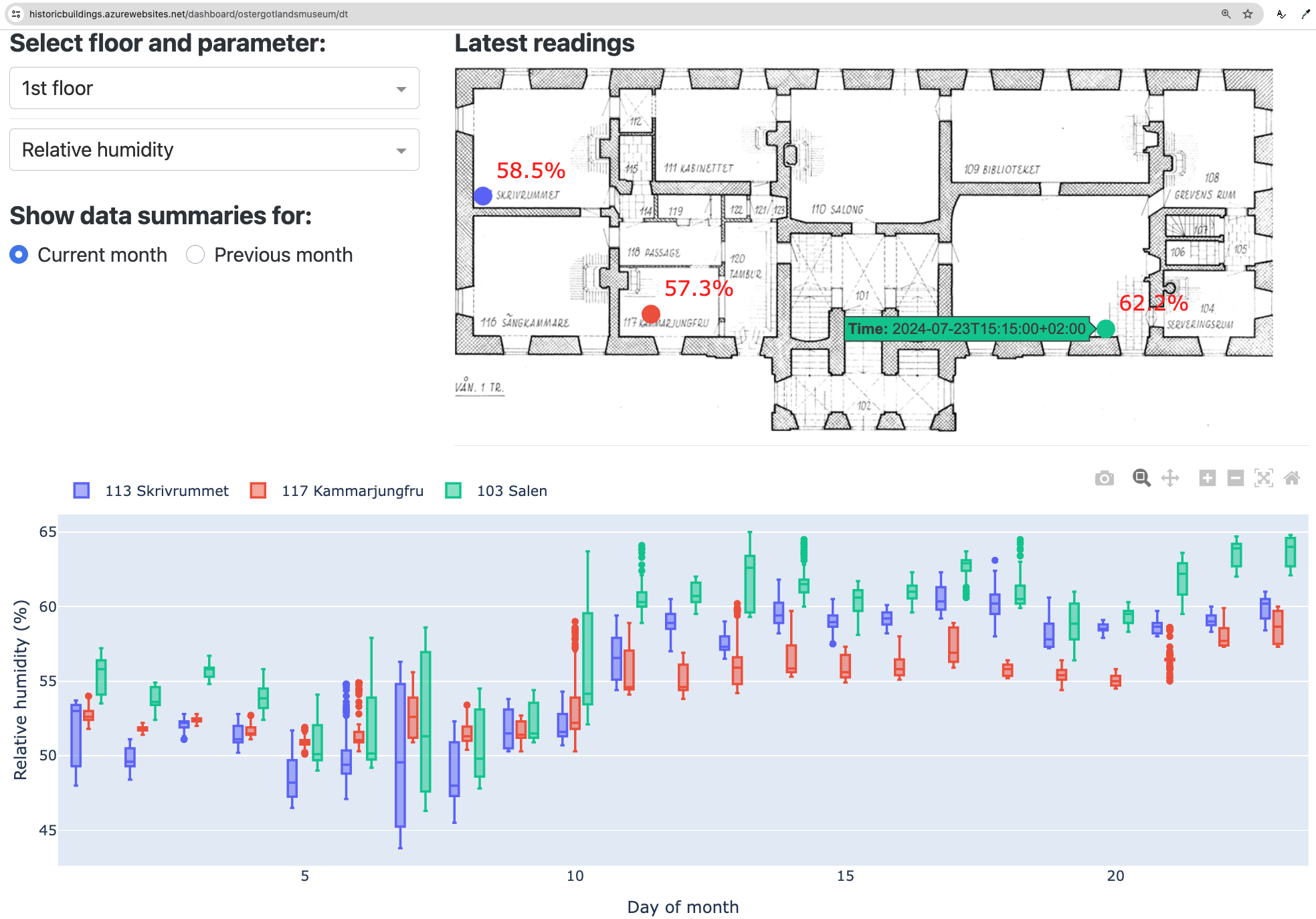}}
\caption{Overview of developed data applications~\cite{ni_parametric_2025}: (\textbf{a}) application for visualizing time series of indoor environmental data by selected room and time range; (\textbf{b}) applications for displaying current values of a chosen parameter across a selected floor, alongside daily distributions for a specified month.}
\label{fig:phd_parametric_fig18}
\end{figure*}

The first application offers interactive visualization of time series data from selected sensor boxes. As illustrated in Fig.~\ref{fig:phd_parametric_fig18}{a}, users can specify a date range and simultaneously inspect multiple environmental parameters, such as temperature, RH, and CO\textsubscript{2} concentration. This tool enables detailed analysis of trends, fluctuations, and correlations at specific monitoring points. For example, data collected from Room 103 over one week revealed a pronounced daily temperature cycle. Additionally, six distinct peaks in CO\textsubscript{2} levels reflected varying occupancy patterns. During the five-day period from 4 to 8 July, the Pearson correlation coefficient between temperature and CO\textsubscript{2} concentration was 0.64, indicating a strong positive correlation likely associated with occupant presence.

The second and third applications complement time series plots by providing spatial and statistical overviews of sensor data. As shown in Fig.~\ref{fig:phd_parametric_fig18}{b}, the upper panel displays real-time sensor readings overlaid on a digital floor plan, enabling intuitive spatial comparisons across rooms. The lower panel presents a boxplot summarizing the daily distribution of a selected environmental parameter for a specified month. This visualization supports the identification of typical value ranges, detection of outliers, and assessment of environmental stability over time.

The fourth application facilitates open data access for project participants and external researchers (see https://historicbuildings.azurewebsites.net/dow\\nload). Users can specify sampling intervals, environmental parameters, and time periods for data export. The downloadable datasets are preprocessed to improve quality and usability. Two types of anomalies are addressed: outliers resulting from sensor noise or malfunction are removed using threshold-based rules based on sensor specifications and typical indoor conditions; short missing data segments (less than two hours) are interpolated linearly to preserve continuity, while longer gaps are retained to avoid introducing bias. The cleaned datasets are provided in comma-separated values (CSV) format to ensure compatibility with a wide range of analysis tools.

\chapter{Parametric Digital Twins for Preventive Conservation}
\label{cha:parametric}

This chapter presents selected applications of parametric digital twins developed to support preventive conservation in historic buildings. Parametric digital twins enable the integration of contextual information and real-time sensor data, providing a consistent and extensible representation of building conditions. While a range of case studies has been conducted as part of this research, this chapter highlights three representative examples. First, it illustrates the creation of a parametric digital twin for the City Theatre, including how physical components and sensor data are structured and queried. Second, it explores how the digital twin helps analyze occupancy-related impacts on the indoor climate during live shows. Third, it presents an in-depth investigation of high humidity in Löfstad Castle, revealing moisture migration patterns and guiding conservation strategies. These selected examples demonstrate the practical value of parametric digital twins for heritage conservation. Further details are available in Papers II and III.

\section{The Created Digital Twin of the City Theatre}

\begin{figure}[!tb]
\includegraphics{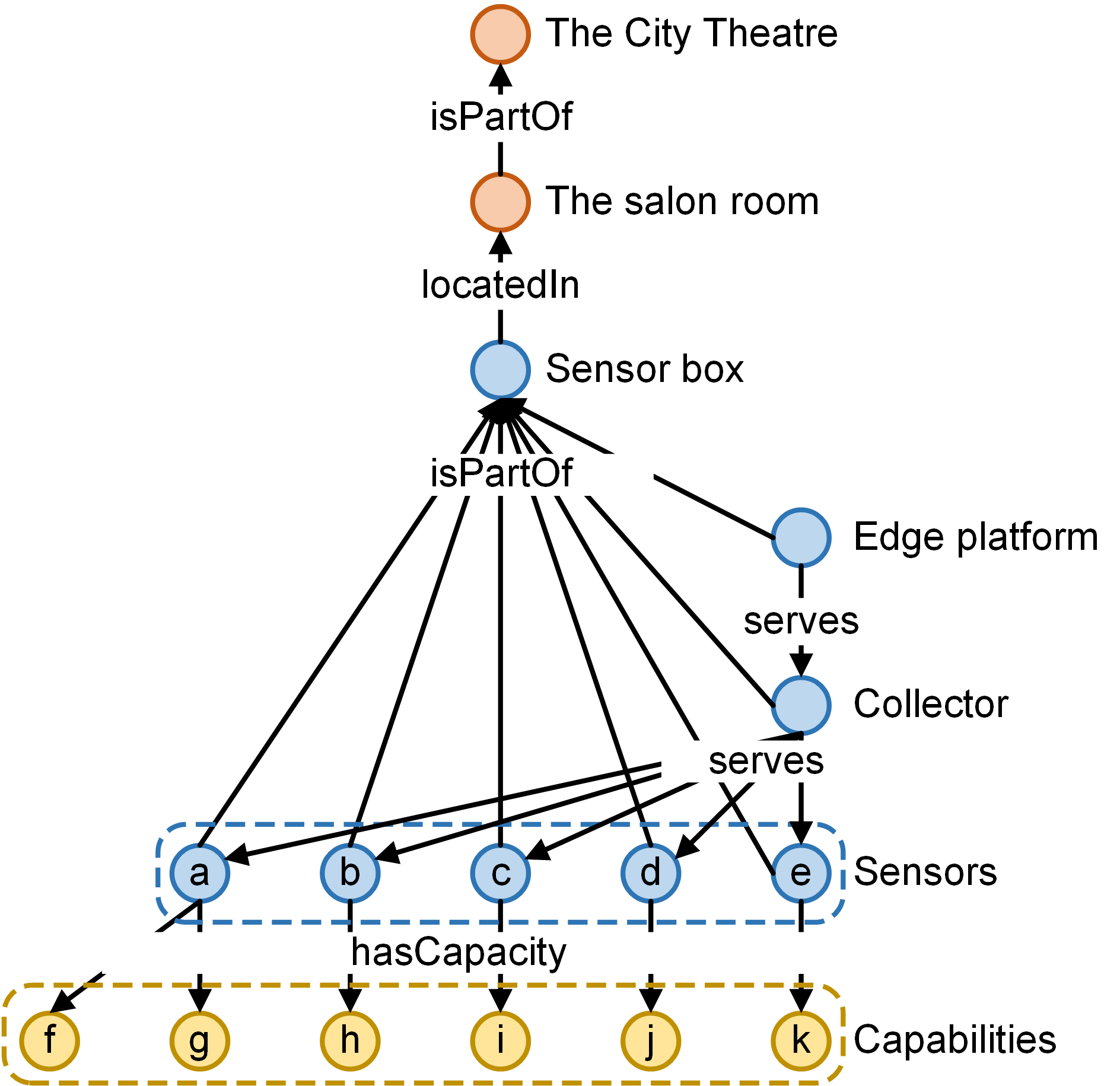}
\centering
\caption{Twin graph representing the digital twin of the City Theatre~\cite{ni_enabling_2022}. (\textbf{a})--(\textbf{e}) denote sensor devices: temperature and humidity sensor, CO\textsubscript{2} sensor, dust sensor, air quality sensor, and vibration sensor, respectively. (\textbf{f})--(\textbf{k}) indicate the corresponding measurement functions: temperature, relative humidity, CO\textsubscript{2} concentration, dust concentration, harmful gas concentration, and vibration.}\label{fig:phd_enabling_fig10}
\end{figure}

A limited parametric digital twin of the City Theatre was developed using data from the locally deployed sensor box. As illustrated in Fig.~\ref{fig:phd_enabling_fig10}, the digital twin is represented as a graph, where circular nodes denote virtual models of physical entities and directed edges represent semantic relationships between them. Each node corresponds one-to-one with a real-world object, and each edge reflects a meaningful physical or spatial relationship. For example, the top-level node \textit{The City Theatre} represents the entire building and is linked to the node \textit{The salon room} via an \textit{isPartOf} relationship, indicating that the salon is a component of the City Theatre.

The digital twin is designed to be modular, extensible, and adaptable. New virtual entities and relationships can be added to represent newly modeled rooms, installed sensors, or building components. Existing entities can be enriched with additional properties to more accurately reflect evolving characteristics of their physical counterparts.

\begin{figure}[!tb]
\includegraphics[width=3.3in]{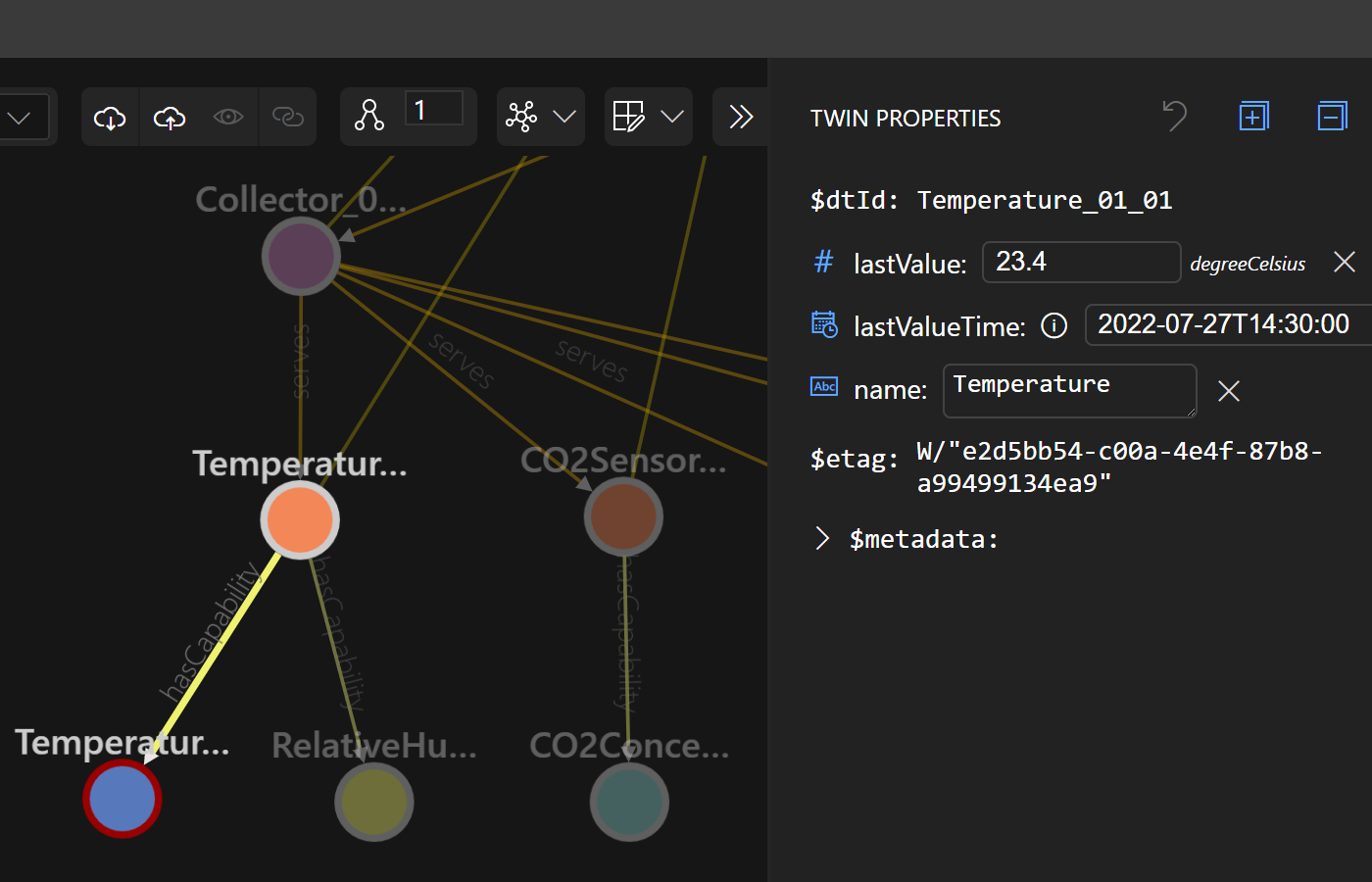}
\centering
\caption{Interface of the digital twin explorer used to inspect the properties of virtual entities~\cite{ni_enabling_2022}.}\label{fig:phd_enabling_fig11}
\end{figure}

The created digital twin supports three core capabilities. First, it enables near real-time monitoring by integrating streaming data from IoT sensors into the virtual model. Users can interact with the graph to inspect the current status of specific physical entities. For example, as shown in Fig.~\ref{fig:phd_enabling_fig11}, selecting the node corresponding to a temperature sensor displays the current indoor temperature in the salon room.

Second, the digital twin functions as a structured knowledge base for the historic building. The underlying ontology-based graph can be queried using a custom query language similar to structured query language (SQL), allowing users to retrieve information about entities and their current states. For instance, users can identify all rooms in the City Theatre where the CO\textsubscript{2} concentration is below 500 parts per million (ppm) using the below statement. This semantic querying capability supports informed decision-making and efficient exploration of the system.


\begin{figure}[!htb]
\includegraphics[width=\textwidth]{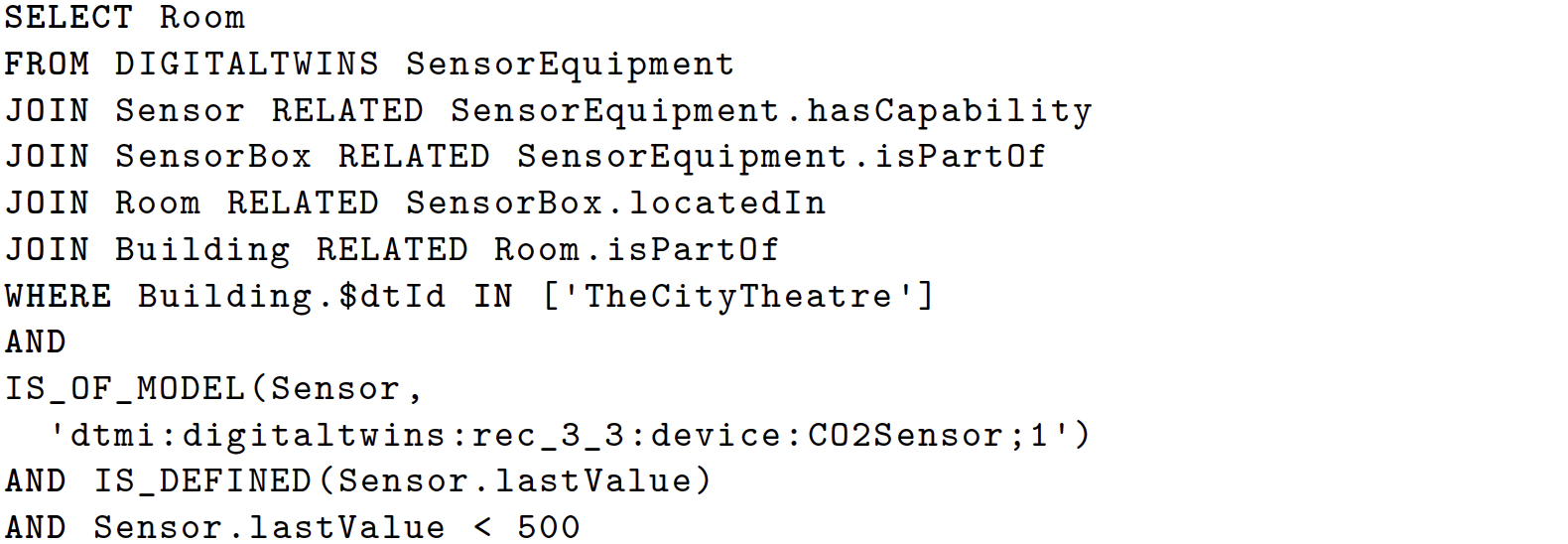}
\centering
\end{figure}

Third, the digital twin facilitates integration with data analytics to support preventive conservation strategies. Both historical and real-time data stored in the twin can be analyzed to detect patterns, correlations, and anomalies. As described in the data applications (see Chapter~\ref{cha:iot}), this integration enables users to gain insights into environmental dynamics and optimize building operations. Analytics-driven decision support empowers facility managers and conservation professionals to implement proactive measures that preserve the building and maintain stable indoor environmental conditions.

\section{The Impact of Occupants on Indoor Environment in the City Theatre}

\begin{figure}[!tb]
\includegraphics{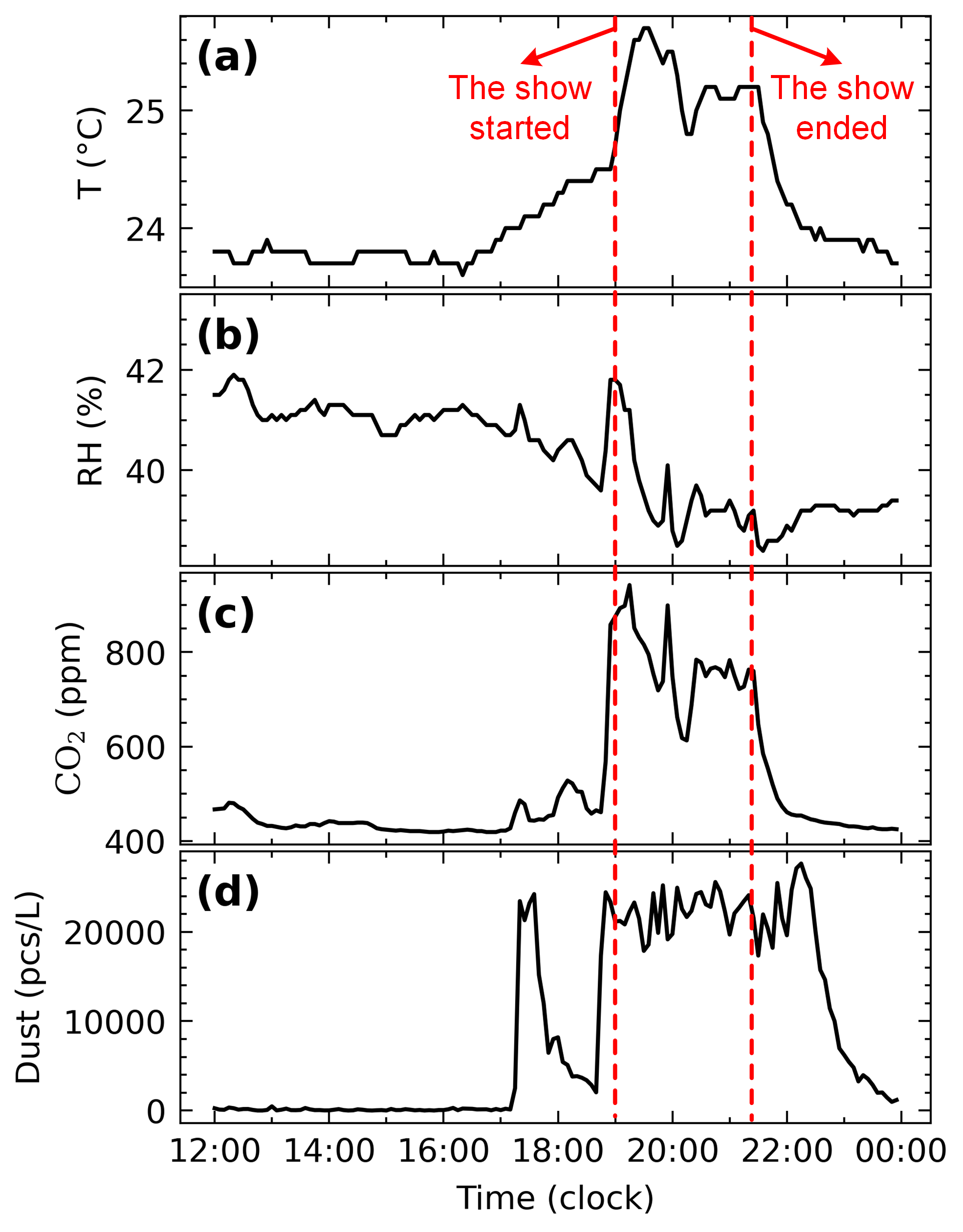}
\centering
\caption{Indoor environmental conditions in the salon room on October 7, 2021, from 12:00 to 23:59~\cite{ni_enabling_2022}: (\textbf{a}) temperature (T), (\textbf{b}) RH, (\textbf{c}) CO\textsubscript{2} concentration, and (\textbf{d}) dust concentration.}\label{fig:phd_enabling_fig13}
\end{figure}

Occupant presence during live shows has a clear impact on the indoor environment of the salon room in the City Theatre. Fig.~\ref{fig:phd_enabling_fig13} presents a representative example from a show held between 19:00 and approximately 21:20.

As shown in Fig.~\ref{fig:phd_enabling_fig13}{a}, the indoor temperature began to rise around 19:00 as the audience entered the salon and reached a peak of 25.6~\textdegree{}C around 19:30. The temperature then remained elevated throughout the show. A brief drop followed by a rise after 20:00 corresponds to a short intermission, during which some audience temporarily left the room and later returned. After the show ended, the temperature gradually declined to below 24~\textdegree{}C. A similar trend was observed in CO\textsubscript{2} concentration (Fig.~\ref{fig:phd_enabling_fig13}{c}), which increased sharply as the audience arrived and decreased after the show ended.

The RH, depicted in Fig.~\ref{fig:phd_enabling_fig13}{b}, reflects a combined effect of moisture generation and temperature change. RH levels fluctuated more during the show compared to non-occupied periods, highlighting the dynamic environmental conditions associated with occupancy.

Dust concentration patterns (Fig.~\ref{fig:phd_enabling_fig13}{d}) display two distinct peaks around 20,000 pieces (pcs)/L. The first occurred between 17:00 and 18:00, likely due to pre-show preparations by staff and performers. The second, more sustained peak coincided with the show and extended for about an hour afterward. These increases are primarily attributed to enhanced air circulation from the ventilation system, which stirs up particulates.

\begin{figure}[!tb]
\includegraphics{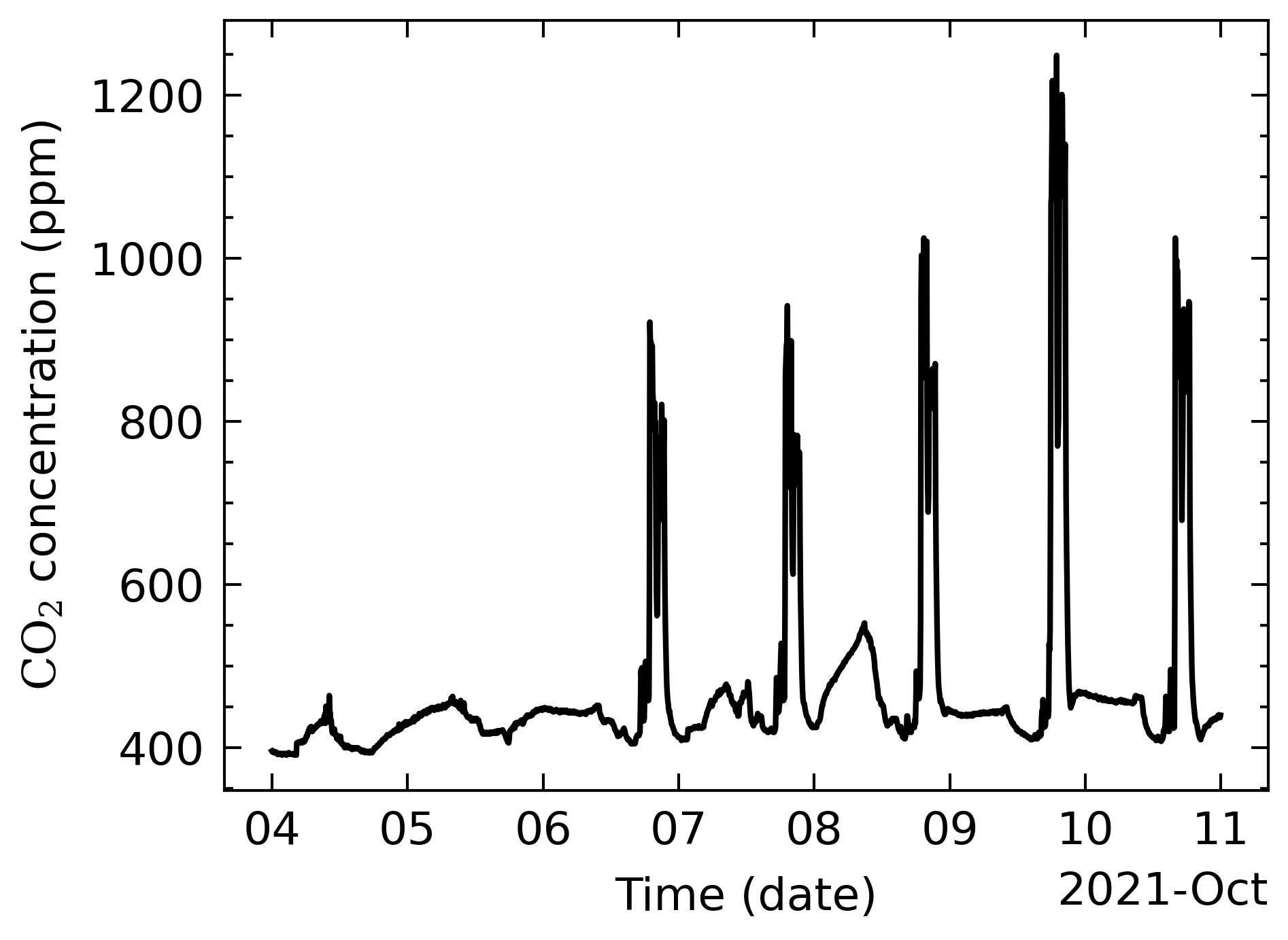}
\centering
\caption{Weekly CO\textsubscript{2} concentration profile in the salon room of the City Theatre, recorded from October 4 (Monday) to October 10 (Sunday), 2021~\cite{ni_enabling_2022}.}\label{fig:phd_enabling_fig14}
\end{figure}

Among all measured parameters, CO\textsubscript{2} concentration was the most sensitive to occupant presence. This is evident in two ways. First, CO\textsubscript{2} levels respond more rapidly than temperature, often increasing immediately upon entry of the audience, before any noticeable thermal change is observed. Second, peak CO\textsubscript{2} levels correlate with the number of occupants. As illustrated in Fig.~\ref{fig:phd_enabling_fig14}, which displays CO\textsubscript{2} concentrations from October 4 to 10, 2021, values remained below 500 ppm on Monday and Tuesday, when no shows were scheduled. On the other five days with shows, peak CO\textsubscript{2} levels exceeded 900 ppm. Saturday recorded the highest value, over 1,200 ppm, indicating the highest number of audience. Such high concentrations may degrade indoor air quality and occupant comfort. To address this, increasing ventilation airflow during high-attendance shows is recommended to keep CO\textsubscript{2} levels below 1,200 ppm.

\begin{figure}[!tb]
\includegraphics[width=\textwidth]{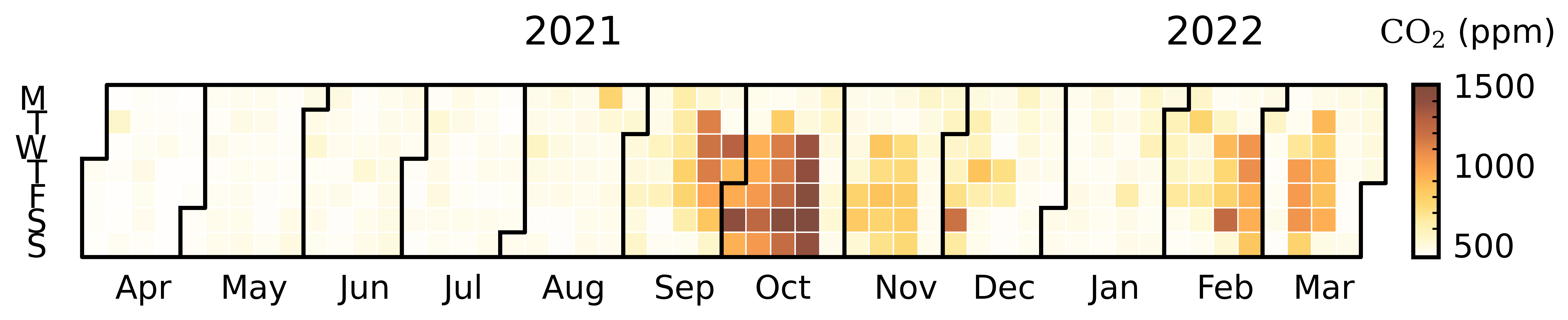}
\centering
\caption{Daily maximum CO\textsubscript{2} concentration in the salon room of the City Theatre from April 1, 2021 to March 31, 2022~\cite{ni_enabling_2022}. Days of the week are abbreviated as M (Monday), T (Tuesday), W (Wednesday), T (Thursday), F (Friday), S (Saturday), and S (Sunday).}\label{fig:phd_enabling_fig15}
\end{figure}

Fig.~\ref{fig:phd_enabling_fig15} shows the daily maximum CO\textsubscript{2} concentration over an entire calendar year, offering insights into occupancy patterns. Periods without shows, especially during COVID-19 restrictions prior to September 2021, are characterized by baseline CO\textsubscript{2} levels below 500 ppm. In contrast, CO\textsubscript{2} levels increased obviously during months with regular public events, particularly in October.

These findings demonstrate how environmental sensing can effectively capture occupancy-related variations in indoor environmental conditions, supporting the development of informed conservation and maintenance strategies.

\section{High Humidity Issue in Löfstad Castle}

Excessively high RH has been observed in the basement and in Room 5 on the ground floor of the main building at Löfstad Castle. This issue is primarily attributed to the direct soil flooring in the basement, which allows continuous evaporation of groundwater into the indoor air. The resulting moisture is suspected to migrate upward through the masonry walls, creating a persistent internal moisture source that not only affects the basement but also influences the indoor environment of rooms on the upper floors.

\begin{figure}[!tb] 
\centering
\subfloat[]{\includegraphics{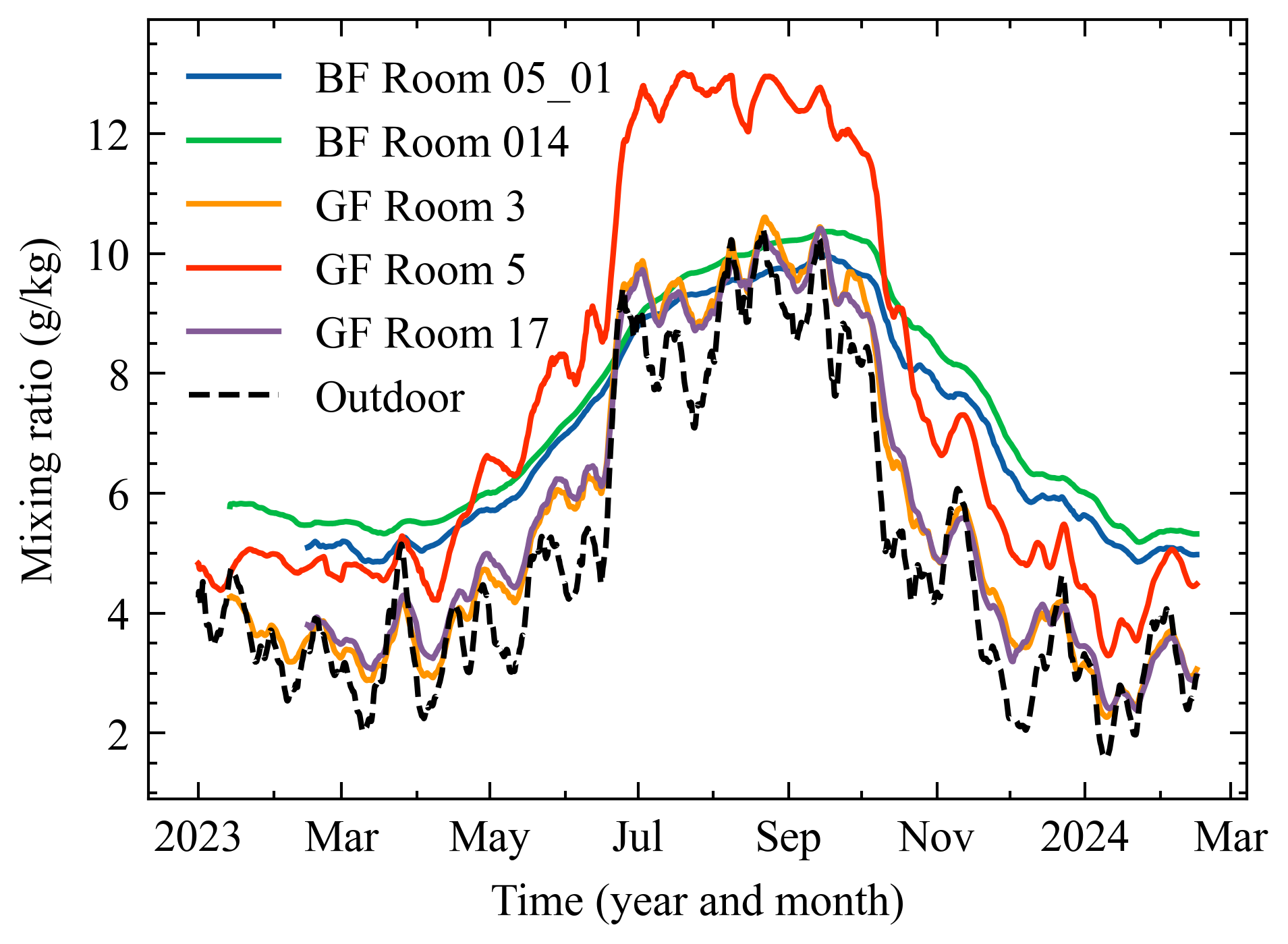}}
\hfill
\subfloat[]{\includegraphics{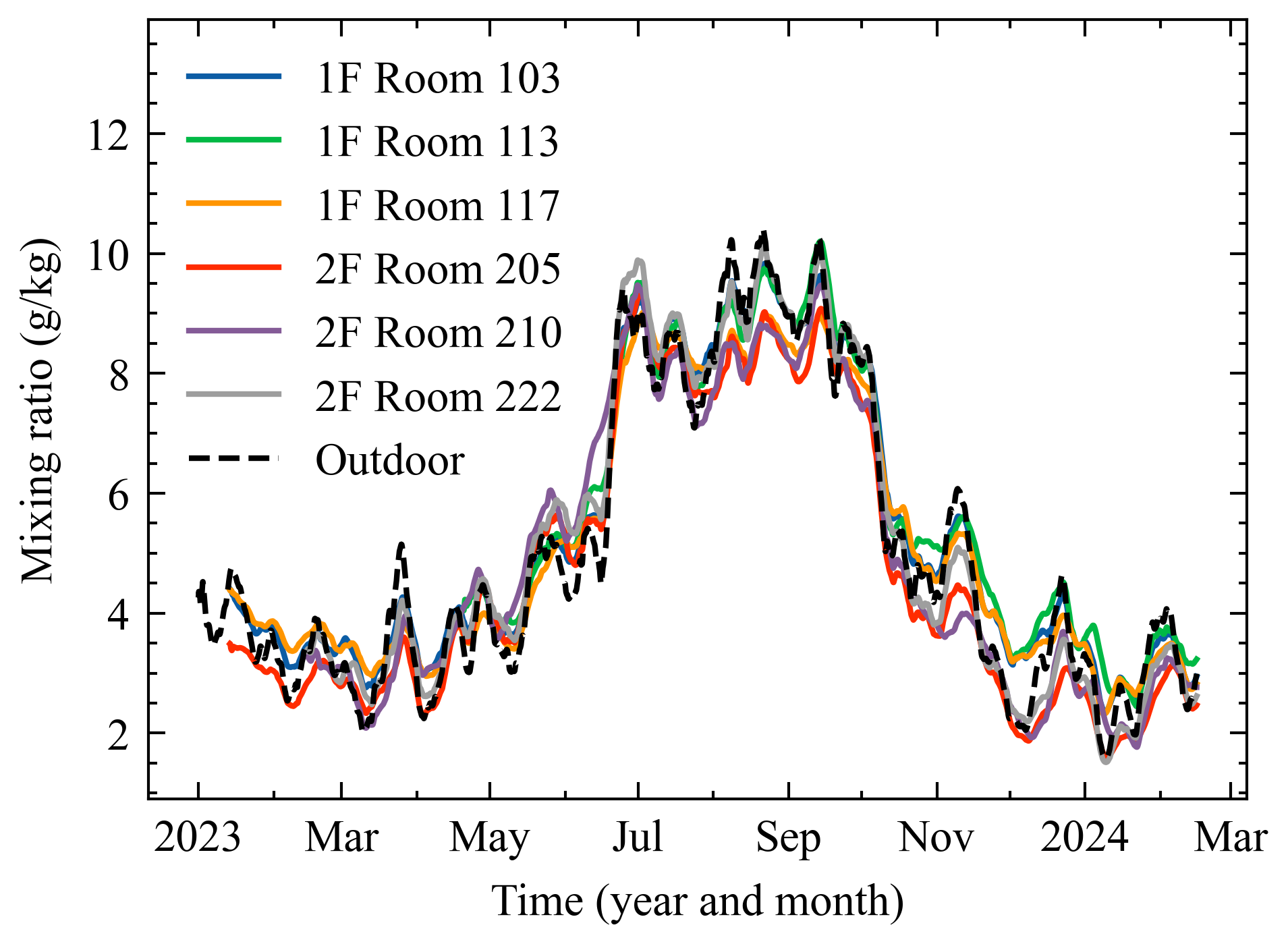}}
\caption{Seven-day moving average of hourly humidity mixing ratios (MRs) from January 2023 to February 2024~\cite{ni_parametric_2025}: (\textbf{a}) outdoor and indoor MRs in rooms located in the basement and on the ground floor; (\textbf{b}) outdoor and indoor MRs in rooms on the first and second floors.}
\label{fig:phd_parametric_fig10}
\end{figure}

This phenomenon is evident when comparing indoor and outdoor humidity MRs. As shown in Fig.~\ref{fig:phd_parametric_fig10}{a}, the MRs in the basement and ground floor rooms consistently exceeded outdoor values throughout the year. This persistent difference indicates the presence of an internal moisture source. The problem is particularly severe in Room 5, where the sensor box is located near the floor and wall. In this area, rising dampness has led to sustained elevated RH levels. A logical mitigation strategy involves installing vapor barriers in the basement to reduce evaporation and prevent vertical moisture migration.

Rooms on the first and second floors are less affected by this internal moisture source. As illustrated in Fig.~\ref{fig:phd_parametric_fig10}{b}, the MR difference between indoor and outdoor air on these floors is considerably smaller. For most of the year, the second-floor MRs are equal to or lower than outdoor levels, suggesting minimal upward moisture transfer from the basement.

\begin{figure}[!tb]
\includegraphics{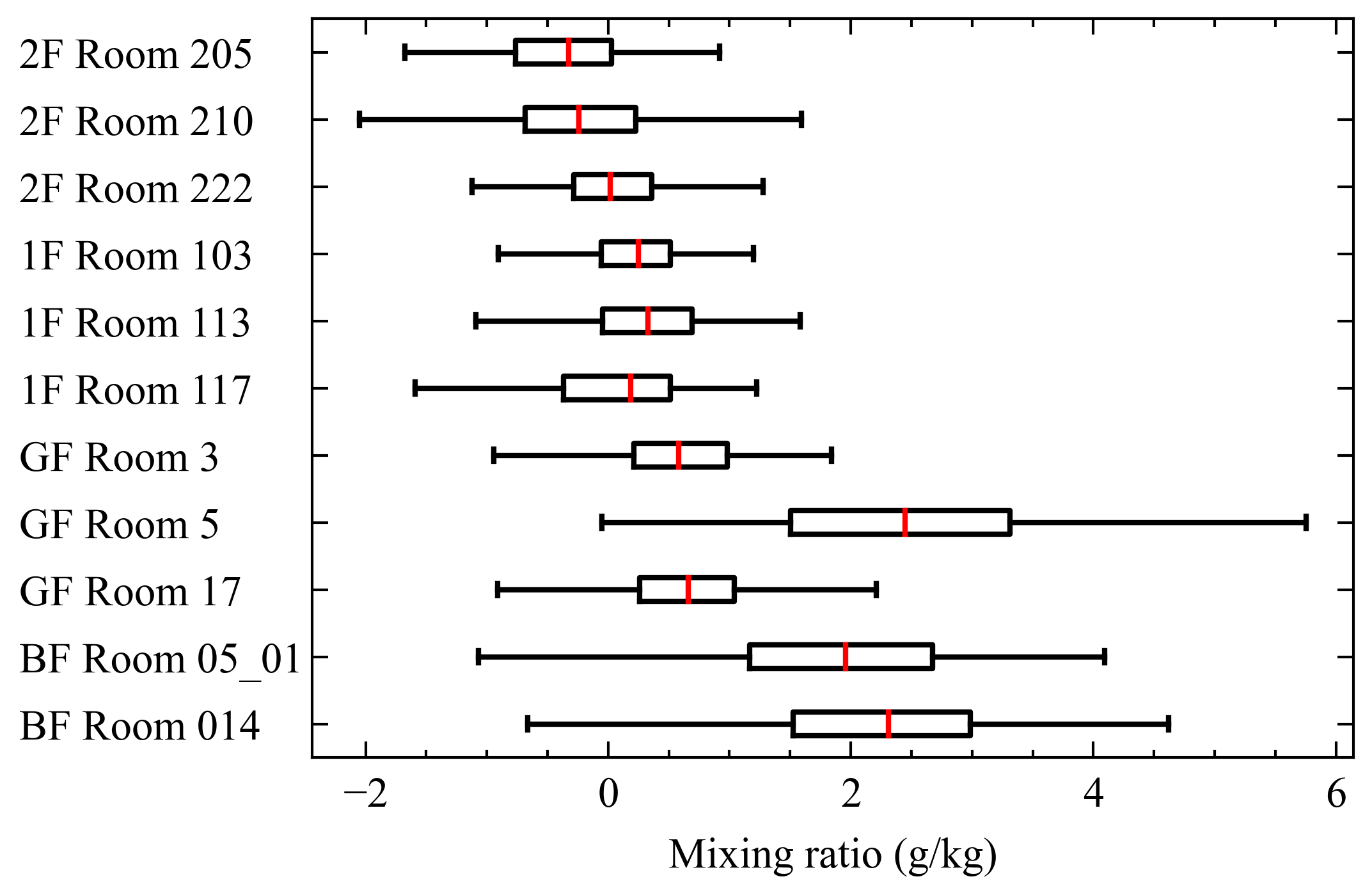}
\centering
\caption{Boxplot showing the difference between indoor and outdoor humidity MRs from January 2023 to February 2024~\cite{ni_parametric_2025}. Each value represents the 7-day moving average of hourly indoor MR minus the corresponding outdoor MR. Outliers are excluded, as they are not relevant to the focus of this study.}\label{fig:phd_parametric_fig11}
\end{figure}

The vertical distribution of moisture levels is further analyzed in Fig.~\ref{fig:phd_parametric_fig11}, which presents a boxplot of MR differences by floor level, based on a 7-day moving average from January 2023 to February 2024. Each data point represents the average hourly difference between indoor and outdoor MRs, with outliers removed to emphasize consistent trends. The results clearly show that rooms in the basement and on the ground floor had higher indoor MRs than outdoors for approximately 75\% of the monitoring period. This proportion decreases on the first floor and drops to about 50\% on the second floor. This vertical gradient supports the hypothesis of rising damp originating from the basement, with diminishing influence at higher levels.

These findings highlight the importance of localized environmental monitoring and the value of parametric digital twin models for identifying and understanding conservation risks. They also emphasize the need for targeted interventions, such as installing vapor barriers, to mitigate humidity-related damage in historic buildings like Löfstad Castle.

\chapter{Deep Learning-based Building Energy Forecasting}
\label{cha:deep_learning}

This chapter presents selected results on multi-horizon point and probabilistic forecasting of energy use in public historic buildings. The focus is on forecasting hourly electricity consumption and heating load 24 hours ahead for the City Museum and the City Theatre in Norrköping, Sweden. These buildings exhibit seasonal energy patterns due to their public functions and variable occupancy, highlighting the need for advanced deep learning-based approaches. Forecasting models were trained under four experimental cases to evaluate the effects of different input features and modeling objectives. These include case 1: multi-horizon point forecasting using only past predictors; case 2: point forecasting incorporating both past and future predictors; case 3: probabilistic forecasting via quantile regression; and case 4: a sensitivity analysis that excludes building operation-related features. This chapter highlights the quantitative results of the first three cases. A more detailed evaluation, including both quantitative and qualitative analyses, is provided in Paper IV. One-step-ahead forecasting and its integration with digital twins are discussed in Paper VI.

\section{Comparison of Predictability of Electricity and Heating}

Both electricity consumption and heating load in the City Museum and the City Theatre exhibit strong daily seasonality, as indicated by the superior performance of the SN-24 model compared to the SN-168 model across all evaluation metrics (see Table~\ref{tab:phd_deep_no_future}). These two seasonal naïve (SN) models serve as baselines, reflecting daily (SN-24) and weekly (SN-168) seasonality patterns~\cite{hyndman_forecasting_2018}. The results suggest that energy use in both buildings is more influenced by daily cycles than by weekly ones. Heating load, in particular, displays a more consistent daily pattern than electricity consumption. The SN-24 model yielded lower CV-RMSE values for heating load in both buildings, indicating higher predictability.

\begin{table}[!tb]
\centering
\caption{Prediction accuracy of point forecasts from different models on the test set without future information (Case 1)~\cite{ni_deep_2024}. Lower absolute values of CV-RMSE and NMBE indicate higher accuracy. Electricity is abbreviated as El.. }
\label{tab:phd_deep_no_future}
\end{table}
\begin{figure}[!tb]
\includegraphics[width=\textwidth]{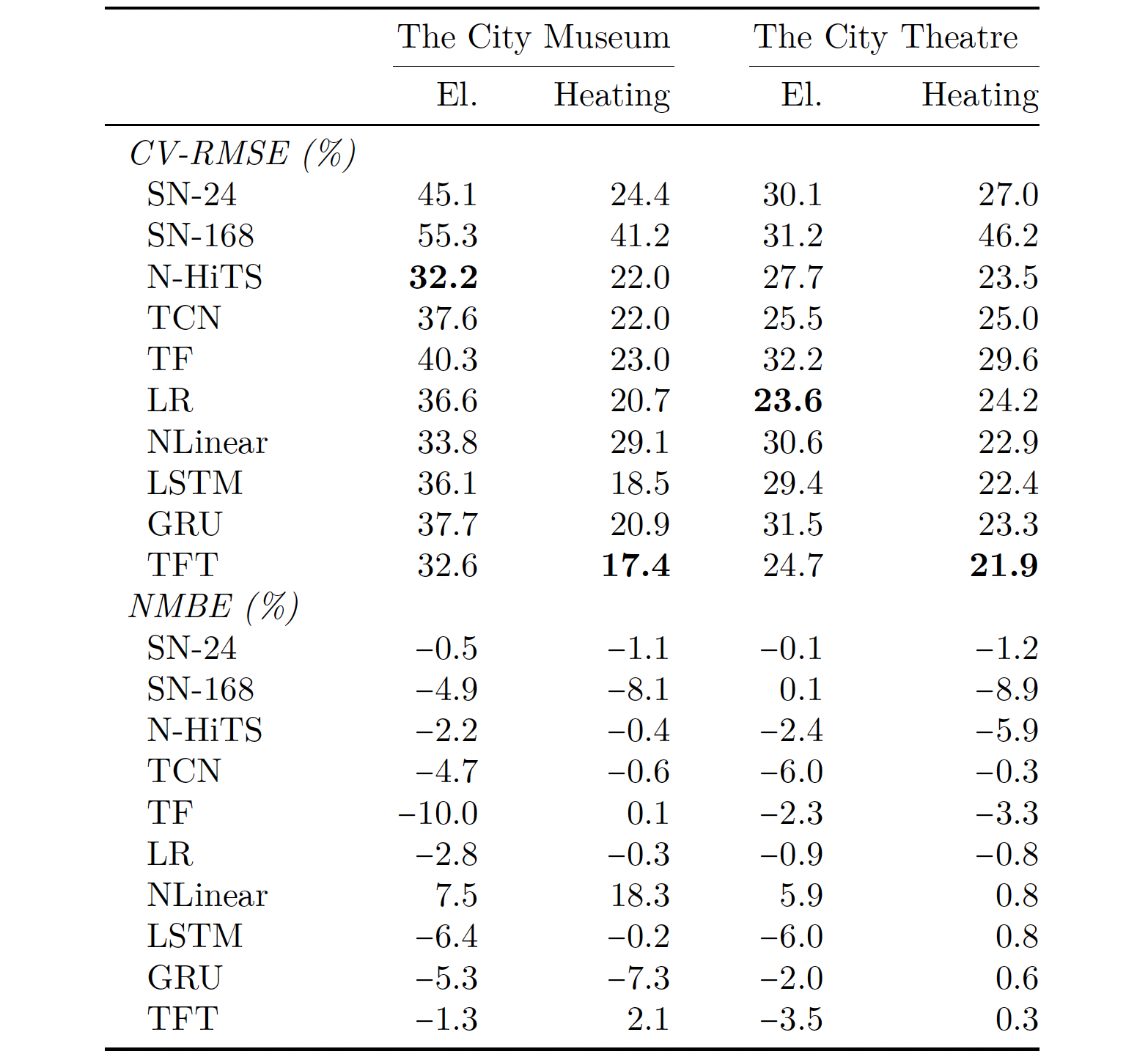}
\centering
\end{figure}

Overall, electricity consumption is less predictable than heating load. Although the SN-24 model provided a strong baseline for heating load, where it met ASHRAE Guideline 14-2014 criteria (CV-RMSE $\leq$ 30\%, NMBE within $\pm$10\%)~\cite{ashare_measurement_2014}, its performance was limited for electricity consumption. This was especially evident for the City Museum, where the model failed to achieve acceptable forecasting accuracy. In contrast, it met ASHRAE criteria for heating load in both buildings, highlighting the relative ease of forecasting this variable using seasonal patterns alone.

The evaluation of the eight deep learning models also supports the conclusion that heating load is generally more predictable. With the exception of the linear regression (LR) model on the City Theatre, all models achieved lower CV-RMSE values for heating load than for electricity consumption. All eight models met ASHRAE accuracy criteria for heating load prediction in the City Museum, and all but NLinear did so for the City Theatre. However, none of the models reached the CV-RMSE threshold of 30\% for electricity consumption in the City Museum. In the City Theatre, five of the eight models met the criterion, suggesting a modest improvement in predictability.

These differences are attributed to operational characteristics. Heating is controlled adaptively based on the temperature differential between indoor and outdoor environments, leading to a more deterministic and regular usage pattern. In contrast, electricity consumption is influenced by more variable and less observable factors, such as equipment use and irregular events.

When comparing the same energy type across buildings, electricity consumption in the City Museum was more difficult to predict than in the City Theatre. Conversely, heating load was easier to forecast in the City Museum. All eight models recorded higher CV-RMSE values for electricity use in the City Museum, while seven of the eight models achieved lower CV-RMSE values for heating load in the same building. This difference is likely due to operational scheduling: in the City Theatre, shows of the same production are often scheduled on consecutive days, creating consistent patterns. However, the associated increase in internal heat gains introduces greater variability in heating demand~\cite{ni_enabling_2022}.

Despite the use of advanced deep learning architectures, most models did not apparently outperform the SN-24 baseline in the absence of future predictors. For electricity forecasting, the best-performing model for the City Museum was N-HiTS (CV-RMSE 32.2\%), while LR achieved the best result for the City Theatre (CV-RMSE 23.6\%). For heating load, TFT outperformed all other models in both buildings with CV-RMSE values of 17.4\% for the City Museum and 21.9\% for the City Theatre. However, several models, such as NLinear for predicting heating load of the City Museum and TF for predicting electricity consumption of the City Theatre, did not outperform the SN-24 baseline, emphasizing the challenge of improving forecasting accuracy without incorporating additional information.

\section{The Impact of Incorporating Future Information}

Incorporating future values of predictor variables within the forecast horizon improves the predictive performance of deep learning models by providing additional information about factors that influence energy use. As shown in Table~\ref{tab:phd_deep_with_future}, all five evaluated models achieved lower CV-RMSE values for both electricity consumption and heating load in both buildings when future information---specifically outdoor weather conditions and opening hours---was included. This highlights the practical value of using known or expected circumstances that can be predicted, such as building schedules, weather forecasts, and operational modes, in multi-horizon energy prediction.

\begin{table}[!tb]
\centering
\caption{Prediction accuracy of point forecasts from different models on the test set with future input features included (Case 2)~\cite{ni_deep_2024}. Values in brackets indicate the change in performance relative to Table~\ref{tab:phd_deep_no_future}, where negative values reflect improvements. }
\label{tab:phd_deep_with_future}
\end{table}
\begin{figure}[!tb]
\includegraphics[width=\textwidth]{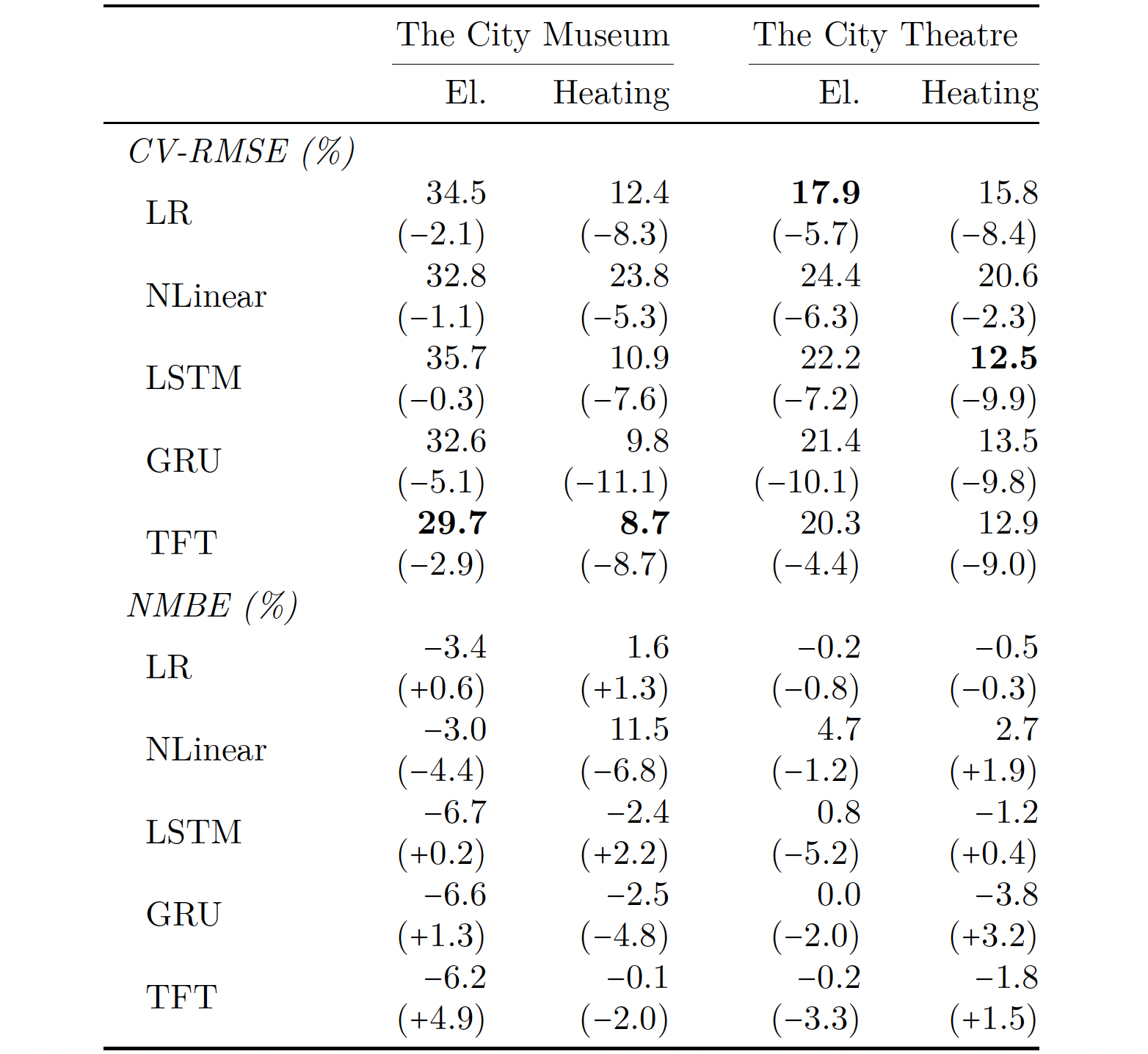}
\centering
\end{figure}

The improvements were particularly notable for heating load forecasts. For instance, the TFT model achieved the best overall performance in the City Museum, with a CV-RMSE of 29.7\% for electricity and only 8.7\% for heating load. This result demonstrates the model’s ability to capture complex interactions among variables when provided with both historical and future contextual information. In the City Theatre, the LR model achieved the best performance for electricity consumption (CV-RMSE 17.9\%), while the LSTM model performed best for heating load (CV-RMSE 12.5\%).

The strong performance of the LR model for electricity consumption in the City Theatre suggests that, in some cases, simple models may be sufficient, particularly when strong linear relationships exist between predictor variables and the target.

These findings confirm the benefits of including future predictors across all model types and energy variables, with the greatest improvement observed in heating load forecasts. They also emphasize the importance of feature engineering and indicate that simpler models can be highly effective when predictive relationships are well-defined.

\section{The Performance of Probabilistic Forecasting}

Building on the results from Cases 1 and 2, six models were identified as top performers for multi-horizon point forecasting: N-HiTS, LR, NLinear, LSTM, GRU, and TFT. Among them, the TFT model consistently delivered the best performance in probabilistic forecasting, demonstrating strong capability in capturing both central tendencies and upper quantiles of energy use distributions. As shown in Table~\ref{tab:phd_deep_rho_risk}, TFT achieved the lowest $\rho$-risk values at the 0.5th and 0.9th quantiles for heating load in the City Museum and for both electricity consumption and heating load in the City Theatre. These results indicate that TFT is particularly well-suited for modeling uncertainty and generating reliable predictive intervals in complex environments.


For electricity consumption in the City Museum, the GRU model produced the most accurate median forecasts, achieving the lowest $\rho$-risk at the 0.5th quantile ($\rho$-risk$(0.5) = 0.182$). In contrast, the N-HiTS model performed best at the upper tail, yielding the lowest $\rho$-risk at the 0.9th quantile ($\rho$-risk$(0.9) = 0.142$). These findings highlight the complementary strengths of the two architectures: GRU is effective in capturing sequential patterns, while N-HiTS excels at detecting high-impact outliers and modeling the upper bounds of energy use.

In contrast, the LR model consistently performed worst in probabilistic forecasting across all cases. This can be attributed to its assumption of normally distributed residuals, which is often violated when estimating quantiles. In the presence of skewed or heavy-tailed error distributions, linear models struggle to provide accurate quantile estimates.

\begin{table}[!htb]
\centering
\caption{$\rho$-risk at the 0.5th and 0.9th quantiles of probabilistic forecasts for different models on the test set (Case 3)~\cite{ni_deep_2024}. Lower values indicate better predictive performance for each metric.}
\label{tab:phd_deep_rho_risk}
\end{table}
\begin{figure}[!htb]
\includegraphics[width=\textwidth]{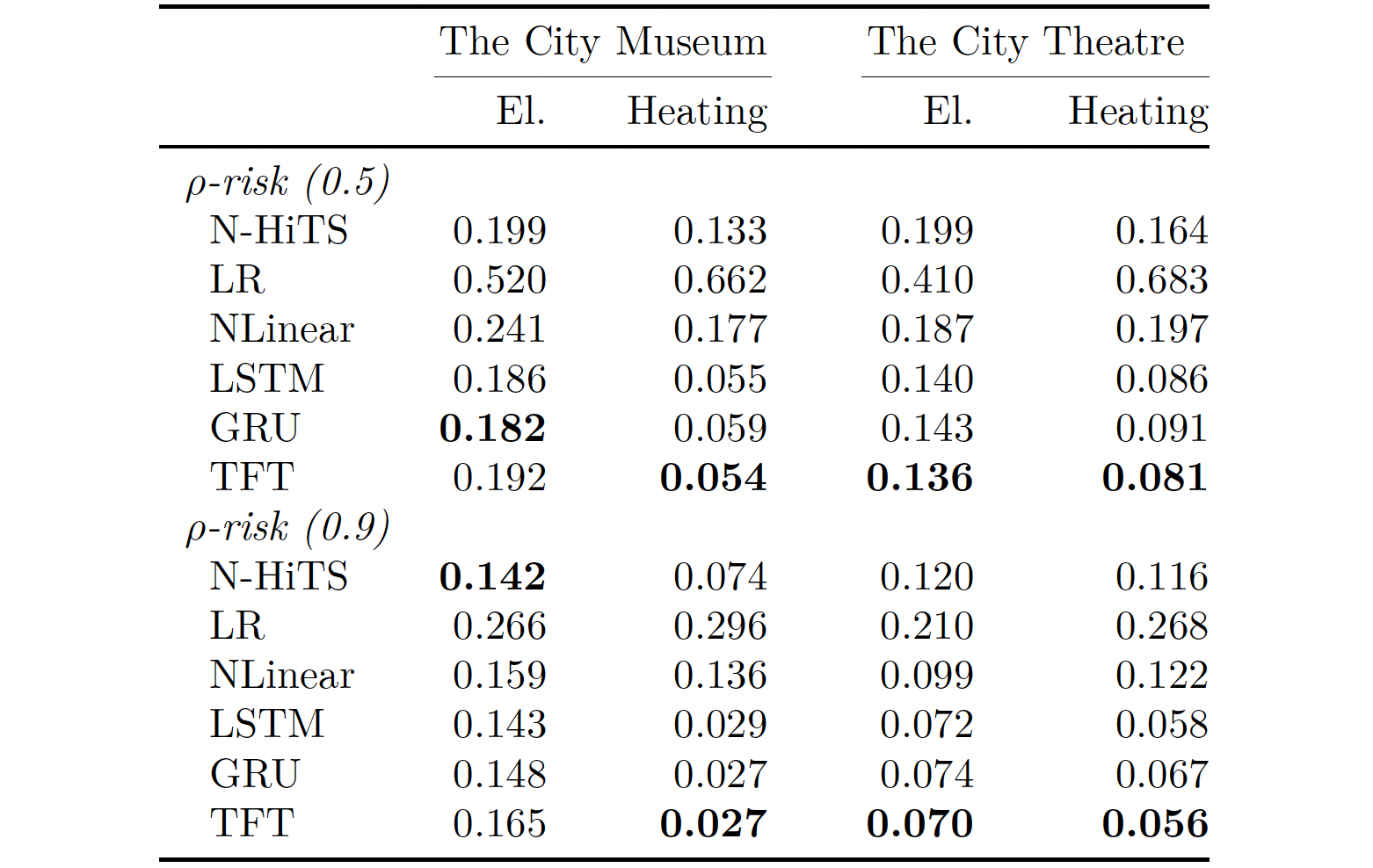}
\centering
\end{figure}

The results further support the observation that heating load is more predictable than electricity consumption. Except for the NLinear model on heating load in the City Theatre and the LR model across all tasks, all evaluated models achieved lower $\rho$-risk values at the 0.5th quantile when forecasting heating load. A similar trend was observed at the 0.9th quantile, with most models exhibiting reduced uncertainty in heating predictions. This reflects the relative stability and stronger seasonal regularity of heating demand, which is primarily driven by outdoor temperature and governed by well-defined heating control logic.

Conversely, electricity consumption forecasts were characterized by greater uncertainty, indicating higher sensitivity to short-term operational changes and less predictable usage patterns. While this presents a challenge for forecasting, it also underscores opportunities for improving energy management. Reducing variability through scheduling optimization or adaptive control could enhance both predictability and system efficiency. Additionally, incorporating more detailed building operation-related features, such as equipment usage schedules or occupancy levels, may further improve forecasting accuracy for electricity consumption.

\chapter{Edge-centric and Federated Indoor Climate Forecasting}
\label{cha:edge_federated}

This chapter presents selected results on the development and deployment of deep learning-based models for indoor climate forecasting in historic buildings. It focuses on two complementary approaches: inference on edge devices and FL across distributed measurement points. These methods address key challenges in heritage conservation. They support low-latency predictions, protect data privacy, and improve system scalability. This is especially important in settings where constant cloud connectivity or centralized data aggregation is not feasible. The first part evaluates the computational feasibility of executing trained models directly on embedded edge platforms. The second part investigates how FL enables collaborative model training across spatially distributed measurement points without transferring raw data. Together, these results demonstrate practical strategies for real-world implementation of intelligent predictive models. More comprehensive descriptions and evaluations of the edge deployment and FL experiments are provided in Paper VII and Paper V, respectively.

\section{Inference Cost on the Edge}

To assess the feasibility of deploying predictive models on resource-constrained devices, the inference cost of selected deep learning models was evaluated under an edge-centric setting. Models were developed to forecast indoor temperature and RH for several rooms in the main building of Löfstad Castle, using a 24-hour forecast horizon and a 168-hour lookback window.

Five deep learning models---LSTM, TCN, TFT, N-HiTS, and TiDE---were trained using historical indoor climate and outdoor weather data. Model training was performed on a desktop equipped with an NVIDIA GeForce GTX 1080 GPU. After training, the models were deployed on the edge platform integrated into the prototype sensor box. This edge device features a quad-core Cortex-A72 64-bit system-on-chip (SoC) running at 1.8~GHz, 1~GB of RAM, and operates on Debian GNU/Linux 12.

As illustrated in Table~\ref{tab:phd_edge_inference_time}, inference times varied greatly across model architectures. Models with parallel processing capabilities, such as TCN and TiDE, achieved the shortest inference times. In contrast, the LSTM model, which processes sequences sequentially, exhibited slower performance. N-HiTS required slightly more computation than LSTM due to its hierarchical structure, while TFT had the highest inference cost. The self-attention mechanism in TFT introduces quadratic time complexity with respect to input length, making it less efficient for devices with limited processing resources.

\begin{table}[!tb]
\centering
\caption{Inference time (mean$\pm$standard deviation) of different models, measured in milliseconds (ms)~\cite{ni_edge_2024}.}
\label{tab:phd_edge_inference_time}
\end{table}
\begin{figure}[!tb]
\includegraphics[width=\textwidth]{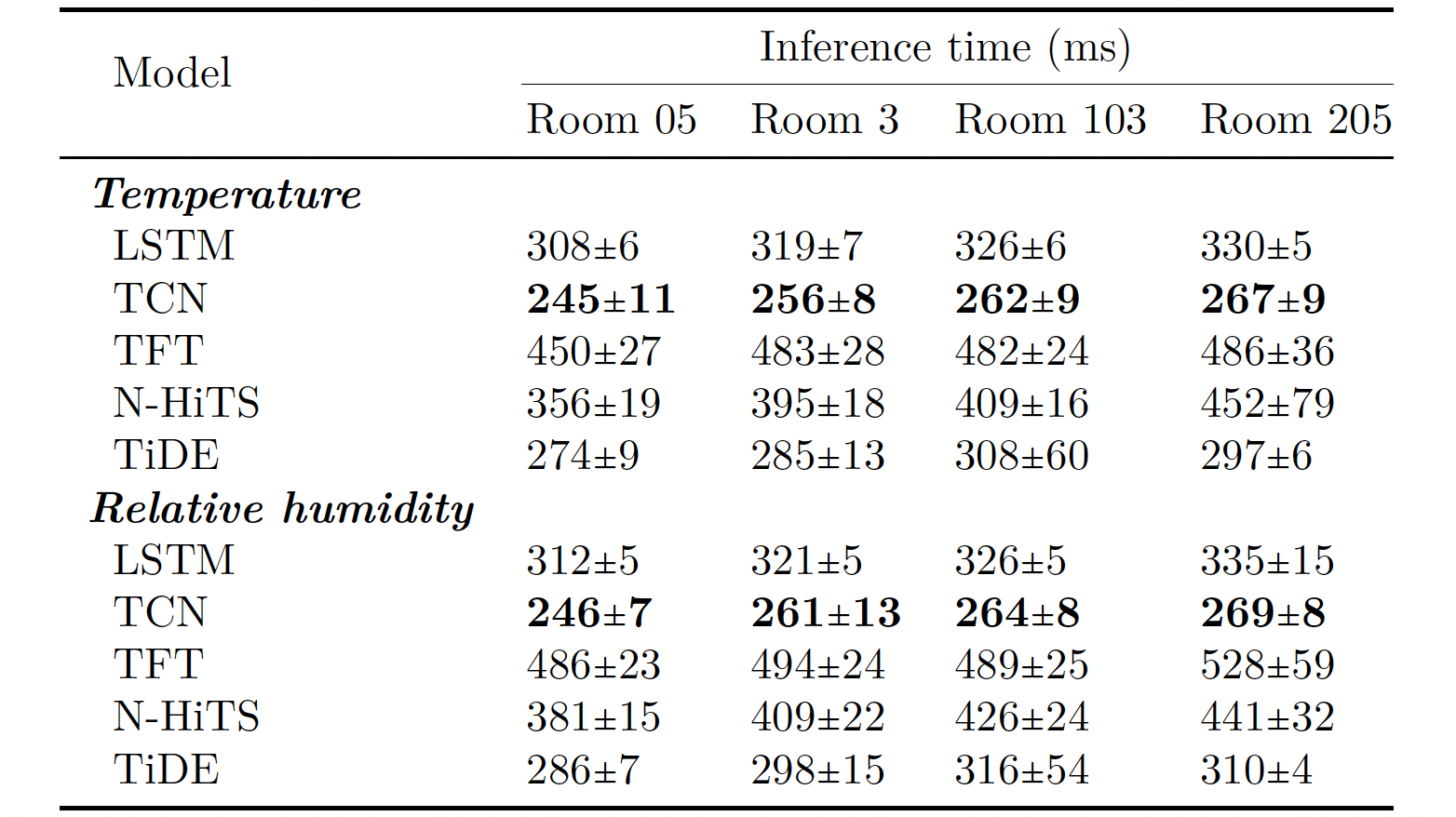}
\centering
\end{figure}

These results highlight the importance of balancing prediction accuracy with computational efficiency when selecting models for edge deployment. Lightweight architectures such as TCN and TiDE offer an effective trade-off between performance and responsiveness, making them well-suited for real-time indoor climate forecasting in historic buildings.

\section{Federated Predictive Models}

To evaluate the feasibility and performance of FL for indoor climate forecasting, predictive models were developed to generate six-hour-ahead forecasts of indoor temperature, RH, and CO\textsubscript{2} concentration at each measurement point. A SN-24 model~\cite{hyndman_forecasting_2018} was used as a baseline, where each forecast was set to the value observed 24 hours earlier, leveraging daily seasonality.

To compare the effectiveness of FL with traditional learning paradigms, three experimental cases were designed:
\begin{itemize}
  \item Case 1 (local learning): Models were trained and evaluated independently using data from each measurement point, with no data sharing.
  \item Case 2 (centralized learning): A single model was trained and evaluated on the combined dataset from all measurement points, assuming full access to data.
  \item Case 3 (federated learning): The most promising model architecture, selected based on prior performance in the local and centralized learning cases, was trained under a federated setup, where only model updates were exchanged between clients and a central server, and raw data remained local. 
\end{itemize}

This experimental design enabled a fair and consistent comparison of learning paradigms using identical model architectures. It also allowed for assessing whether FL can achieve a suitable balance between predictive accuracy, scalability, and privacy in decentralized heritage building environments.

\subsection{Local Learning}

In general, CO\textsubscript{2} is more difficult to predict than temperature and RH. As shown in Table~\ref{tab:phd_federated_local_learning}, all eight deep learning models yielded higher CV-RMSE values for CO\textsubscript{2} than for temperature or RH across all measurement points. This variation in predictability is linked to the underlying physical dynamics of each variable. Temperature is typically the most predictable due to the thermal inertia of building materials, which buffers short-term fluctuations. This effect is especially pronounced in historic buildings with thick masonry walls, such as Löfstad Castle, where the main building has walls approximately 1.1 meters thick. RH is moderately predictable, as it is influenced by both temperature and moisture exchange, making it sensitive to factors such as ventilation, infiltration, and occupancy. CO\textsubscript{2} is the least predictable, as it responds rapidly and nonlinearly to occupant activity and ventilation dynamics. Understanding the nature of each variable also helps in interpreting the causes of observed changes. For example, sharp increases in CO\textsubscript{2} concentration are often linked to occupant presence, making it a useful indicator for detecting occupancy patterns. Among the three case study buildings, CO\textsubscript{2} concentration in the City Theatre was the most challenging to forecast due to irregular occupancy patterns associated with public shows. In contrast, the City Museum and Löfstad Castle have more consistent opening hours, resulting in more regular CO\textsubscript{2} patterns and improved predictability. For temperature and RH, even the baseline SN-24 model performed relatively well, reflecting stable environmental conditions over the forecast horizon.

\begin{table}[!tb]
\centering
\caption{Prediction accuracy of different models on the test set under local learning (Case 1)~\cite{ni_federated_2025}. Lower absolute values of CV-RMSE and NMBE indicate higher accuracy. For each target variable, the best result is shown in bold and the second-best is underlined. Temperature is abbreviated as T.}
\label{tab:phd_federated_local_learning}
\end{table}
\begin{figure}[!tb]
\includegraphics[width=\textwidth]{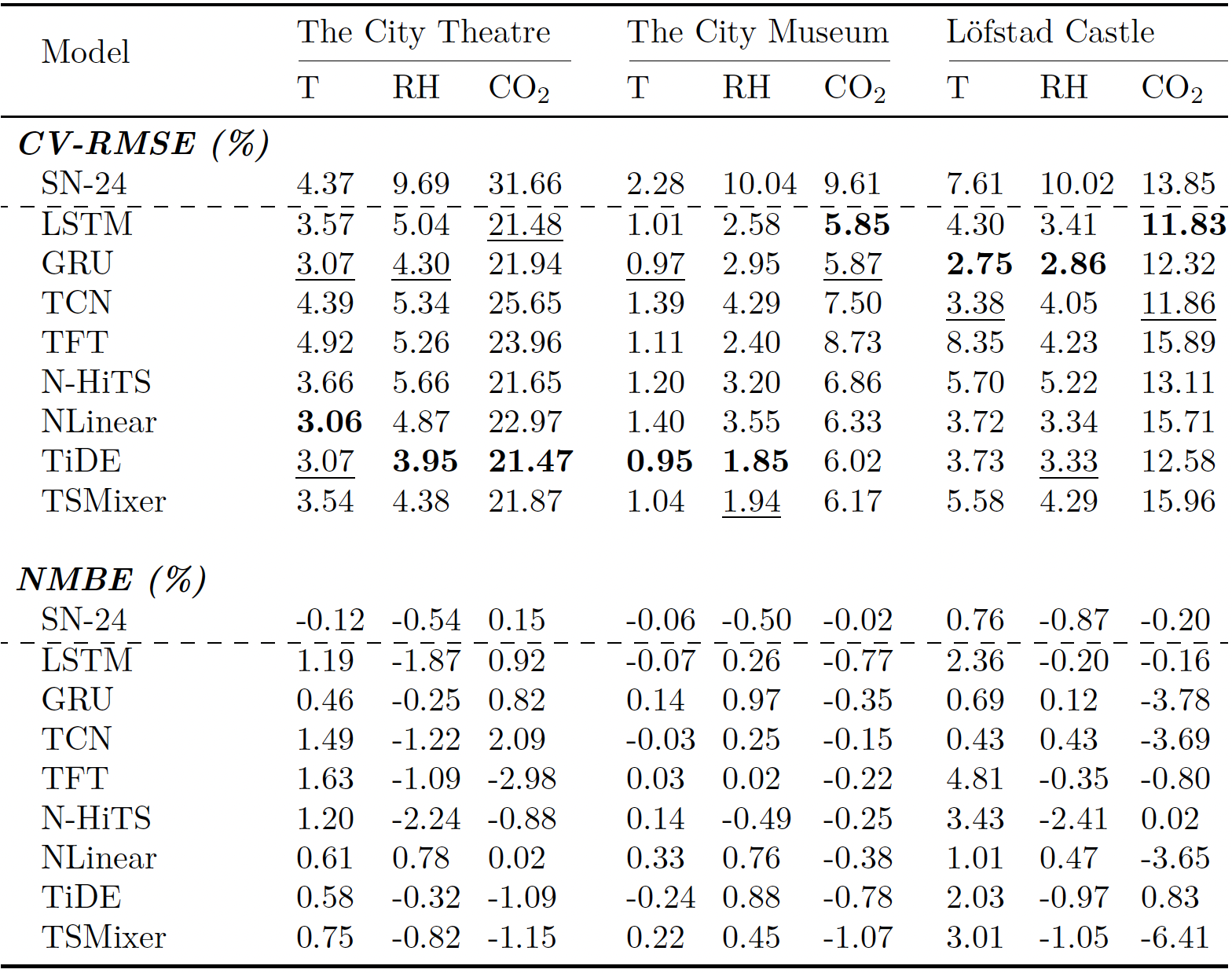}
\centering
\end{figure}

Under the local learning setting, the TiDE and GRU models consistently achieved strong predictive performance across different buildings and variables. In the City Theatre, TiDE yielded the lowest CV-RMSE values for CO\textsubscript{2} (21.47\%) and RH (3.95\%). For temperature, NLinear performed best with a CV-RMSE of 3.06\%, slightly outperforming TiDE (3.07\%). Most models exhibited minimal bias, and GRU in particular achieved absolute NMBE values below 1\%.

In the City Museum, TiDE produced the most accurate forecasts for temperature and RH, with CV-RMSE values of 0.95\% and 1.85\%, respectively. LSTM performed best for CO\textsubscript{2} prediction, with a CV-RMSE of 5.85\%, closely followed by GRU (5.87\%). GRU also achieved competitive performance for temperature (CV-RMSE 0.97\%).

At Löfstad Castle, GRU outperformed other models in predicting temperature and RH, with CV-RMSE values of 2.75\% and 2.86\%, respectively. TiDE also performed well for RH (CV-RMSE 3.33\%).

These findings highlight the robustness and adaptability of TiDE and GRU across different environmental conditions and operational contexts. Their strong performance across multiple variables and buildings demonstrates their potential for indoor climate forecasting in heritage conservation.

\subsection{Centralized Learning}

A comparison between local and centralized learning revealed several key insights. Centralized learning generally improved the prediction accuracy for temperature and RH at the City Theatre, as reflected in reduced CV-RMSE values across most models (see Table~\ref{tab:phd_federated_centralized_learning}). For example, TiDE’s accuracy for temperature improved from 3.07\% (local) to 2.81\% (centralized), while RH prediction improved from 3.95\% to 3.50\%. These results highlight the advantage of aggregating data across sites when environmental patterns are sufficiently similar.

In contrast, the benefits of centralized learning were less consistent at the City Museum and Löfstad Castle. At the City Museum, temperature prediction performance deteriorated across all models, with higher CV-RMSE values than those obtained under local learning. RH predictions at the City Museum, along with both temperature and RH forecasts at Löfstad Castle, showed mixed results. Some models improved slightly, e.g., TiDE’s CV-RMSE at Löfstad Castle decreased from 3.73\% to 3.48\% for temperature, and from 3.33\% to 2.83\% for RH. Others, such as TSMixer, exhibited increased prediction errors in the centralized learning case.

For CO\textsubscript{2} forecasting, the impact of centralized learning varied by building. Prediction accuracy declined at the City Theatre and the City Museum, with all models showing higher CV-RMSE values. However, performance improved at Löfstad Castle, where seven of eight models achieved lower CV-RMSE values. For example, TSMixer’s CV-RMSE for CO\textsubscript{2} dropped from 15.96\% under local learning to 10.80\% in the centralized setting.

Bias, measured by NMBE, was generally reduced in centralized learning, particularly for temperature and RH, indicating more stable and less variable predictions.

These results suggest that the effectiveness of centralized learning depends on both the model and the building context. In some cases, models benefit from generalized knowledge derived from shared environmental patterns. In others, centralized learning may obscure building-specific dynamics, reducing prediction accuracy. This highlights a key challenge in generalization, especially for variables affected by localized operational or behavioral factors.

\begin{table}[!htb]
\centering
\caption{Prediction accuracy of different models on the test set under centralized learning (Case 2)~\cite{ni_federated_2025}.}
\label{tab:phd_federated_centralized_learning}
\end{table}
\begin{figure}[!htb]
\includegraphics[width=\textwidth]{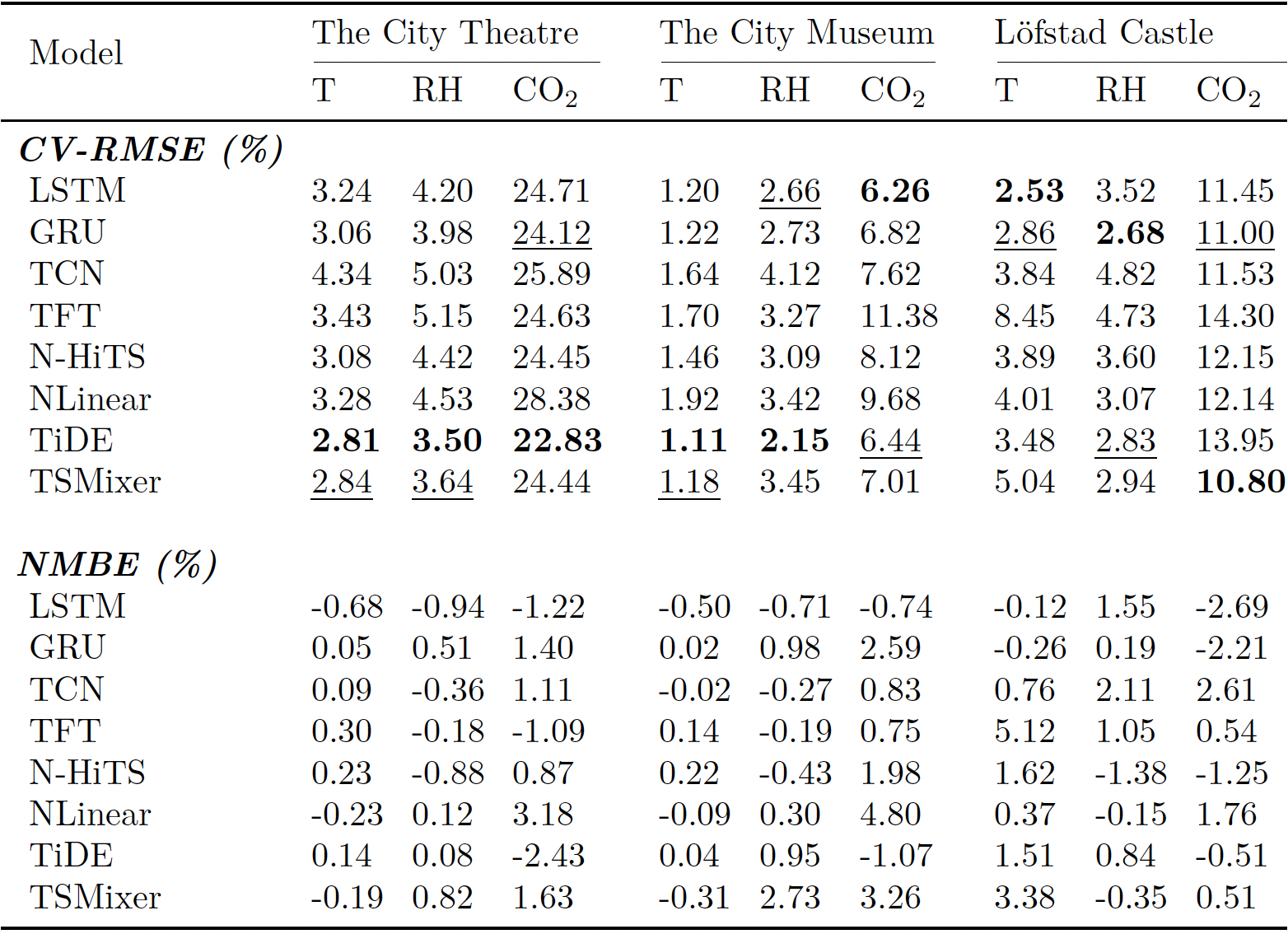}
\centering
\end{figure}

Among the evaluated models, TiDE consistently achieved the highest accuracy for all three indoor environmental parameters in the City Theatre, with CV-RMSE values of 2.81\% for temperature, 3.50\% for RH, and 22.83\% for CO\textsubscript{2}. TSMixer demonstrated similar performance in temperature (2.84\%) and RH (3.64\%), while GRU performed comparably for CO\textsubscript{2} (24.12\%).

At the City Museum, TiDE again outperformed other models, achieving the lowest CV-RMSE for temperature (1.11\%) and RH (2.15\%), along with strong performance for CO\textsubscript{2} (6.44\%). GRU and LSTM also performed well, particularly in RH and CO\textsubscript{2} forecasting, although slightly behind TiDE.

At Löfstad Castle, LSTM achieved the most accurate temperature prediction (CV-RMSE 2.53\%), while GRU provided the best RH forecast (2.68\%). TSMixer delivered the most accurate CO\textsubscript{2} prediction (CV-RMSE 10.80\%). TiDE remained highly competitive in RH prediction (CV-RMSE 2.83\%), reinforcing its consistence across different environmental conditions and building types.

\subsection{Federated Learning}

TiDE, GRU, and TSMixer were evaluated in combination with six FL algorithms to forecast indoor temperature, RH, and CO\textsubscript{2} concentration across the measurement points in the three historic buildings. In the City Theatre, TiDE combined with either FedAvg or FedYogi generally achieved the highest accuracy for temperature and CO\textsubscript{2} (see Table~\ref{tab:phd_federated_federated_learning}). TiDE also performed well with FedAvg for RH, indicating the robustness of FedAvg in this setting. While GRU and TSMixer produced comparable results under certain algorithms, models trained with FedAvgM and FedAdagrad typically showed higher prediction errors.

In the City Museum, TiDE paired with FedAdam consistently outperformed other model-algorithm combinations across all three variables. Although GRU and TSMixer occasionally approached TiDE’s performance, their accuracy was more sensitive to the choice of FL algorithm. Across all models, FedAvg, FedMedian, and FedYogi often produced lower NMBE values, reflecting reduced bias, though performance varied depending on the specific variable.

At Löfstad Castle, FedAvg and FedYogi were particularly effective. TiDE combined with FedAvg achieved strong performance in temperature prediction, while FedYogi delivered the best results for RH and CO\textsubscript{2}. GRU and TSMixer showed similar levels of accuracy in some cases, but their performance declined under algorithms such as FedAvgM and FedAdagrad.

Overall, TiDE emerged as the most consistent architecture across all buildings and target variables. Among the FL algorithms, FedAvg, FedMedian, and FedYogi consistently achieved reliable convergence and low errors, while FedAvgM and FedAdagrad showed less stable performance. A holistic comparison of local, centralized, and federated learning approaches highlights FL as a practical and privacy-preserving alternative to centralized training. While FL does not always outperform centralized models in terms of accuracy, it frequently delivers comparable results, particularly for temperature and RH, while addressing critical concerns related to data privacy, communication constraints, and scalability. The effectiveness of FL depends on the combination of model architecture and optimization strategy. When these are appropriately selected, FL enables accurate forecasting across heterogeneous environments without requiring centralized access to sensitive data. These findings demonstrate the potential of FL to support distributed, real-time predictive modeling in historic buildings, offering a robust solution for smart maintenance.

\newpage

\begin{table}[!htb]
\centering
\caption{Prediction accuracy of different models on the test set under federated learning (Case 3)~\cite{ni_federated_2025}. For each model architecture, the best performance for each target variable is shown in bold, and the second-best is underlined. }
\label{tab:phd_federated_federated_learning}
\end{table}
\begin{figure}[!htb]
\includegraphics[width=\textwidth]{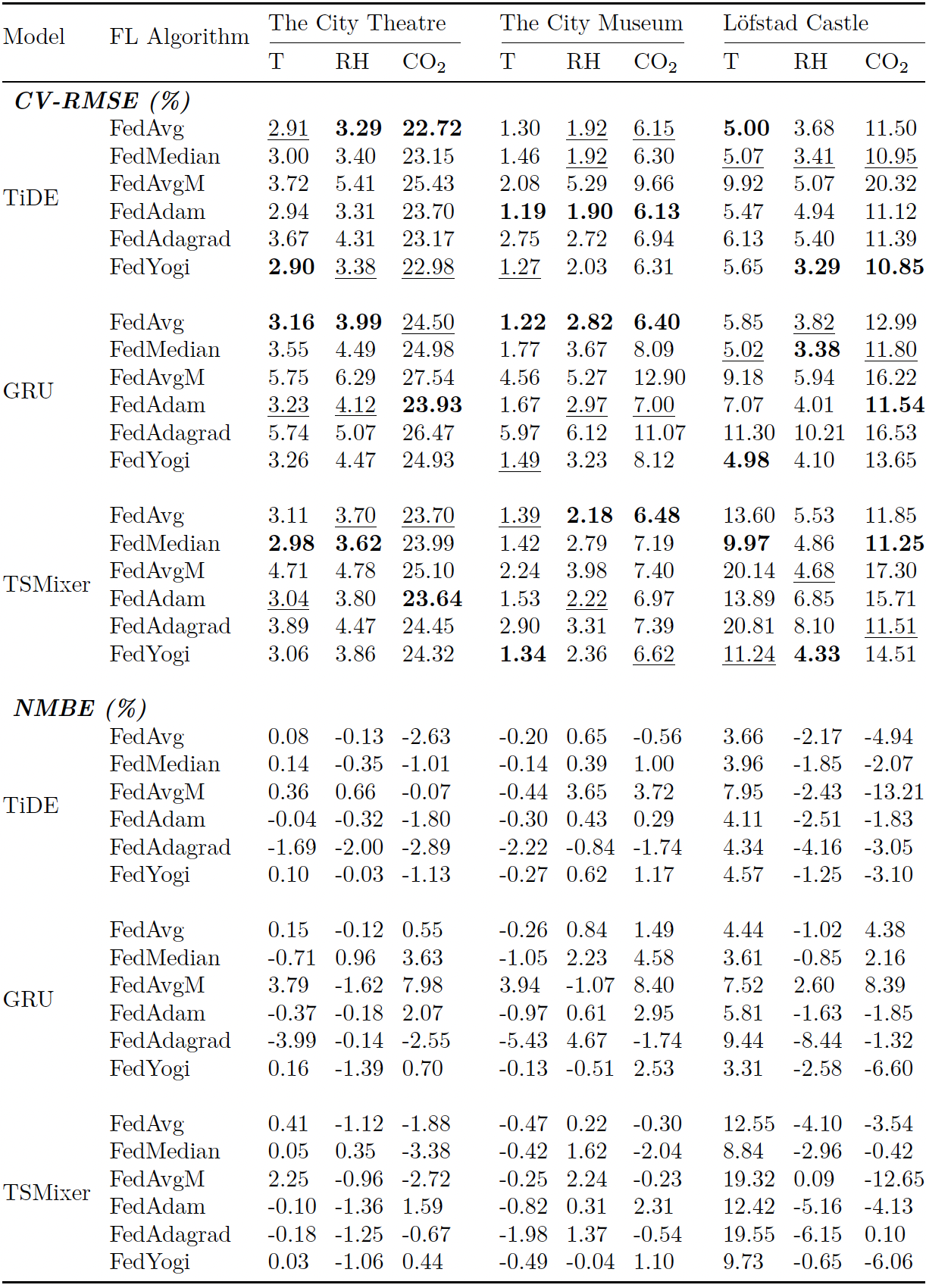}
\centering
\end{figure}

\chapter{Summary of Included Papers}
\label{cha:summary}

This chapter summarizes the included research papers and highlights the author's main contributions to each work.

\section{Paper I}

\textbf{Z. Ni}, Y. Liu, M. Karlsson, and S. Gong, “A Sensing System Based on Public Cloud to Monitor Indoor Environment of Historic Buildings,” \textit{Sensors}, vol. 21, no. 16, 2021.

\textbf{Summary}: This paper presents a cloud-based IoT system for real-time monitoring of indoor environmental conditions. The system was field-tested in three historic buildings in Norrköping, Sweden. Field tests demonstrated reliable data collection, transmission, and cloud storage. Preliminary analysis of RH fluctuations revealed potential risks to both the buildings and housed collections.

\textbf{The author's contributions}: I conducted the research and proposed an adaptable digitalization framework for preserving historic buildings. The framework was designed with a modular architecture to support scalability and future extension. I developed and implemented a sensing system aligned with this framework, using public cloud services, open-source software, and low-cost hardware. The system supports data collection, transmission, storage, and visualization, and is designed for future functional upgrades. For validation, I deployed the system in three historic buildings in Norrköping, Sweden, and carried out field testing to evaluate its long-term operational stability. I also conducted a preliminary indoor environmental analysis based on relevant industry standards to demonstrate the system’s functionality. I wrote the initial manuscript, performed data analysis and visualization, and revised the text in response to feedback from co-authors.

\section{Paper II}

\textbf{Z. Ni}, Y. Liu, M. Karlsson, and S. Gong, “Enabling Preventive Conservation of Historic Buildings Through Cloud-based Digital Twins: A Case Study in the City Theatre, Norrköping,” \textit{IEEE Access}, vol. 10, pp. 90924--90939, 2022.

\textbf{Summary}: This paper addresses the challenges of preventive conservation in historic buildings and the creation of digital twins that accurately reflect their dynamic conditions. To overcome these challenges, it proposes a solution that integrates IoT with ontology-based data modeling to create parametric digital twins. This approach ensures consistent data representation and enables real-time monitoring. The paper also presents a reference implementation using hardware, open-source software, and public cloud services to support replication and reuse in other heritage contexts. A case study at the City Theatre in Norrköping, Sweden, has demonstrated the functionality and benefits of the proposed approach.

\textbf{The author's contributions}: I conducted the study by conceptualizing a solution for consistent data representation and analysis in historic buildings. The proposed approach integrates IoT with ontology-based modeling to create digital twins that reflect real-time operating conditions and support advanced data analytics for preventive conservation. I designed and implemented a reference IoT system using hardware, open-source software libraries, and Microsoft Azure cloud services. This architecture reduces redundancy and enhances reproducibility, supporting transferability to other heritage buildings. To validate the approach, I conducted a case study at the City Theatre in Norrköping, Sweden, demonstrating the system's practical functionality and benefits. I developed a methodology to assess the impact of occupancy on the indoor climate, using both qualitative and quantitative analyses. I carried out data interpretation and visualization, integrated the findings into the results, and prepared the manuscript draft, incorporating revisions in response to co-authors' feedback.

\section{Paper III}

\textbf{Z. Ni}, J. Hupkes, P. Eriksson, G. Leijonhufvud, M. Karlsson, and S. Gong, “Parametric Digital Twins for Preserving Historic Buildings: A Case Study at Löfstad Castle in Östergötland, Sweden,” \textit{IEEE Access}, vol. 13, pp. 3371--3389, 2025.

\textbf{Summary}: This paper presents a comprehensive digitalization approach for Löfstad Castle in Sweden to support heritage conservation through continuous environmental monitoring and data-driven decision-making. Thirteen sensor boxes equipped with 84 sensors were deployed across the main building, enabling the collection of high-resolution environmental data. These data support the creation of a parametric digital twin that integrates contextual information and time series data using a combination of graphical and relational data models. An extended Brick ontology was used to ensure semantic consistency and interoperability across systems. Findings from the case study identified critical issues, such as high humidity in the basement and on the ground floor, leading to practical recommendations, including the installation of vapor barriers and improvements to heating strategies. The proposed solution provides a transferable and scalable framework for preventive conservation in other similar historic buildings.

\textbf{The author's contributions}: I conducted the research by developing a cloud-based solution for creating parametric digital twins to support the conservation of historic buildings. The proposed approach integrates graphical and relational data models to enable real-time monitoring and data-driven analysis of building operations. I implemented the full system and demonstrated its applicability through a comprehensive case study at Löfstad Castle, a historic building in Östergötland, Sweden. I deployed thirteen cloud-connected sensor boxes equipped with 84 sensors across all floors to continuously monitor indoor environmental parameters. I designed the methodology to analyze spatial and temporal patterns in temperature and humidity, assess the influence of high RH in the basement on upper floors, examine groundwater level variations across basement locations, and evaluate the impact of occupancy on indoor climate. Through formal data analysis and visualization, I identified the need for mitigation measures, including the installation of vapor barriers to address moisture-related risks. I wrote the initial manuscript and revised it according to comments from co-authors.

\section{Paper IV}

\textbf{Z. Ni}, C. Zhang, M. Karlsson, and S. Gong, “A study of deep learning-based multi-horizon building energy forecasting,” \textit{Energy and Buildings}, vol. 303, 2024.

\textbf{Summary}: This paper adapts and applies advanced deep learning methods to the task of multi-horizon building energy forecasting. Eight forecasting models, including seven deep learning architectures, namely N-HiTS, TCN, Transformer, NLinear, LSTM, GRU, and TFT, were developed and evaluated for two public historic buildings in Norrköping, Sweden. The buildings serve different functions and exhibit distinct energy use patterns. Both point and probabilistic forecasts were performed to assess model performance under varying operational conditions. The results show that incorporating future exogenous variables obviously improves forecasting accuracy. However, fluctuations in operating modes and activity schedules introduce uncertainty that reduces predictive performance. Among all tested models, the TFT model consistently achieved the best results for both point and probabilistic forecasting.

\textbf{The author's contributions}: I conceived and designed the study, formulating the research objectives and methodology. I adapted and applied seven deep learning architectures to the task of multi-horizon building energy forecasting and extended two of them to support probabilistic forecasting using quantile regression. To validate the models, I conducted a comprehensive case study in two public historic buildings with different operating modes. I performed data analysis and visualization to interpret the results and assess model performance. The findings offer insights for forecasting energy use in historic buildings with similar operational characteristics. I wrote the initial manuscript and revised it in response to feedback from co-authors.

\section{Paper V}

\textbf{Z. Ni}, M. Karlsson, and S. Gong, “A study of federated deep learning for building indoor climate forecasting,” \textit{Manuscript}, 2025.

\textbf{Summary}: This paper explores the use of federated deep learning for multi-horizon indoor climate forecasting in public historic buildings to balance predictive accuracy and data privacy. The study compares eight deep learning architectures and six FL algorithms across three learning paradigms: local, centralized, and federated. Using multi-year datasets from three historic buildings in Östergötland, Sweden, the models were trained to forecast indoor temperature, RH, and CO\textsubscript{2} concentration based on indoor environmental data and outdoor weather conditions. The results show that the TiDE and GRU architectures consistently delivered strong forecasting performance across all cases. Among the FL algorithms, FedAvg and FedMedian demonstrated the most robust aggregation performance. Compared to centralized learning, FL achieved comparable forecasting accuracy while offering stronger privacy protection. These findings demonstrate FL’s potential to handle data heterogeneity across multiple buildings and support scalable, privacy-preserving indoor climate forecasting for sustainable management of heritage buildings.

\textbf{The author's contributions}: I conceived and designed the study, defining the research objectives and methodological approach. I conducted a comprehensive evaluation of eight deep learning architectures and six FL algorithms across three learning cases: local, centralized, and federated learning. This analysis provided insights into the predictive performance of different architectures for indoor environmental forecasting across multiple buildings. For investigation and validation, I carried out a case study involving three public historic buildings with distinct operational characteristics and unbalanced data availability. I performed formal analysis through data interpretation and visualization to derive meaningful insights from the results. The findings underscore the potential of FL as a transformative approach for sustainable indoor climate management, linking theoretical development with practical application. I drafted the initial manuscript and revised it based on feedback from co-authors.

\section{Paper VI}

\textbf{Z. Ni}, C. Zhang, M. Karlsson, and S. Gong, “Leveraging Deep Learning and Digital Twins to Improve Energy Performance of Buildings,” in \textit{2023 IEEE 3rd International Conference on Industrial Electronics for Sustainable Energy Systems (IESES)}, 2023.

\textbf{Summary}: This paper presents a solution that integrates deep learning with digital twins to analyze building energy use and identify opportunities for optimization. An ontology-based approach was used to develop parametric digital twins, ensuring a consistent data structure across heterogeneous building systems. Deep learning techniques were applied for energy data analysis. A case study was conducted to evaluate the performance of five deep learning architectures in predicting energy use and quantifying uncertainty. The results show that these models effectively capture both the trends and uncertainties in building energy use. This approach can support facility managers in gaining deeper insights into energy consumption, enabling cost savings, enhancing occupant comfort, and promoting sustainability in the built environment.

\textbf{The author's contributions}: I conceived and designed the study, formulating the research objectives and methodological framework. I proposed a solution for analyzing building energy use and identifying optimization opportunities through predictive modeling. To investigate and validate the approach, I conducted a comprehensive case study assessing the performance of five deep learning methods for one-step-ahead energy forecasting and uncertainty quantification. I trained and evaluated the models, examining their predictive accuracy and computational efficiency. I carried out data interpretation and visualization to derive insights from the results. I drafted the initial manuscript and revised it according to feedback from co-authors. I also contributed to preparing the presentation materials for the conference at which the paper was presented.

\section{Paper VII}

\textbf{Z. Ni}, C. Zhang, M. Karlsson, and S. Gong, “Edge-based Parametric Digital Twins for Intelligent Building Indoor Climate Modeling,” in \textit{2024 IEEE 20th International Conference on Factory Communication Systems (WFCS)}, 2024.

\textbf{Summary}: This paper presents an edge-centric approach that combines digital twins and deep learning for intelligent indoor climate modeling in buildings. By deploying parametric digital twins and predictive models at the edge, the solution enables low-latency operation, improved data privacy, and reduced reliance on cloud infrastructure. The digital twin was developed using an ontology-based data schema to ensure consistent and interoperable representations of building systems and indoor environmental parameters. Deep learning techniques were employed to model complex temporal dynamics in sensor data, supporting accurate multi-horizon forecasts of indoor temperature and relative humidity. A case study conducted in a historic building in Östergötland, Sweden, has demonstrated the effectiveness of the proposed approach. Among the five deep learning architectures evaluated, the TiDE model achieved high forecasting accuracy with low computational cost. The findings confirm the feasibility of real-time, privacy-preserving indoor climate forecasting and lay the groundwork for integrating predictive control strategies in heritage building management.

\textbf{The author's contributions}: I conceived and designed the study, defining the research objectives and methodological framework. I proposed an integrated edge-centric solution for intelligent indoor climate modeling in buildings, offering an alternative to cloud-centric approaches. For investigation and validation, I conducted a comprehensive case study to assess the performance of five deep learning architectures. I was responsible for data preprocessing and the development of predictive models for multi-horizon indoor climate forecasting. I trained and evaluated the models, analyzing their forecasting accuracy and computational efficiency for inference on edge devices. I performed formal analysis through data interpretation and visualization to extract meaningful insights from the results. I drafted the initial manuscript and revised it according to comments from co-authors. I also contributed to preparing the presentation materials for the conference.

\chapter{Conclusion} 
\label{cha:conclusion}

This thesis set out to develop and validate data-driven approaches for smart maintenance of historic buildings. By integrating IoT, cloud and edge computing, ontology-based data modeling, deep learning, and FL, the research addresses the challenges in long-term preservation of historic buildings. The work was guided by four sub-goals: G1 design and implement an IoT system for long-term preservation of historic buildings; G2 develop parametric digital twins using ontology-based data modeling; G3 investigate deep learning techniques for building energy forecasting; and G4 enhance indoor climate forecasting by applying deep learning in combination with privacy-preserving methods. The findings presented in the seven appended papers collectively demonstrate the feasibility and effectiveness of data-driven approaches.

Paper I introduces a scalable and robust IoT sensing system, implemented using low-cost hardware, open-source software, and Microsoft Azure cloud services. Field tests in several public historic buildings in Sweden have demonstrated low end-to-end data loss and confirmed the suitability of the system for long-term monitoring. Paper II builds upon this system by integrating real-time data acquisition with ontology-based modeling to create a parametric digital twin that reflects a building’s operational status and supports data analytics and visualization. Paper III extends the validation through a detailed case study at Löfstad Castle, where 84 sensors were deployed across five floors. The results revealed key insights into moisture accumulation, air quality issues, and heating requirements. Together, these papers demonstrate the effectiveness of a flexible and adaptable digitalization solution suitable for diverse heritage contexts.

Paper IV evaluates several deep learning architectures for point and probabilistic forecasting of electricity use and heating load in historic buildings. The findings highlight the influence of operational variables, such as occupancy and activity schedules, on prediction accuracy. Models including TFT, LSTM, and GRU performed well, with TFT excelling in probabilistic forecasting. Paper VI integrates deep learning with parametric digital twins, showing how predictive analytics embedded within a unified data representation can support intelligent energy management. These studies confirm the applicability of deep learning for modeling complex energy dynamics in historic buildings and generating actionable insights for sustainable operation.

Paper V investigates eight deep learning models and six FL algorithms across three learning paradigms. The results show that FL, particularly with aggregation strategies such as FedAvg and FedMedian, can achieve forecasting accuracy comparable to centralized models while preserving data privacy. Paper VII proposes an edge-centric solution that deploys parametric digital twins and predictive models directly on edge devices, enabling low-latency analytics and improved data privacy. A case study confirmed that the TiDE model provided accurate multi-horizon forecasts with minimal computational cost. These contributions affirm the potential of FL and edge computing to enable decentralized, privacy-aware predictive analytics in heritage building contexts.

In conclusion, this thesis demonstrates the practical application of data-driven methods for the smart maintenance of historic buildings. The proposed solutions have been validated through real-world case studies across Sweden, covering diverse architectural styles, usage patterns, and conservation challenges. The outcomes offer practical tools and strategies for facility managers, conservation professionals, and researchers working toward the sustainable management and digital transformation of built heritage.

